\documentclass[aps,rmp,reprint,amsmath,amssymb,showkeys,floatfix]{revtex4-1}

\usepackage{amsmath}
\usepackage{amssymb}

\usepackage[T1]{fontenc} 
\usepackage{lmodern} 

\usepackage{graphicx}
\usepackage{algpseudocode}
\usepackage{algorithm}
\usepackage{hyperref}
\usepackage{color}

\newcommand*{\Figref}[1]{Fig.~\ref{#1}}
\renewcommand*{\eqref}[1]{Eq.~(\ref{#1})}

\newcommand*{\anglebrackets}[1]{\langle #1 \rangle}  

\newcommand*{\eqimage}[2]{\vcenter{\hbox{\includegraphics[height=#2]{#1}}}}
\newcommand*{\picbracket}[1]{\Bigg\langle\vcenter{\hbox{\includegraphics[height=8ex]{#1}}}\Bigg\rangle}
\newcommand*{\picroundbracket}[1]{\Bigg(\vcenter{\hbox{\includegraphics[height=8ex]{#1}}}\Bigg)}
\newcommand*{\abs}[1]{\left\vert #1 \right\vert}
\newcommand*{\bra}[1]{\langle #1 \vert}
\newcommand*{\ket}[1]{\vert #1 \rangle}
\renewcommand{\Re}{\operatorname{Re}} 
\renewcommand{\Im}{\operatorname{Im}}

\begin{document}
\title{Introduction to topological quantum computation with non-Abelian anyons}
\date{\today}
\author{Bernard Field}
\author{Tapio Simula}
\affiliation{School of Physics and Astronomy, Monash University, Victoria 3800, Australia}

\begin{abstract}
	Topological quantum computers promise a fault tolerant means to perform quantum computation. Topological quantum computers use particles with exotic exchange statistics called non-Abelian anyons, and the simplest anyon model which allows for universal quantum computation by particle exchange or braiding alone is the Fibonacci anyon model. One classically hard problem that can be solved efficiently using quantum computation is finding the value of the Jones polynomial of knots at roots of unity. We aim to provide a pedagogical, self-contained, review of topological quantum computation with Fibonacci anyons, from the braiding statistics and matrices to the layout of such a computer and the compiling of braids to perform specific operations. Then we use a simulation of a topological quantum computer to explicitly demonstrate a quantum computation using Fibonacci anyons, evaluating the Jones polynomial of a selection of simple knots. In addition to simulating a modular circuit-style quantum algorithm, we also show how the magnitude of the Jones polynomial at specific points could be obtained exactly using Fibonacci or Ising anyons. Such an exact algorithm seems ideally suited for a proof of concept demonstration of a topological quantum computer.
\end{abstract}

\keywords{Aharonov-Jones-Landau algorithm; Kauffman bracket polynomial; braid; Fibonacci anyons; fusion; Hadamard test; Ising anyons; Jones polynomial; knot; link; Majorana zero mode; non-Abelian vortex; quantum circuit; quantum computer; quantum dimension; superfluid; topological quantum computing; topological qubit}

\maketitle
\tableofcontents
\section{Introduction}

The exponential growth observed over the past decades in information processing capacity of digital computers, and as quantified by Moore's law, is unsustainable and will eventually be complemented or surpassed by quantum technologies \cite{Milburn1996a,Kauffman2007a}. Quantum computing is a field of much interest because it promises to outperform regular, classical computing for many otherwise intractable problems. While classical computers perform Boolean operations on a register of bits, quantum computers perform unitary operations on an exponentially large vector space, typically composed from many quantum bits, or qubits \cite{Quantum_Textbook,Quantum_Basics,Galindo2002a}. Using this exponentially large computation space, it is possible, at least in principle, for quantum computers to efficiently solve classically difficult problems such as prime factorisation of large numbers  \cite{Shor} or the simulation of complex quantum systems \cite{Feynman1982,Quantum_Simulators}.

Another example of a classically hard algorithm, which can benefit from quantum computation, is the determination of the Jones polynomial of knots \cite{Jones1985a}. The Jones polynomial is a knot invariant with connections to topological quantum field theory  \cite{witten1989,Freedman_TQC_Jones_Polynomial} and other knot-like systems. It is also, in general, exponentially difficult to compute by classical means. However, a quantum algorithm developed by Aharonov, Jones and Landau (AJL)  \cite{Aharonov_Jones_Algorithm1} can be used to efficiently estimate the value of the Jones polynomial at the roots of unity, by first reducing the problem to finding the diagonal elements of the product of certain matrices. The resource of nonclassical correlations required in such evaluation of the Jones polynomial \cite{Shor2008a} may be quantified by quantum discord \cite{Zurek2003a,Datta2008a,Datta2011a,Modi2012a}.

Most implementations of a quantum computer are highly susceptible to errors. A major source of error in quantum computation is decoherence, caused by interactions between the quantum state and the environment, which causes uncontrolled randomness in the system \cite{Zurek2003a,Pachos_TQC_Book}.  Local perturbations can also cause errors in many quantum systems, as can imperfections in the execution of quantum operations \cite{Preskill_Fault_Tolerant}. This results in notable overheads devoted to error correction schemes, which only work in computers with a sufficiently low basic error rate, which makes implementing such a quantum computer very difficult.

One way to mitigate the effect of these errors is in using topological quantum computing \cite{Kitaev_Anyon_Computing,Freedman_TQC_Jones_Polynomial,Collins2006a,Wang2010a,Pachos_TQC_Book,Stanescu2017a,Anyon_Computing_Nayak}. In contrast to locally encoding information and computation using, for example, the spin of an electron \cite{Kane1998a,Loss1998a,Reilly2008a,Castelvecchi2018a}, the energy levels of an ion \cite{Cirac1995a,Blatt2003a}, optical modes containing one photon \cite{Milburn2001a}, or superconducting Josephson junctions \cite{Shnirman1997a}, topological quantum computers encode information using global, topological properties of a quantum system, which are resilient to local perturbations \cite{Kitaev_Anyon_Computing,Anyon_Computing_Nayak,Pachos2014a,Bombin2008a,Bombin2011a}. These topological quantum computers can be implemented using non-Abelian anyons, which are quasiparticles in two-dimensional systems which exhibit exotic exchange statistics, beyond a simple phase change \cite{Pachos_TQC_Book}. Considering the anyons in 2+1 dimensions (where the third dimension is time), the motion of these anyons traces worldlines in this 2+1 dimensional space, and exchanging the anyons results in braiding the worldlines \cite{Anyon_Computing_Brennen}. Exchanging non-Abelian anyons results in a unitary operation determined solely by the topology of this braid, and for certain models of anyon, such as the Fibonacci model, it is possible to reproduce any unitary operation to arbitrary accuracy by choosing the right braid to perform, making them universal for quantum computation \cite{Anyon_Computing_Nayak,Preskill_Lecture}. Because the operations are determined by topology alone, they are far more resistant to decoherence and errors. This makes topological quantum computers an area of significant interest and investment \cite{SciAm2006a,MicrosoftTQCNews}.

In the case of topological quantum computers made from Fibonacci anyons, compiling more useful operations from the elementary braiding operations available with Fibonacci anyons \cite{Bonesteel_Braid_Topologies,Bonesteel_Fibonacci_General,Bonesteel_Weave,Carnahan_Diagonal_Weaves,Vadym_Compiling,Improved_Injection_Weave,Hormozi_Compiling,No_Exact_Braids}, and testing of various error correction codes for Fibonacci anyon-based quantum computers \cite{ErrorCorrectionThesis,Burton_Fibonacci_Error,Wootton_Error_Correction}, as well as simulation of the physics involved with Fibonacci anyons \cite{Fibonacci_Simulation_Physics} have been investigated. There has also been considerable study into candidate physical systems which could contain non-Abelian anyons. Most notable candidate for finding Fibonacci anyons is the fractional quantum Hall effect at $\nu=12/5$ \cite{Anyon_Computing_Nayak,Anyon_Intro_Trebst,Anyon_Computing_Brennen,Anyons_Hall_Effect,Sarma_TQC,Fibonacci_Anyons_Hall1,RezayiRead1,Fibonacci_Wavefunctions,Anyon_Interferometry,Fibonacci_Hall_Size}, although other candidates exist \cite{Anyon_Computing_Brennen,BEC_Anyons,Fibonacci_Anyons_1D,Fibonacci_Net}. Meanwhile, significant effort is directed toward finding Ising anyons in nanowires hosting Majorana zero modes \cite{Zhang2018a,Sarma2015a,Alicea2012a}

In this work, we have explicitly carried out a quantum algorithm, specifically the AJL algorithm, by simulating the braiding of Fibonacci anyons. In doing so, we have demonstrated from first principles how Fibonacci anyons can be used for quantum computation, and provided an explicit recipe for the actions that would need to be performed on a system of Fibonacci anyons to perform such computations. We have also presented and performed an exact algorithm, which demonstrates the direct connection between Fibonacci and Ising anyons and the value of the Jones polynomial at a specific point.

In Section \ref{sec:topology}, we review the relevant components of knot theory and topology, including the definition of knots (Sec.~\ref{sec:knots}), braids (Sec.~\ref{sec:braids}) and the Jones polynomial (Sec.~\ref{sec:invariants}). Section \ref{sec:quantum_basics} provides a brief review of conventional quantum computation. In Section \ref{sec:quantum_computing}, we cover the theoretical basis for the Fibonacci anyon topological computer starting with a discussion on Fibonacci anyons (Sec.~\ref{sec:anyons}), followed by the derivation of the elementary braiding matrices (Sec.~\ref{sec:fib_braid}) and an explanation of how we can perform quantum computation with Fibonacci anyons (Sec.~\ref{sec:fib_compute}). Section \ref{sec:compile} illustrates how braids which approximate desired operations can be formed. Section \ref{sec:algorithm} covers the details of the AJL algorithm, including the Hadamard test (Sec.~\ref{sec:hadamard}) that can be performed on a quantum computer. Section \ref{sec:exact_ajl} contains a discussion on how non-Abelian anyons could be used to exactly calculate the magnitude of the Jones polynomial. Intermediate results demonstrating the rate of convergence of braids approximating matrices and the Hadamard test are presented in Sec.~\ref{sec:braid_convergence} and Sec.~\ref{sec:hadamard_convergence}, respectively. Finally, our simulation of the topological quantum computer is presented in Section \ref{sec:simulation}. We also provide a qualitative summary of the main points of this work in Section \ref{sec:overview} for ease of reference.

\section{Overview} \label{sec:overview}

\subsection{Principles of Topological Quantum Computation}

A quantum computer uses the principles of quantum mechanics to manipulate a quantum state in such a way as to perform a useful computation. A topological quantum computer uses quantum states which are encoded by the topology of the system rather than in any local properties.

\begin{figure}
	\includegraphics[width=\linewidth]{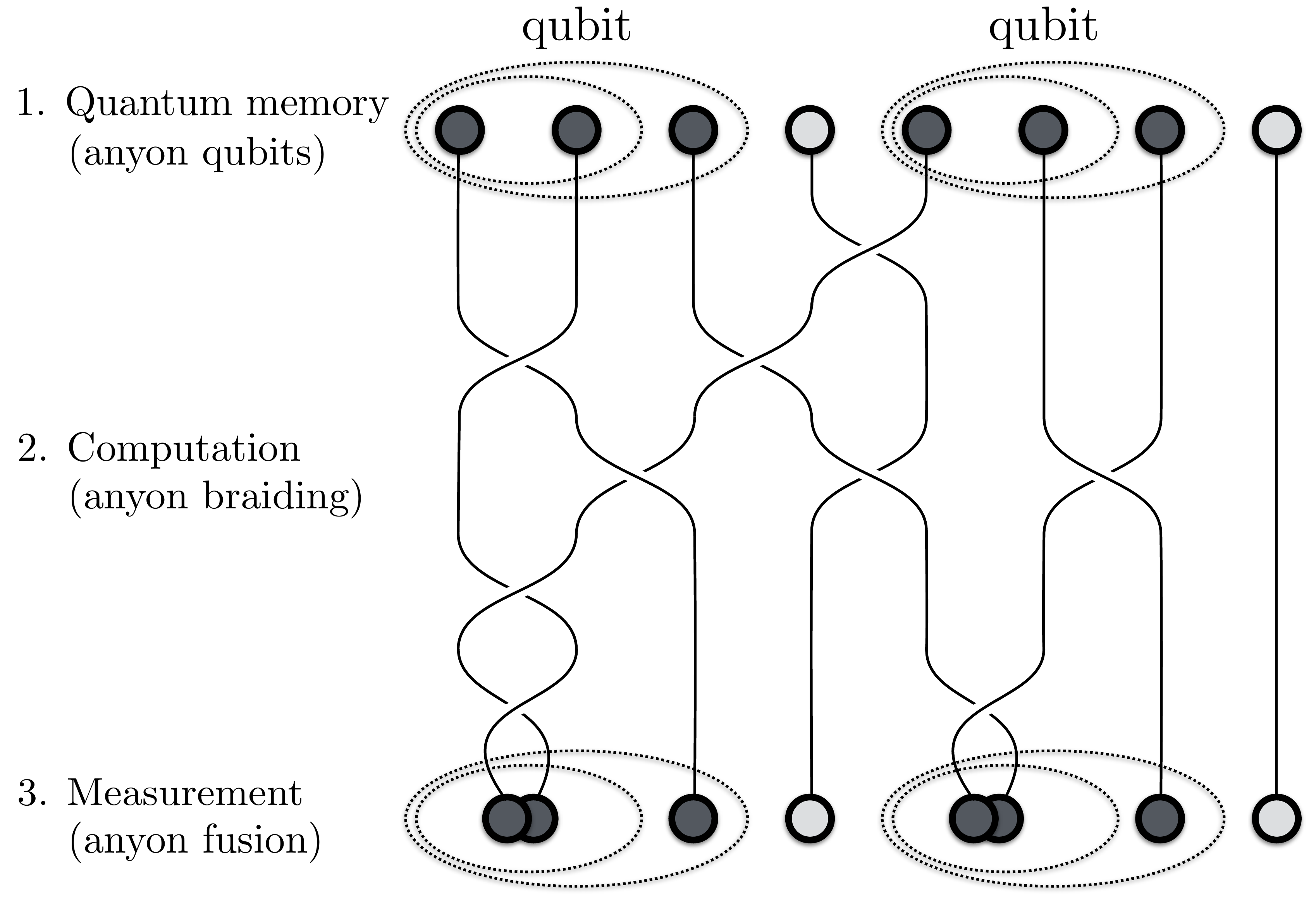}
	\caption{A demonstration of braiding anyons in a topological quantum computer. Time points downwards in this diagram. This computer has two qubits composed of four anyons each, where the ellipses group the anyons into qubits. Some braiding is performed with the anyons, then the anyons are fused to measure the state of the qubits. The light grey, inert, anyons do not participate in any non-trivial braiding, and could potentially be deployed for error correction.}
	\label{Fig:overview_tqc}
\end{figure}

There are three fundamental steps in performing a topological quantum computation, illustrated in \Figref{Fig:overview_tqc}.

\begin{enumerate}
	\item Creating qubits from non-Abelian anyons.
	\item Moving the anyons around---`braiding' them---to perform a computation.
	\item Measuring the state of the anyons by fusion.
\end{enumerate}

Each of these steps is discussed in further detail below.

\subsubsection{Non-Abelian Anyons and Qubits}

Anyons are a type of particle which can exist in two-dimensional quantum systems \cite{Wilczek1982a}. When two anyons are exchanged, the states of those particles may be subjected to an arbitrary phase shift (for Abelian anyons) or even a unitary operation (for non-Abelian anyons) \cite{Anyon_Computing_Brennen,Pachos_TQC_Book}. This is unlike the bosons and fermions which constitute regular three-dimensional particles, where the particle states undergo a multiplication by 1 or $-1$, respectively, upon particle exchange. For non-Abelian anyons, exchanging of particles can perform significant changes to the state of the system, which can be used to perform quantum computation.

The state of a system of anyons is defined by the anyons produced by fusing those anyons together, with each possible set of fusion outcomes representing one state in the Hilbert space of the quantum system of anyons. The dimension of this Hilbert space, or the number of different possible fusion outcomes, grows by a factor called the quantum dimension when more anyons are added, on average and in the limit of many anyons. For Abelian anyons, because each fusion gives a definite outcome, the quantum dimension is 1, because adding more anyons does not add more possible fusion outcomes. Non-Abelian anyons have a quantum dimension greater than 1. The quantum dimension does not need to be an integer, or even rational number \cite{Anyon_Intro_Trebst}.

A qubit is a quantum system which can be in two possible states, and forms the basic unit of most quantum computers \cite{Quantum_Basics}. Multiple qubits are brought together to form a register of qubits. For topological quantum computers, each qubit is composed of a number of anyons. In the Fibonacci model, a qubit can be constructed from four Fibonacci anyons, Fig.~\ref{Fig:overview_tqc}, with zero net overall `charge' or `spin' (i.e. the four anyons will annihilate when all of them are fused) \cite{Anyon_Computing_Brennen}. As such, the first step in performing a topological quantum computation is to create anyons to form a register of qubits.

For the sake of concreteness, we focus on the model of Fibonacci anyons. However, the concepts explored are directly applicable to generic non-Abelian anyon models.

\subsubsection{Braiding Anyons}

Exchanging two non-Abelian anyons performs a unitary operation on the quantum state, which can change the relative phases and probability densities of the basis states corresponding to each fusion outcome.

The anyons exist in two-dimensional space. Consider a 2+1 dimensional space, where the third dimension is time. The worldlines that thread through the time dimension as the anyons move around each other are strands which are braided, as in \Figref{Fig:overview_tqc}. Hence, exchanging anyons is referred to as braiding, because the operation braids their worldlines. Furthermore, the operation performed on the quantum state is dependent solely on the topology of the braid, meaning that the braid can be stretched and deformed in almost any manner but still perform the same operation. This topological robustness provides the key advantage of topological quantum computers over other quantum computers, which is tolerance to errors from local perturbations \cite{Anyon_Computing_Nayak,Kitaev_Anyon_Computing}.

By braiding anyons within a qubit, the probabilities of the fusion outcomes within that qubit can be changed. This puts the qubits into a superposition of states. By braiding anyons between two qubits, the states of the qubits in general become dependent on each other, such that it is not possible to measure the state of one qubit without affecting the other qubit. Thus performing a braid which literally entangles two qubits will also induce quantum entanglement between those two qubits.

Before performing any braiding, it is essential to know what braids are necessary to perform the desired operation. In quantum computation, the quantum algorithms are composed of several quantum gates, which each enact a predetermined operation. It is necessary to determine what braid enacts the required gates to within a desired accuracy, and this is performed using classical computation with a combination of exhaustive search  \cite{Bonesteel_Braid_Topologies} and iterative methods  \cite{Kitaev_Theorem,Solovay_Kitaev_Explained,Vadym_Compiling,Compiling_Hashing}. Once the braid corresponding to a given gate has been determined, that braid can be recorded for later use during quantum computation. Here we rely on the simpler exhaustive search method, which is adequate for first-order approximations of a small number of quantum gates.

\subsubsection{Measuring Anyons}

After the computation is complete, it is necessary to measure the state of the system. This is performed by fusing two of the anyons in each qubit and observing the outcome of each fusion. Each set of fusion outcomes corresponds to a unique basis state \cite{Pachos_TQC_Book}. Because the anyons are a quantum system, the probability of each set of fusion outcomes is determined by the amplitudes of the basis states in the quantum system.

The state of the system after the braiding encodes the result of the computation. However, the full state cannot be measured directly. As such, it is often necessary to perform repeated identical computations and measurements to statistically determine the probability distribution and thus the state of the system. However, due to the embarrassing parallelism of such repeated measurements, this task can be completed efficiently and simultaneously by deploying multiple copies of the same system.

\subsubsection{Physical Realization}

A variety of physical systems have been suggested for implementing topological quantum computation using non-Abelian anyons \cite{Anyon_Computing_Nayak,Sarma2015a}. Hence, complementing the generic but abstract notion of anyons, braiding their worldlines, and their eventual fusion as illustrated in Fig.~\ref{Fig:overview_tqc}, it may be useful to have a concrete mental picture of the physical entities and processes comprising such a topological quantum computer. For this purpose, we may choose to consider the anyons to be (quasiparticles associated with) vortices nucleated in a quasi-two-dimensional superfluid. Such vortices are the quantum mechanical counterpart to the familiar bathtub vortices and are ubiquitous in quantum liquids including superfluid helium-4 \cite{Yarmchuk1979a,Fonda2014a}, superfluid helium-3 \cite{Hakonen1982a,Autti2016a}, superconductors \cite{Essmann1967a,Abrikosov2004a}, Bose--Einstein condensates \cite{Matthews1999a,Madison2000a,Abo-Shaeer2001a,Fetter2009a} and superfluid Fermi gases \cite{Zwierlein2015a}. The particular types of non-Abelian anyons that may be realised depend on the physical details of the vortices. For example, in chiral $p$-wave Fermi systems the vortices may host Majorana zero modes \cite{Volovik1999a,Gurarie2007a,Mizushima2008a}, the topological properties of which correspond to the Majorana zero modes found in solid state systems leading to Ising anyons \cite{Sarma2015a,Zhang2018a}.

In a Bose--Einstein condensate, vortex-antivortex pairs can be spawned from vacuum controllably using e.g. steerable laser beams \cite{Samson2016a}. Similar techniques could be developed for preparing non-Abelian vortex anyons in spinor Bose--Einstein condensates or chiral $p$-wave Fermi gases to initialise the topological qubits.

Vortices in quantum gases such as Bose-Einstein condensates can be pinned using focused laser beams \cite{Tung2006a,Simula2008a,Samson2016a} and using optical tweezer technology positions of individual optical pinning sites can be controllably steered \cite{Roberts2014a,Samson2016a} to move vortices around adiabatically \cite{Simula2018a,Virtanen2001a}. When such vortices are braided, their worldlines trace out three-dimensional vortex structures familiar from, e.g., studies of quantum turbulence \cite{Barenghi2014a}, propagating singular optical fields \cite{Dennis2010a,Tempone2016a,Taylor2016a}, fluid knots \cite{Kleckner2013a}, and electron vortices \cite{Petersen2013a}. 

\begin{figure}
	\includegraphics[width=\linewidth]{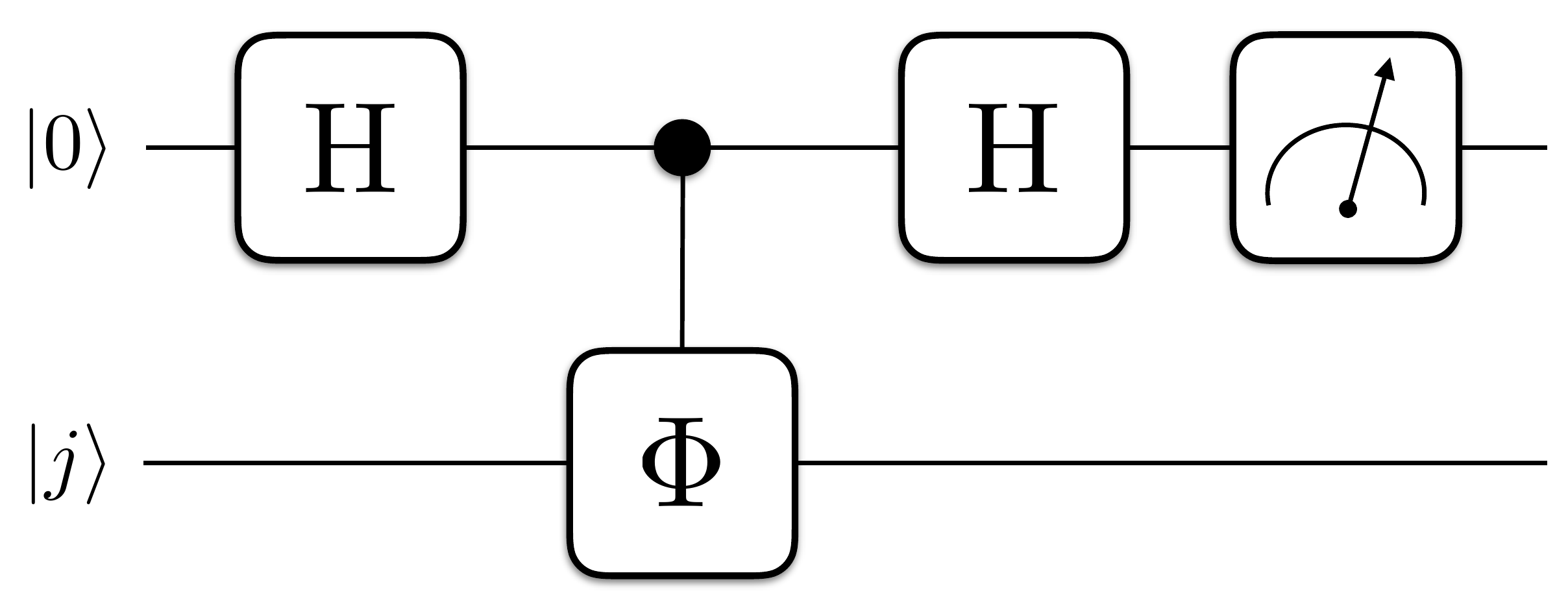}
	\caption{Quantum circuit diagram for the Hadamard test for evaluating the real component of a matrix element, as required by the AJL algorithm. The Hadamard gate is denoted by $H$ and the gate $\Phi$ is a controlled gate, where the circle marks the control qubit and the box is the operation acting on the target qubit, with this operation specific to the knot being investigated by the AJL algorithm.}
	\label{Fig:overview_quantum_circuit}
\end{figure}

Fusion of vortices in quantum gases could be achieved by simply overlapping the optical pinning potentials, closing the worldlines such that the vortices will either annihilate or form another topological defect. From the topological quantum computation perspective the most important aspect of such vortices is that they must be governed by non-Abelian exchange statistics. For this purpose spinor Bose-Einstein condensates \cite{Kawaguchi2012a,Kurn2013a} seem to be a promising platform for searching non-Abelian vortex anyons \cite{Mawson2018a}. Many such high-spin Bose-Einstein condensates, including rubidium \cite{Kurn2013a}, chromium \cite{Griesmaier2005a}, erbium \cite{Aikawa2012a}, strontium \cite{Stellmer2009a}, ytterbium \cite{Takasu2003a} and dysprosium \cite{Lu2011a,Lian2012a} atoms have already been produced. Such spinor Bose-Einstein condensates may host non-Abelian fractional vortices \cite{Semenoff2007a,Kobayashi2009a,Huhtamaki2009a,Kobayashi2016a,Borgh2016a,Mawson2017a} whose topological invariants  \cite{Mermin1979a,Thouless1998a} are described by finite non-Abelian symmetry groups. Notwithstanding the finiteness of their underlying symmetry groups, such condensates may possess experimentally realizable ground states with non-Abelian vortex anyons with the capacity to be harnessed for topological quantum computation. Indeed, such vortices in spinor Bose--Einstein condensates have been proposed for realisation of a variety of non-Abelian anyon models  \cite{Mawson2018a}.

\subsection{Simulation of a Topological Quantum Computer}

We have simulated a topological quantum computer by performing matrix multiplication in MatLab, where each matrix corresponds to an elementary braiding operation of the anyons. State measurement was performed in this simulated quantum computer by multiplying the overall braid matrix with a vector, then using that vector to construct a probability distribution, from which the measured state of the qubits was randomly selected.

\begin{figure}
	\includegraphics[width=\linewidth]{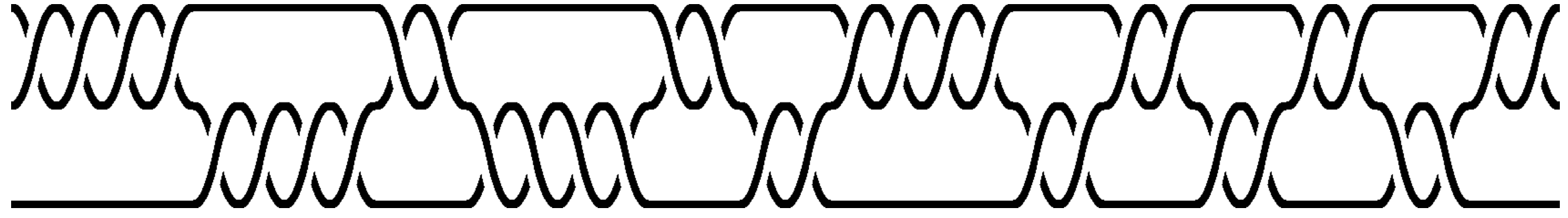}
	\caption{A weave of non-Abelian anyons approximating the Hadamard gate, ${\rm H}=\tfrac{1}{\sqrt{2}}\left(\begin{smallmatrix}1&1\\1&-1\end{smallmatrix}\right)$, with an error of 0.003. Time points to the right in this diagram.}
	\label{Fig:overview_hadamard_weave}
\end{figure}

With this simulator we performed the Aharonov Jones Landau (AJL) algorithm  \cite{Aharonov_Jones_Algorithm1} for approximating the Jones polynomial at the complex roots of unity ($e^{2\pi i/k}$). The Jones polynomial is a knot invariant, which can be used to distinguish inequivalent knots. The AJL algorithm involves quantum circuits such as those shown in \Figref{Fig:overview_quantum_circuit}. Implementing the algorithm in a topological quantum computer required finding braids such as that in \Figref{Fig:overview_hadamard_weave} and constructing controlled operations by the method described by  \cite{Bonesteel_Braid_Topologies}.

\begin{figure}
	\includegraphics[width=\linewidth]{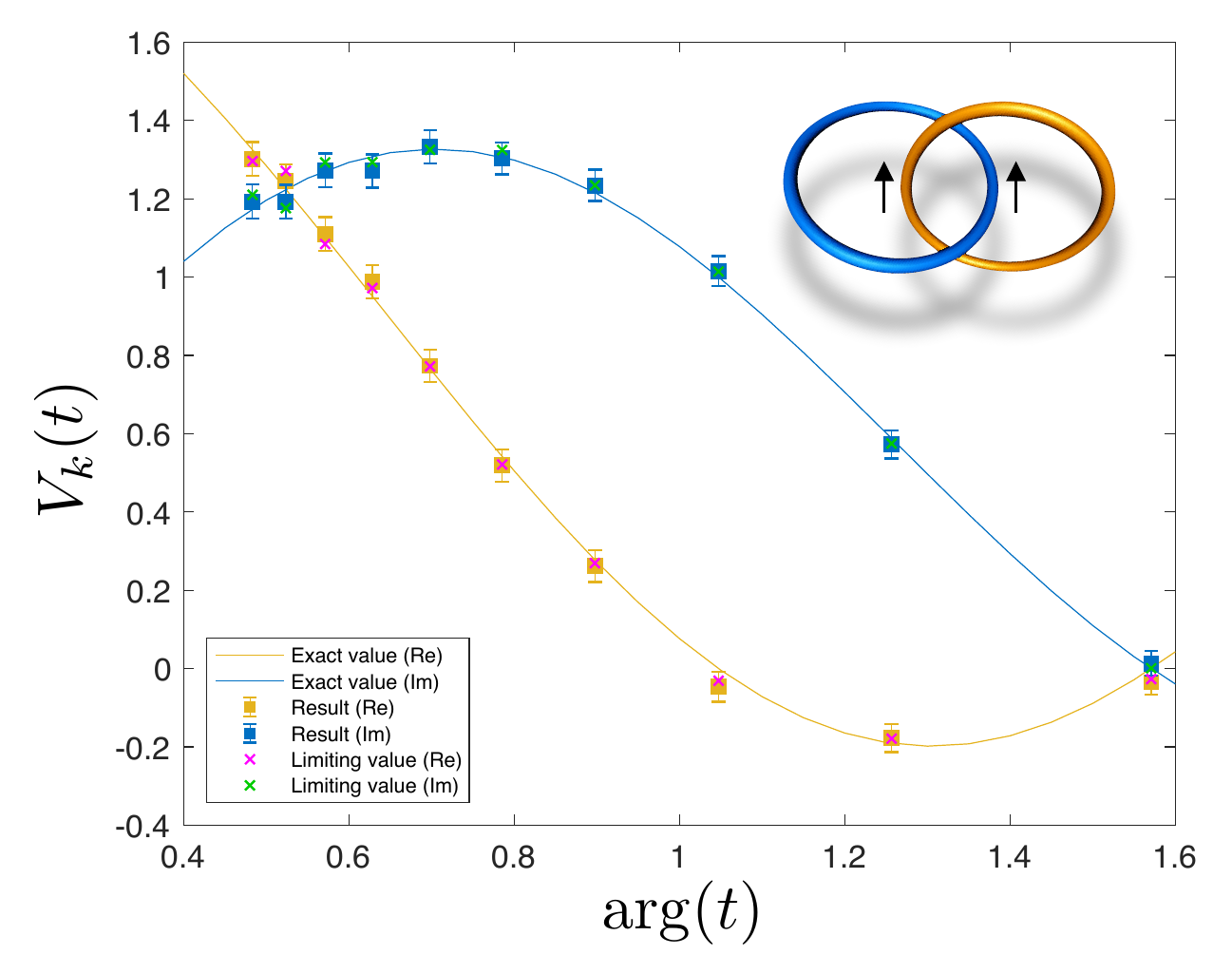}
	\caption{Results from the determination of the Jones polynomial $V_k(t)$ of the positive Hopf link. The horizontal axis shows the complex phase of the point $t$ where the Jones polynomial is being evaluated. Square markers with error bars are the results obtained stochastically from the Hadamard test. Real and imaginary components are evaluated separately. The limiting values of these stochastic measurements are also shown, see the legend.}
	\label{Fig:overview_jones_polynomial}
\end{figure}

By compiling weaves and performing the AJL algorithm in our simulated quantum computer, we were able to approximate the Jones polynomial of simple knots at the complex roots of unity, as shown in \Figref{Fig:overview_jones_polynomial}. Evaluations of the Jones polynomial to this precision took on the order of $10^8$ elementary braiding operations for these simple knots. The time complexity of these evaluations in the quantum computer with respect to the desired error $\epsilon$, measured by the number of elementary braiding operation required, scales as $\mathcal{O}((1/\epsilon)^2 \log(1/\epsilon))$.

In a demonstration of the connection between topological quantum field theories and the Jones polynomial, we showed that if the knot under investigation was directly copied by the braided worldlines of Fibonacci anyons, then the probability of measuring the initial state is simply the magnitude of the Jones polynomial at the point $t=e^{2\pi i/5}$ times the quantum dimension to a simple power. An identical result holds for Ising anyons at the point $t=e^{2\pi i/4}=i$, and we conjecture that similar results hold for a countably infinite set of anyon models. This method is facile and involves no approximations, but is limited to obtaining the magnitude of the Jones polynomial at a fixed point. Nevertheless, it facilitates a straightforward proof of concept demonstration of a topological quantum computer.

We note, however, that the elementary methods used here to evaluate the Jones polynomials of knots can be used to simulate any quantum algorithm in a universal topological quantum computer.

\section{Topology and Knot Theory} \label{sec:topology}

\subsection{Knots} \label{sec:knots}

Formally, a knot is a closed loop embedded in three-dimensional space. Intuitively, a knot may be understood as a tangled loop of string or rope. Knots are studied within the field of topology, meaning that we are permitted to stretch, deform and move this loop in a continuous manner, without cutting the loop or allowing it to intersect itself. This type of transformation is referred to as ambient isotopy. If two knots are isotopic to one another, that is, one knot can be deformed into the other, then those two knots are equivalent. Otherwise, the knots are inequivalent.

Knots can be generalised to being made from multiple closed loops. Knots containing multiple loops are called links. All the theory which applies to knots can also be applied to links. In this work we will often use the terms knot and link interchangeably. Knots and links may also be oriented, meaning that their curves have an associated direction.

\begin{figure}
	\includegraphics[width=\linewidth]{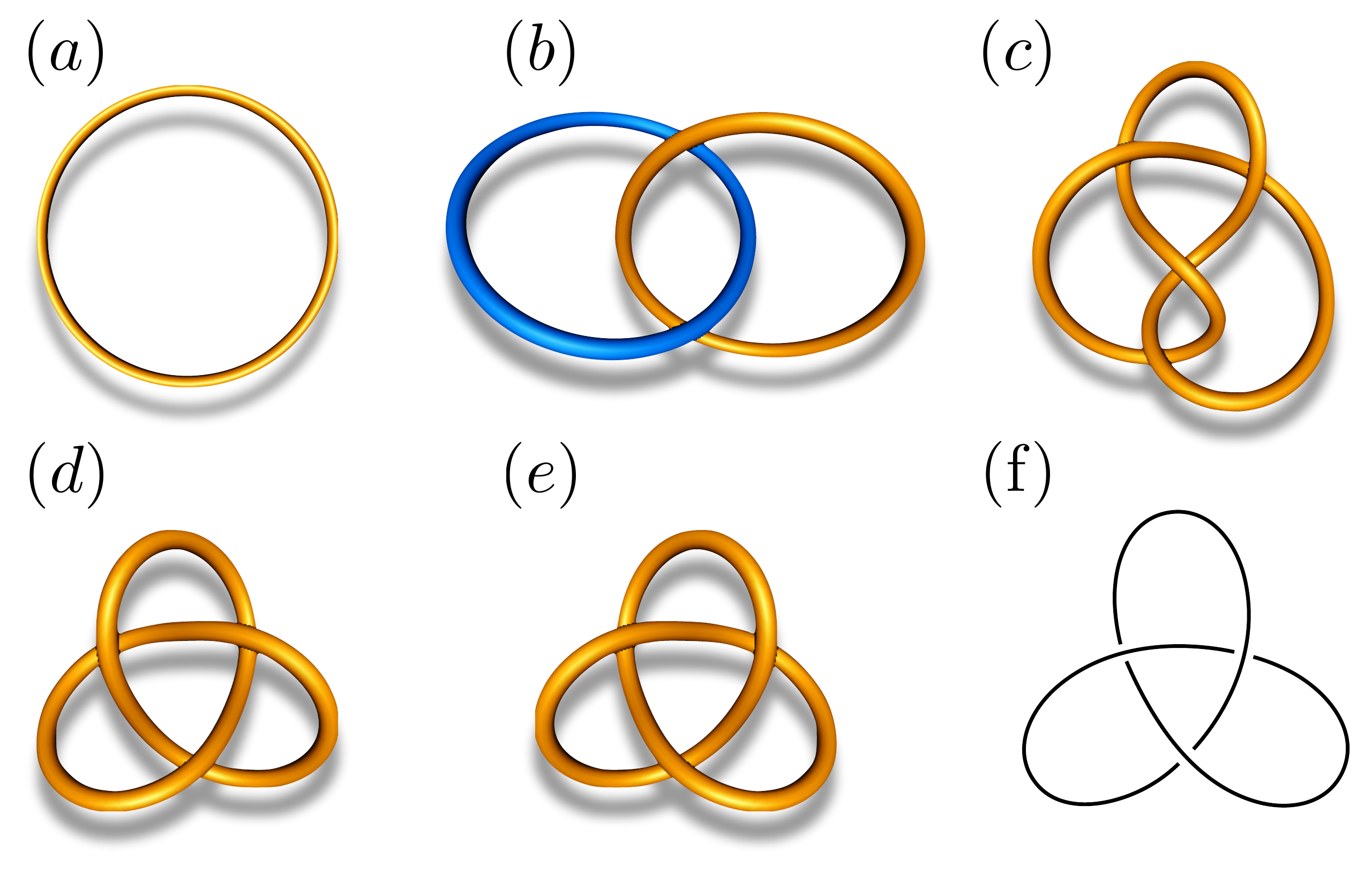}
	\caption{Pictures of (a) the unknot, (b) the Hopf link, (c) the figure-eight knot, (d) and (e), respectively, the left and right trefoils, and (f) knot diagram of the right trefoil.}
	\label{Fig:knots1}
\end{figure}

\begin{figure}
	\includegraphics[width=0.8\linewidth]{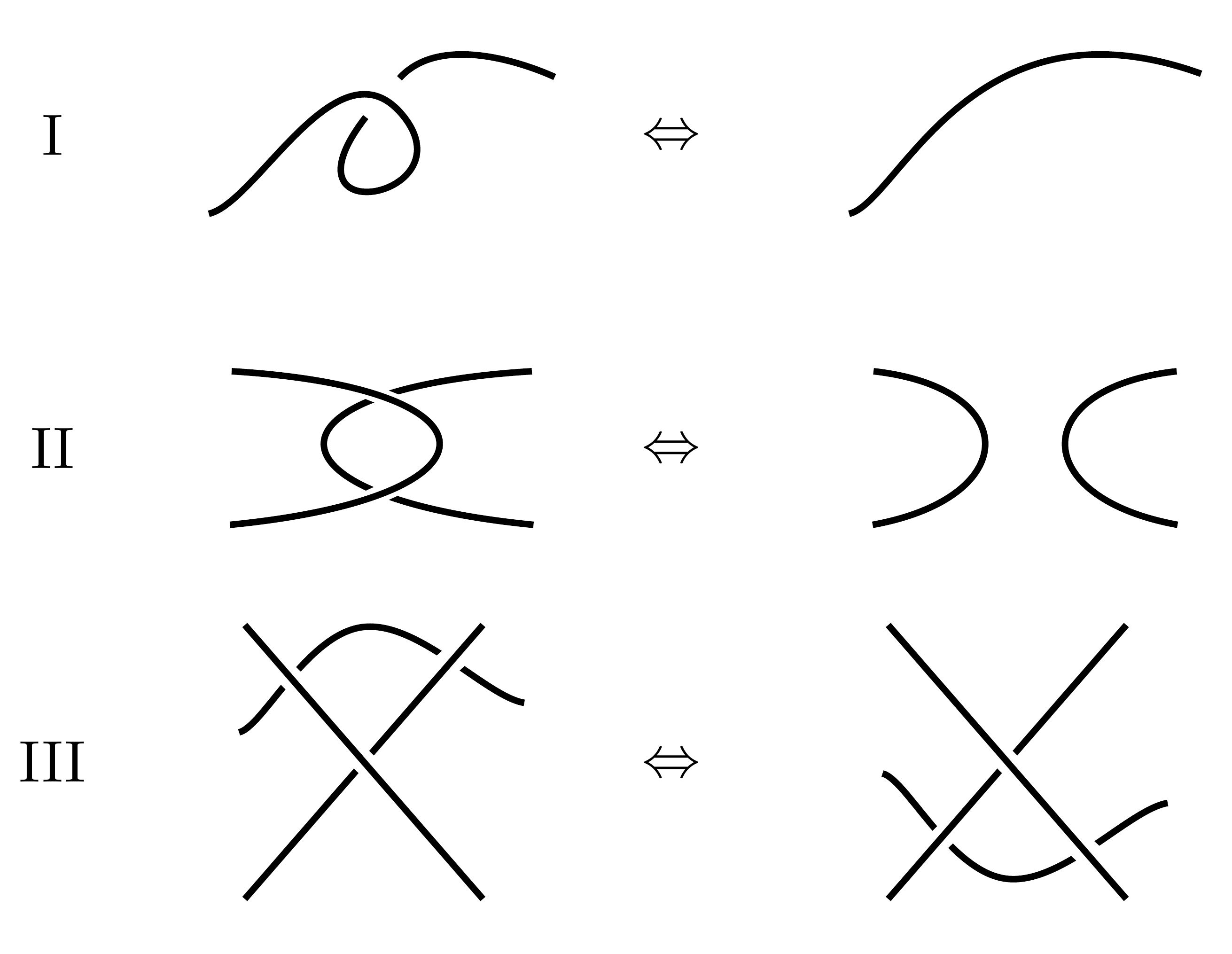}
	\caption{Representations of the three Reidemeister moves.}
	\label{Fig:reidemeister}
\end{figure}

Figure \ref{Fig:knots1} shows pictures of a few simple knots and links. Figure \ref{Fig:knots1}(f) is a knot diagram of the knot in \Figref{Fig:knots1}(e). Knot diagrams are a projection of knots, which are three-dimensional objects, into a two-dimensional drawing. At each intersection on the diagram, the crossing is marked to indicate which arc is above the other in 3D space. The Reidemeister moves, illustrated in \Figref{Fig:reidemeister}, can be used to manipulate a knot diagram while maintaining ambient isotopy. If one knot diagram can be transformed into another via a finite number of Reidemeister moves, then those two knots are equivalent \cite{Knot_Majorana_Kauffman}. Not listed is planar isotopy, where the diagram can be stretched and deformed provided no crossings are modified. Planar isotopy is typically assumed to always be allowed.

Figure \ref{Fig:knots1}(a) shows the simplest possible knot, the unknot. It is regarded as a trivial case (although determining whether an arbitrary knot is equivalent to the unknot can be non-trivial). The simplest link is the unlink, which consists of multiple disjoint unknots. It is also a trivial case. The simplest non-trivial link is the Hopf link, in \Figref{Fig:knots1}(b), and is made of two simple loops, which intersect each other. The simplest non-trivial knot is the trefoil, in \Figref{Fig:knots1}(d) and \Figref{Fig:knots1}(e). The trefoil is chiral, meaning that it is not equivalent to its mirror image. The next simplest knot is the figure-eight knot, in \Figref{Fig:knots1}(c). These simple knots and links will form the test cases for the algorithms explored in this work.

\subsection{Knot Invariants} \label{sec:invariants}

A knot invariant is any quantity associated with a knot or its diagrams which does not change under ambient isotopy. A knot invariant calculated from a knot diagram is unchanged by performing the Reidemeister moves on the knot diagram. One of the uses for invariants is to distinguish between inequivalent knots. If two knot diagrams have different values for an invariant, they must be inequivalent. Note, however, that the converse is not true; two inequivalent knots might have the same value for a particular invariant. Better invariants are better able to distinguish between inequivalent knots.

The Jones polynomial is a particularly powerful knot invariant. It has connections to topological quantum field theory  \cite{witten1989,Freedman_TQC_Jones_Polynomial} and statistical mechanics \cite{Knot_Majorana_Kauffman}. One way to define the Jones polynomial is using the Kauffman bracket polynomial.

The Kauffman bracket polynomial $\anglebrackets{K}$ of a knot or link $K$ is computed from the knot diagram of $K$. A recursive relationship called a skein relation, illustrated in \eqref{eqn:Bracket_Rule_1}, is applied locally at each crossing, until the knot diagram has been reduced to a linear superposition of unlinks\footnote{The skein relationship in \eqref{eqn:Bracket_Rule_1} is the convention used in  \cite{Knot_Majorana_Kauffman,Aharonov_Jones_Algorithm1,Anyon_Computing_Brennen}. However, some sources such as  \cite{Pachos_TQC_Book} have $A$ and $A^{-1}$ swapped around. In such a case, the resulting polynomials have the signs of the powers reversed, and \eqref{eqn:Jones1} would need to be modified to use $-A^{-3}$ instead.}. 
Disjoint unknots are then removed via \eqref{eqn:Bracket_Rule_3}, effectively counting the number of unknots. This is done until the base case of the unknot ($O$) is reached, which has a bracket polynomial of one, as per \eqref{eqn:Bracket_Rule_2}. The end result is a polynomial with the variable $A$. Equations (\ref{eqn:Bracket_Rule_1})-(\ref{eqn:Bracket_Rule_2}) are listed below:
\begin{gather}
\picbracket{"Bracket_Crossing"} = A \picbracket{"Bracket_Horizontal_Smooth"} + A^{-1} \picbracket{"Bracket_Vertical_Smooth"} \label{eqn:Bracket_Rule_1} \\
\anglebrackets{K \sqcup O} = -(A^2+A^{-2})\:\anglebrackets{K} = d\:\anglebrackets{K} \label{eqn:Bracket_Rule_3} \\
\anglebrackets{O}=1 \label{eqn:Bracket_Rule_2}.
\end{gather}

The Kauffman bracket polynomial is invariant under Reidemeister moves II and III. However, it is not invariant under Reidemeister move I, instead obtaining the relationships in \eqref{eqn:Bracket_R1}:
\begin{gather}
\begin{split}
\picbracket{"Bracket_R1_right"} = -A^3 \picbracket{"Bracket_Line"} \\
\picbracket{"Bracket_R1_left"} = -A^{-3} \picbracket{"Bracket_Line"}.
\label{eqn:Bracket_R1}
\end{split}
\end{gather}
As such, the Kauffman bracket polynomial is not an invariant.

\begin{figure}[h!]
\includegraphics[height=10ex]{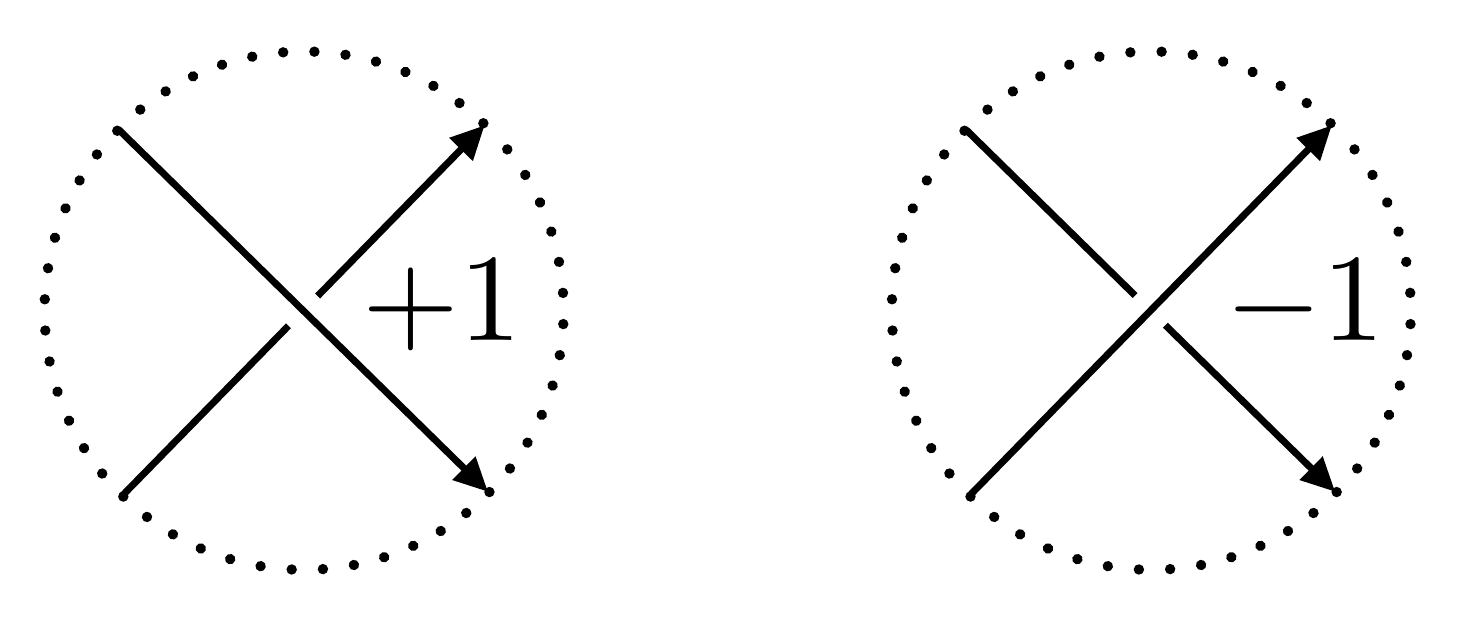}
	\caption{Rules for the contribution to the writhe from each oriented crossing.}
	\label{Fig:writhe}
\end{figure}

Another property which can be calculated, corresponding to oriented knots and links, is the writhe $w(K)$. The writhe assigns a value of $+1$ or $-1$ to each crossing, depending on the orientation of the arcs involved in the crossing, as in \Figref{Fig:writhe}, and the value of the writhe is the sum over all the crossings in a knot diagram.

The writhe is also invariant under Reidemeister moves II and III but varies by $\pm1$ under Reidemeister move I. As such, we can use the writhe to define a normalised version of the Kauffman bracket polynomial which is invariant;

\begin{equation}
f_K(A) = (-A^3)^{-w(K)} \: \anglebrackets{K}. \label{eqn:Jones1}
\end{equation}

This polynomial is a knot invariant. If we make the substitution $A=t^{-1/4}$, then we obtain
\begin{equation}
V_K(t) = (-t^{-3/4})^{-w(K)} \: \anglebrackets{K}, \label{eqn:Jones2}
\end{equation}
which is the Jones polynomial \cite{Knot_Majorana_Kauffman}.

The writhe is simple to calculate, requiring only a linear sum over the crossings. As such, the complexity of computing the Jones polynomial is due to the Kauffman bracket polynomial. The recursive relationship \eqref{eqn:Bracket_Rule_1} results in the complexity of this algorithm scaling by $\mathcal{O}(2^n)$, where $n$ is the number of crossings. As shown in \eqref{eqn:hopf_bracket_derive}, 
\begin{widetext}
\begin{equation}
\begin{split}
&\picbracket{"Hopf_Link"}\\
=& A\picbracket{"Hopf_Step1a"} + A^{-1}\picbracket{"Hopf_Step1b"} \\
=& A \left(A\picbracket{"Hopf_Step2aa"} + A^{-1}\picbracket{"Hopf_Step2ab"} \right) + A^{-1} \left(A\picbracket{"Hopf_Step2ba"} + A^{-1}\picbracket{"Hopf_Step2bb"} \right) \\
=& A \left( A(-A^2-A^{-2}) + A^{-1}(1) \right) + A^{-1} \left( A(1) + A^{-1}(-A^2-A^{-2}) \right) \\
=& A^2(-A^2-A^{-2}) + 1 + 1 + A^{-2}(-A^2-A^{-2}) \\
=& -A^4 - A^{-4}		\label{eqn:hopf_bracket_derive}
\end{split}
\end{equation}
\end{widetext}
the Kauffman bracket polynomial of the simple two-crossing Hopf link expands from one term to four. In the general case, it is not possible to efficiently compute the Jones polynomial, or even approximate it at a point, using a classical computer \cite{Aharonov_Jones_Algorithm1,Freedman_TQC_Jones_Polynomial,Pachos_TQC_Book}. As such, a more efficient quantum algorithm is desirable.

For reference, here we also compute the Jones polynomial of the Hopf link. The writhe of the Hopf link depends on the orientation of the Hopf link:
\begin{align}
w\picroundbracket{"Hopf_Orient_Same"} &= +2 \label{eqn:p_hopf_writhe} \\
w\picroundbracket{"Hopf_Orient_Different"} &= -2.
\end{align}

We shall consider the positive Hopf link, with positive writhe given by \eqref{eqn:p_hopf_writhe}. Using the relationship \eqref{eqn:Jones2} and the Kauffman bracket polynomial \eqref{eqn:hopf_bracket_derive}, its Jones polynomial is then

\begin{equation}
V\picroundbracket{"Hopf_Orient_Same"}(t) = -t^{5/2} - t^{1/2}. \label{eqn:hopf_jones}
\end{equation}

The Jones polynomial of the negative Hopf link is
\begin{equation}
V\picroundbracket{"Hopf_Orient_Different"}(t) = -t^{-5/2} - t^{-1/2}. \label{eqn:hopf_jones_negative}
\end{equation}

We will state that the Kauffman bracket polynomial of the left trefoil is
\begin{equation}
\picbracket{"Left_Trefoil"} = A^{7} - A^{3} - A^{-5}, \label{eqn:Ltrefoil_bracket}
\end{equation}
and the left trefoil has a writhe of $-3$, so it has a Jones polynomial of 
\begin{equation}
V\picroundbracket{"Left_Trefoil"} = -t^{-4} + t^{-3}+t^{-1}. \label{eqn:Ltrefoil_jones}
\end{equation}
The Kauffman bracket polynomial of the right trefoil is
\begin{equation}
\picbracket{"Right_Trefoil"} = A^{-7} - A^{-3} - A^{5}, \label{eqn:Rtrefoil_bracket}
\end{equation}
and the right trefoil has a writhe of $+3$, so it has a Jones polynomial of 
\begin{equation}
V\picroundbracket{"Right_Trefoil"} = -t^{4} + t^{3}+t^{1}. \label{eqn:Rtrefoil_jones}
\end{equation}
Since \eqref{eqn:Ltrefoil_jones} and \eqref{eqn:Rtrefoil_jones} are different, this demonstrates that the left and right trefoils are inequivalent.

The Kauffman bracket polynomial of the figure-eight knot is
\begin{equation}
\picbracket{"Figure_Eight_Knot"} = A^8 - A^4 + 1 - A^{-4} + A^{-8}, \label{eqn:eight_bracket}
\end{equation}
and the figure-eight knot has a writhe of zero, so it has a Jones polynomial of
\begin{equation}
V\picroundbracket{"Figure_Eight_Knot"} = t^2 - t + 1 - t^{-1} + t^{-2}. \label{eqn:eight_jones}
\end{equation}

\subsection{Braids and Closures} \label{sec:braids}

A class of objects related to knots are braids. A braid has $n$ parallel strands, with several twists or crossings. At either end of the braid is a line of $n$ pegs (typically not drawn), each with exactly one strand attached. Each crossing involves two adjacent strands crossing over each other and exchanging places. The strands in the braid always travel in one direction, never looping back on themselves or disappearing. Up to planar isotopy, a braid can be fully defined by a linear sequence of crossings, making them mathematically easy to represent.

\begin{figure}
	\includegraphics[width=\linewidth]{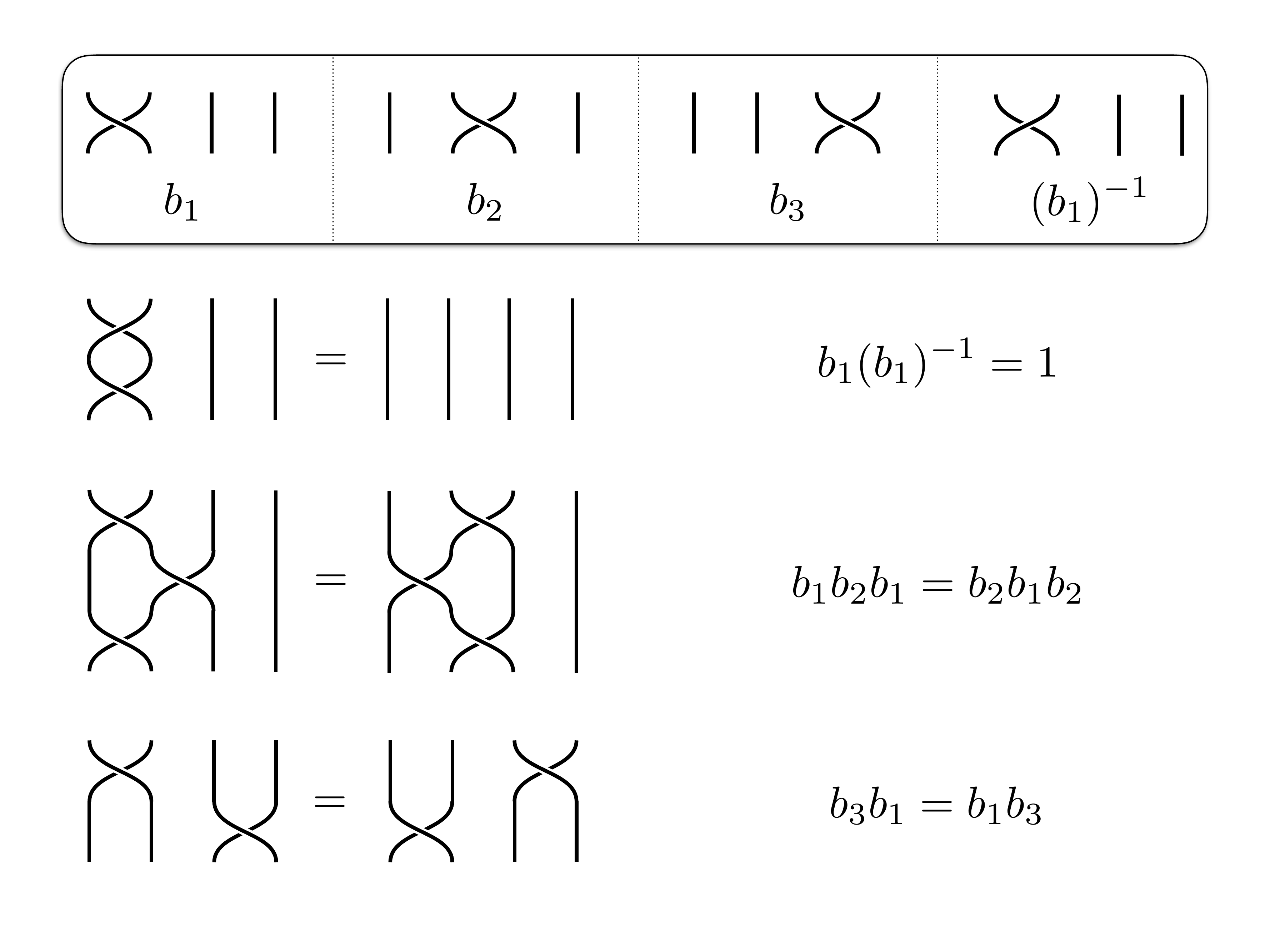}
	\caption{Diagram depicting the braid generators for $B_4$, and demonstrations of the three identities for braid generators \eqref{eqn:braid_identities}.}
	\label{Fig:braid_generators}
\end{figure}

To represent braids algebraically, we define the Artin braid group $B_n$ as the set of all braids with $n$ strands \cite{Knot_Majorana_Kauffman}. The generators of $B_n$ are the identity $\mathbb{I}$, the elementary braids for a crossing in one direction of the $i$'th and $i+1$'th strands $b_i$ for $1\le i < n$, and their inverses for crossings in the opposite direction $b_i^{-1}$ \cite{Anyon_Computing_Nayak}. By taking a product of these non-commutative generators, we can obtain any braid in $B_n$. The sequence of elementary braid operations ($b_i$ and $b_i^{-1}$) defining a braid is called the braidword. When reading a braidword from left to right, the braid is built from bottom to top \cite{Aharonov_Jones_Algorithm1}.\footnote{This is not the only convention for the braid group. Some sources define braiding to go in the opposite direction.} Examples of a few braids and braid generators are provided in \Figref{Fig:braid_generators}. Since a braid can be represented by its braidword, this makes it straightforward to perform algorithms and computations on braids.

Just as knots have the Reidemeister moves, the braid group has the identities \eqref{eqn:braid_identities}. The first two equations are equivalent to Reidemeister moves II and III respectively, whereas the third is equivalent to planar isotopy and is often called far-commutativity.

\begin{equation}
\begin{aligned}
b_i b_i^{-1} &= \mathbb{I} \\
b_i b_{i+1} b_i &= b_{i+1} b_i b_{i+1} \\
b_i b_j &= b_j b_i \text{ if } \abs{i-j} \geq 2 	\label{eqn:braid_identities}
\end{aligned}
\end{equation}

\begin{figure}
	\includegraphics[width=0.7\linewidth]{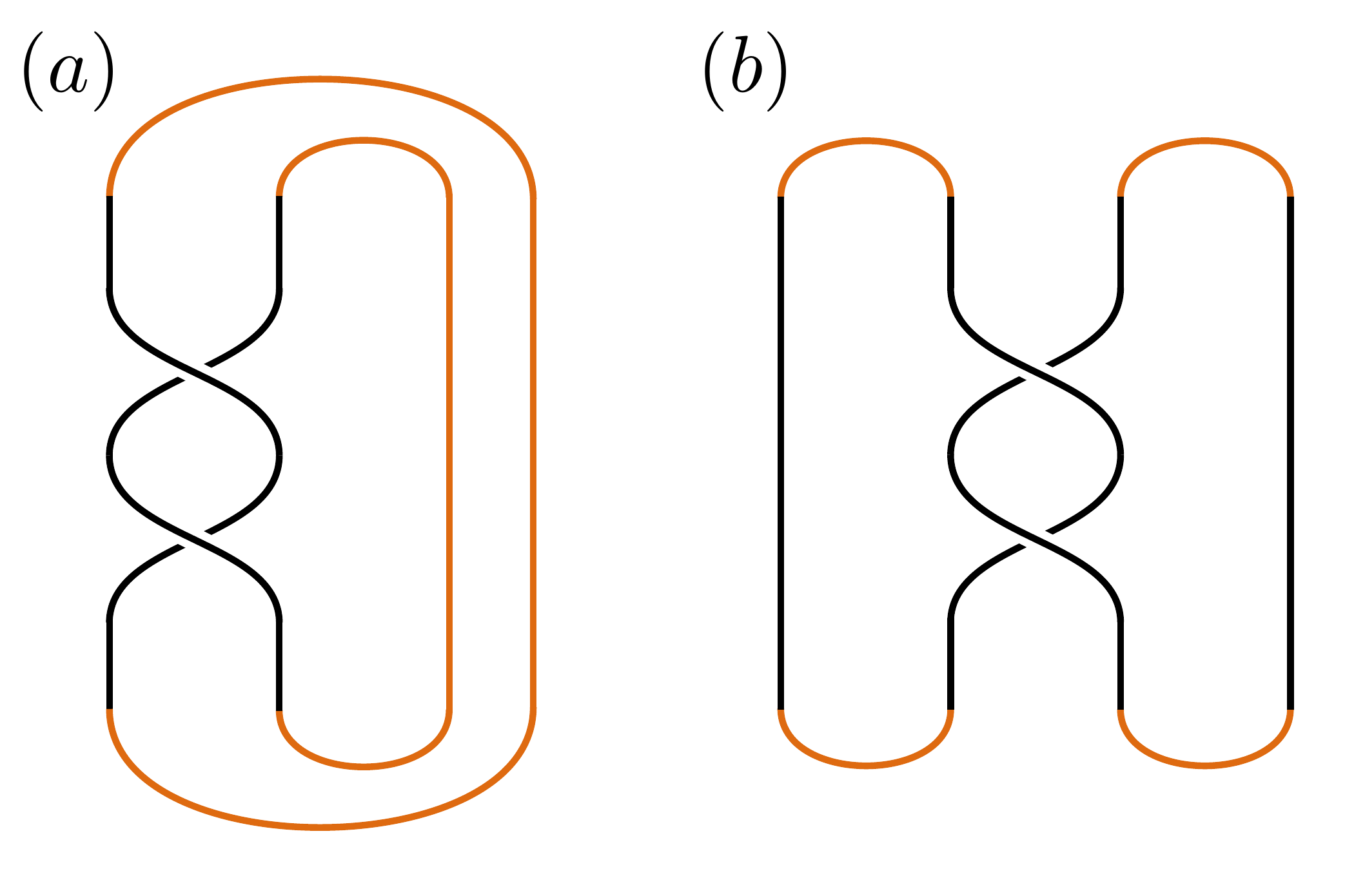}
	\caption{Braid closures yielding the Hopf link. (a) is the trace closure, which is a member of $B_2$ with the braidword $b_1 b_1$. (b) is the plat closure, which is a member of $B_4$ with the braidword $b_2 b_2$. The braid itself is shown in black. The closure is shown in orange.}
	\label{Fig:hopf_braids}
\end{figure}

\begin{figure}
	\includegraphics[width=\linewidth]{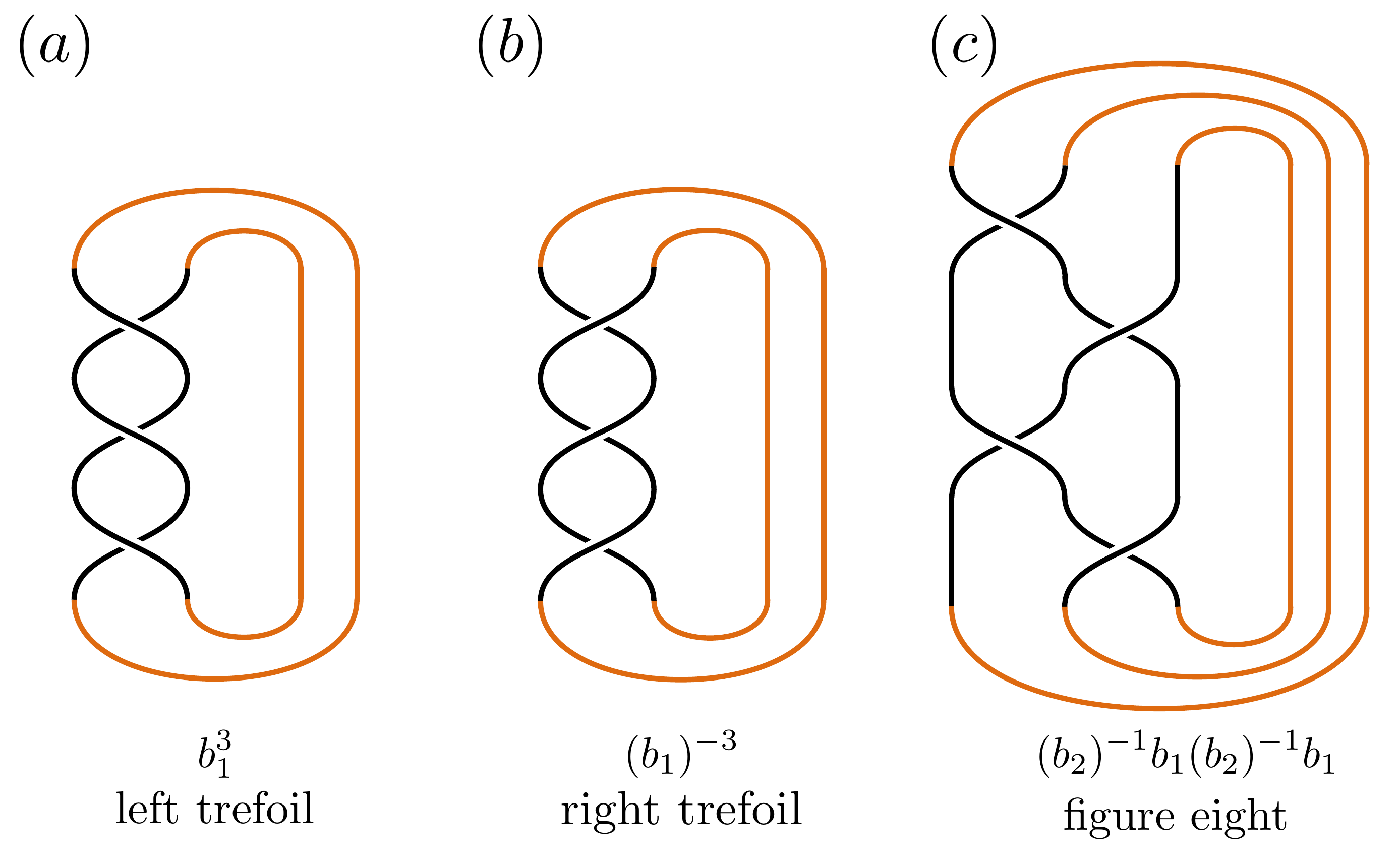}
	\caption{Trace closures and braidwords for braids corresponding to the left and right trefoils and the figure-eight knot.}
	\label{Fig:knot_braids}
\end{figure}

\begin{figure}
	\includegraphics[width=\linewidth]{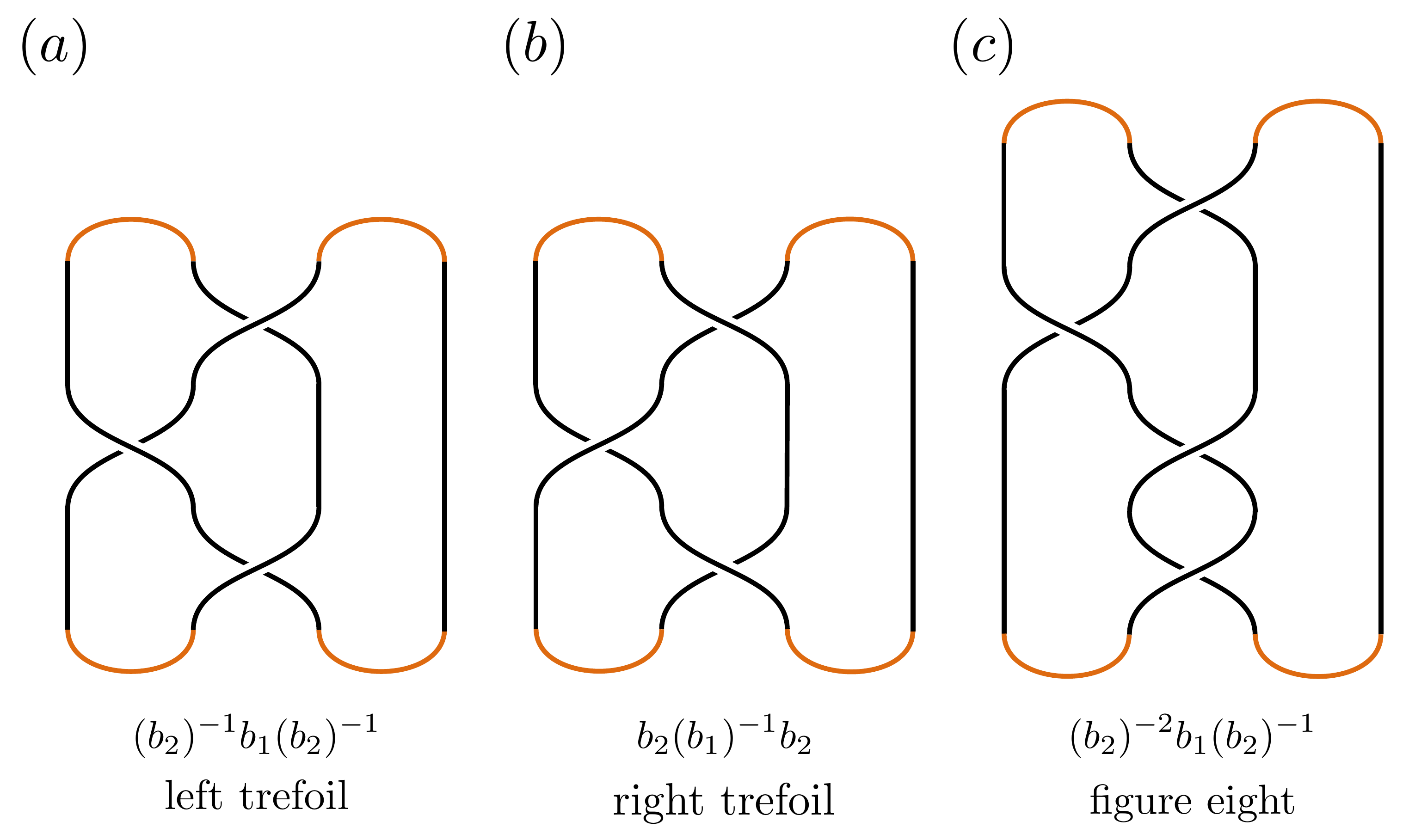}
	\caption{Plat closures and braidwords for braids corresponding to the left and right trefoils and the figure-eight knot.}
	\label{Fig:knot_plat_braids}
\end{figure}

To convert a braid $B$ into a knot or link, it is necessary to close the ends of the braid. There are two common conventions for this: a trace closure and a plat closure. The trace closure connects each peg along the top to the corresponding peg along the bottom, and forms a link denoted $B^{\rm tr}$. The plat closure connects adjacent pegs on the same side, and forms a link denoted as $B^{\rm pl}$ \cite{Aharonov_Jones_Algorithm1}. Note that the plat closure requires an even number of strands. Note also that the knots formed by these two closures for the same braid are in general different. Examples of the closures of braids which yield the Hopf link are shown in \Figref{Fig:hopf_braids}, and closures of braids corresponding to the trefoil and figure-eight knot are shown in \Figref{Fig:knot_braids} and \Figref{Fig:knot_plat_braids}. The process can be reversed to obtain a braid from a knot. In fact, by Alexander's theorem \cite{Alexander93}, every knot and link can be represented as the closure of a braid \cite{Lomonaco_Jones_Algorithm}.

For the purposes of oriented knots, for the trace closure we can consider the braid as having all its strands oriented in the same direction. Note that each element of the braidword has an exponent of $\pm1$. Thus, by comparing the braid elements in \Figref{Fig:braid_generators} to the writhe rule in \Figref{Fig:writhe}, we can state that the writhe of the trace closure of a braid is simply the negative of the sum of the braidword's exponents.

For the plat closure of a braid, it will be necessary to follow each individual strand to determine the orientation at each crossing and the writhe.

\section{Basics of Conventional Quantum Computation} \label{sec:quantum_basics}

Classical computation uses Boolean logic to manipulate ensembles of bits, each of which may be in either the 0 or the 1 state. Quantum computation, in contrast, greatly expands the available computation space by using quantum mechanics to allow the system to be in a superposition of states and even for those states to be entangled, such that there is no classical analogue for the state \cite{Quantum_Basics}.

Quantum computing in the circuit model involves initialising the computer to a state, evolving that state in a way which produces a useful computation, then measuring the resultant quantum state \cite{Quantum_Textbook}. This probabilistic process often needs to be repeated to obtain an average.

A general purpose quantum computer should satisfy the DiVincenzo criteria \cite{DiVincenzo2000a}: 
\begin{itemize}
\item[1.] \emph{A scalable physical system with well characterized qubits}: if the qubits are not uniquely definable entities they cannot be manipulated either and if the computable state space cannot be made sufficiently large, the quantum computer could be outperformed by classical digital computers
\item[2.]  \emph{The ability to initialize the state of the qubits to a simple fiducial state}: for the measured outcome of the quantum computation to be meaningful, the initial state prior to any computational procedures must also be known
\item[3.]  \emph{Long relevant decoherence times}: quantum states are fragile and must not become irreparably broken during the lifetime of the computation 
\item[4.]  \emph{A `universal' set of quantum gates}: a general purpose quantum computer must not be constrained by the types of gate operations it can perform and must be able to implement an arbitrary unitary transformation on the initial state 
\item[5.]  \emph{A qubit-specific measurement capability}: if the state of the qubits, regardless of their physical implementation, cannot be read, determining the outcome of the computation is not possible even if computation could be performed
\end{itemize}
for it to be useful in a practical sense. These criteria are generic and may be applied to all quantum computers irrespective of whether their qubits are conventional or topological. If quantum communication is desired then two further criteria: 6. \emph{The ability to interconvert stationary and flying qubits} and
7. \emph{The ability to faithfully transmit flying qubits between specified locations} should also be satisfied \cite{DiVincenzo2000a}.

There exist methods of quantum computation that differ from the quantum circuit model presented here. However, it is always possible to interconvert between the different models \cite{Pachos_TQC_Book}. Since the quantum circuit model is perhaps the most intuitive model, and can be more directly related with topological quantum computing \cite{Freedman_Quantum_Circuit}, it is the model which will be described here.

\subsection{Qubits}

The mathematics of quantum computation is performed mainly with matrices, within the framework of linear algebra. As such, it shall be necessary to consider the representation of our basis states. Typically, qubits are considered as the fundamental logical unit of a quantum computer. A single qubit is a normalised linear superposition of the orthonormal states $\ket{0}$ and $\ket{1}$. Typically, $\ket{0}$ and $\ket{1}$ are represented as

\begin{equation}
\ket{0}=
\begin{pmatrix}
1 \\ 0
\end{pmatrix},
\quad \ket{1} = \begin{pmatrix} 0\\ 1 \end{pmatrix},
\label{eqn:qubits1}
\end{equation}
although other orthonormal basis vectors may arise in some systems. Note that a single qubit is a member of the two-dimensional complex vector space $\mathbb{C}^2$, which is also a Hilbert space $\mathcal{H}$. Physically, the bases $\ket{0}$ and $\ket{1}$ should be chosen such that they are stationary states or eigenstates of the system, such that when the qubit is measured it will yield one of those two states.

For useful computation, it is necessary to combine multiple qubits. Mathematically, this is done using the direct tensor product \cite{Quantum_Textbook}. Using \eqref{eqn:qubits1} for the single qubit, the bases of a two qubit state are

\begin{equation}
\begin{split}
\ket{0}\otimes\ket{0}=\ket{00}=\begin{pmatrix}1\\0\\0\\0\end{pmatrix}, \quad
\ket{0}\otimes\ket{1}=\ket{01}=\begin{pmatrix}0\\1\\0\\0\end{pmatrix}, \\
\ket{1}\otimes\ket{0}=\ket{10}=\begin{pmatrix}0\\0\\1\\0\end{pmatrix}, \quad
\ket{1}\otimes\ket{1}=\ket{11}=\begin{pmatrix}0\\0\\0\\1\end{pmatrix}.
\end{split}
\label{eqn:qubits2}
\end{equation}

As for the single qubit, a pair of qubits may be in any normalised linear superposition of these four basis states, and is a member of $\mathbb{C}^{4}$. In general, an ensemble of $n$ qubits is a member of $\mathbb{C}^{2^n}$ \cite{Quantum_Basics}.

While superposition is significant, entanglement is also significant. Consider two qubits $\ket{\psi_1}$ and $\ket{\psi_2}$, which are each in a linear superposition of $\ket{0}$ and $\ket{1}$. If a state can be written as $\ket{\psi_1}\otimes\ket{\psi_2}$, then it is called separable. However, the space for the states of $n$ separable qubits is $\mathbb{C}^{2n}$. The remaining states in the general $n$-qubit space $\mathbb{C}^{2^n}$ are non-separable, or entangled \cite{Pachos_TQC_Book}. For an entangled state, there exists no representation of the single qubit for which the states of all $n$ qubits can be specified separately, and measurement of one qubit will also provide information on any qubits it is entangled to \cite{Quantum_Textbook}. Entangled states have no classical analogue, and are where quantum computation derives much of its power \cite{Quantum_Basics}.

It is of note that the qubit is not the only possible unit of a quantum computer. It is possible to have logical units which are more than two-dimensional, containing more than 2 basis states. Such higher-dimensional qubits are often called qudits \cite{Anyon_Computing_Brennen,Qubit_Leakage}.

\subsection{Quantum Gates}

The register of qubits stores the information of the computation and forms the hardware of the quantum computer. After creating an initialised state, it is necessary to modify that state in a way which performs the computation. In quantum systems, the manipulation of states is represented using unitary matrices \cite{Quantum_Textbook}. Given an input state $\ket{\psi_{\rm in}}$, it is necessary to find a series of manipulations which corresponds to a unitary matrix $U$ which produces the desired output $\ket{\psi_{\rm out}}=U\ket{\psi_{\rm in}}$. These unitary operations are referred to as quantum gates \cite{Pachos_TQC_Book}.

These unitary matrices, as for any linear transformation, can be found by considering their action on each of the basis vectors. For single qubit gates, these bases are $\ket{0}$ and $\ket{1}$, as in \eqref{eqn:qubits1}, yielding a $2\times2$ matrix. For two qubit gates those bases are those in \eqref{eqn:qubits2}, yielding a $4\times4$ matrix. If a single qubit gate acts on a qubit within a register of $n$ qubits, then the dimension of the single qubit operation needs to be expanded to fill the whole vector space by taking a tensor product with the identity.

For concreteness, suppose there is a 4-qubit register, and an operation $U$ acts on just the second qubit. Then the transformation acting on the whole register is given by $\mathbb{I}_2 \otimes U \otimes \mathbb{I}_2 \otimes \mathbb{I}_2$, where $\mathbb{I}_2$ is the $2\times2$ identity matrix. The resultant matrix will be a $16\times16$ matrix. A similar process is done for two qubit gates.

A common class of two qubit operations are controlled gates. In a controlled gate, the operation $U$ is applied to the second (target) qubit if and only if the first (control) qubit is in the $\ket{1}$ state \cite{Quantum_Textbook}. Controlled gates are sensitive to the relative phase of the qubits, and typically entangle the two qubits. Controlled gates are of the form

\begin{equation}
\ket{0}\bra{0}\otimes \mathbb{I}_2 + \ket{1}\bra{1}\otimes U =
\begin{pmatrix}
1 & 0 & 0 & 0\\
0 & 1 & 0 & 0\\
0 & 0 & U_{0,0} & U_{0,1} \\
0 & 0 & U_{1,0} & U_{1,1}
\end{pmatrix}.
\label{eqn:controlled_U}
\end{equation}

Because quantum mechanical measurements are sensitive only to relative phases and insensitive to the global phase of a system, any two unitary operations which differ from each other only by a scalar phase factor are equivalent.

\subsection{State Measurement}

After all the computation has been performed, the system needs to be measured. The final state will be a vector, which can be decomposed into a linear combination of the basis vectors. The coefficients for these basis vectors, which correspond to the components of the state vector, will relate to the probability of measuring the system to be in that basis state.

Specifically, if the state vector has the components $a_1, a_2, \dots a_n$, then the probability of measuring the $i$'th basis state $\ket{i}$ is $\abs{a_i}^2=\abs{\bra{i}\psi \rangle}^2$. If the $i$'th basis state is measured, then the state collapses into the $i$'th state (mathematically, it is projected onto $\ket{i}$), unless the state is destroyed by the physical process of measurement \cite{Quantum_Textbook}.

Because quantum computation provides a probabilistic outcome, and the state is effectively destroyed during measurement, it is generally necessary to repeat a quantum computation multiple times to gain adequate statistics to characterise the output of the quantum computation \cite{Quantum_Basics}. To efficiently gather such statistics, the quantum algorithm can be executed simultaneously using multiple quantum processors.

\subsection{Quantum Circuit Diagrams}

Quantum algorithms make use of particular initial states, sequences of quantum gates and measurements to perform a quantum computation. The details of a quantum algorithm are typically depicted using a quantum circuit diagram.

\begin{figure}
	\includegraphics[width=1\linewidth]{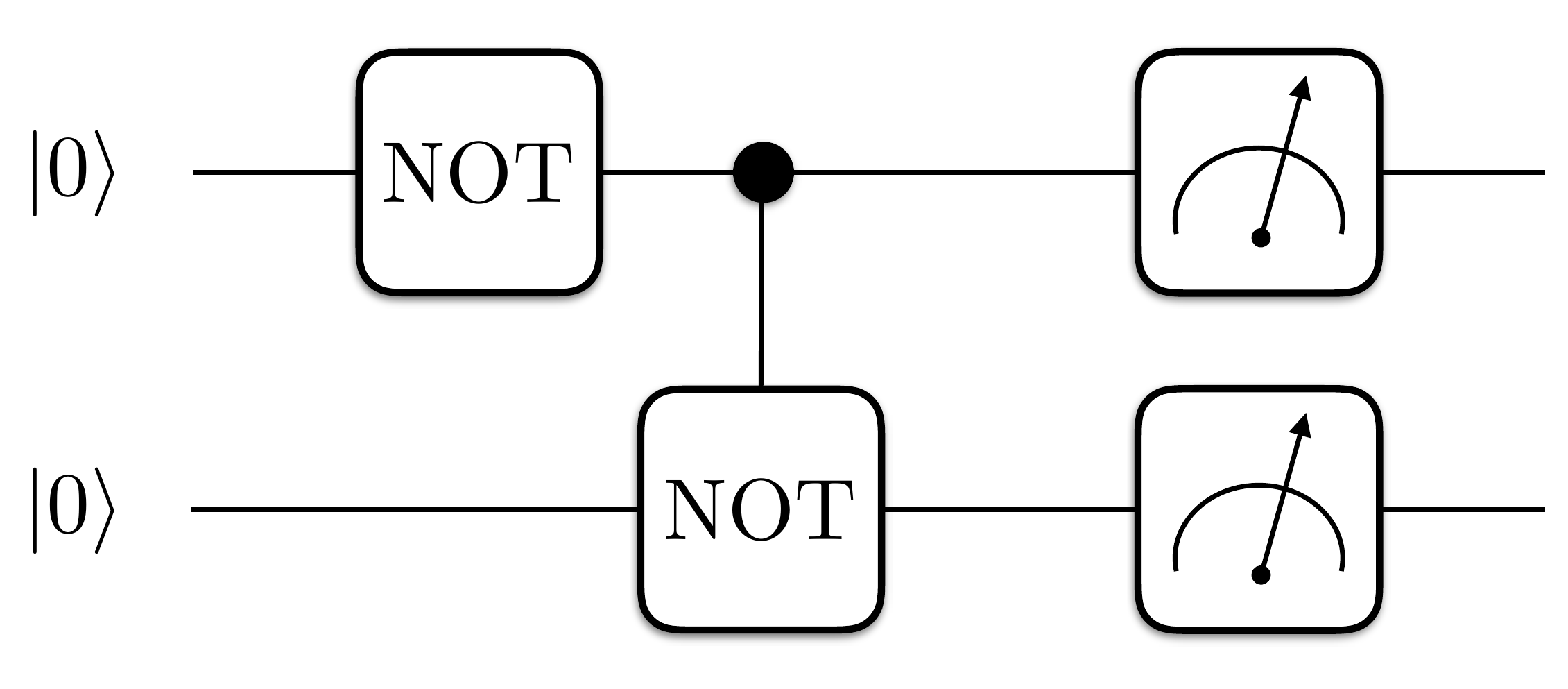}
	\caption{An example of a two qubit quantum circuit diagram, involving a $\rm NOT$ gate $\left(\begin{smallmatrix}0 & 1 \\ 1 & 0\end{smallmatrix}\right)$ and a controlled $\rm NOT$ (or $\rm CNOT$) gate followed by a state measurement. The qubit register starts in the $\ket{00}$ state, and should end in the $\ket{11}$ state.}
	\label{Fig:sample_circuit}
\end{figure}

A sample quantum circuit diagram is provided in \Figref{Fig:sample_circuit}. The initial states of each of the qubits is on the left of the diagram. By convention, the top-most qubit is the first qubit. The horizontal lines correspond to each of the qubits, tracking them through time. Boxes indicate quantum gates. A box with a line and circle attached to another qubit, as in the right-most gate in \Figref{Fig:sample_circuit}, indicates a controlled operation, where the circle marks the control qubit and the box is on the target qubit. Gates are applied (mathematically, left-multiplied) to the qubits in order from left to right. Measurement of the qubits occurs on the far right of the diagram unless indicated otherwise \cite{Quantum_Textbook}.

In this particular example, the qubits both start in the $\ket{0}$ state. Then the first qubit is acted on by a $\rm NOT$ gate, which converts it to the $\ket{1}$ state. Then a controlled operation occurs. Because the first qubit is in the $\ket{1}$ state, the second qubit is acted on by a $\rm NOT$ gate, which converts it to the $\ket{1}$ state. At the end of this quantum circuit, both qubits would be measured (meter symbols) to be in the $\ket{1}$ state.

While it is implied that qubits are measured on the far right of the circuit diagram, sometimes an algorithm is only concerned with the measured states of some of the qubits, or measures a qubit before the end of the algorithm. In such cases, meter symbols such as those shown in \Figref{Fig:sample_circuit} are used to explicitly indicate a measurement.

\subsection{Errors}

Errors can arise during quantum computation, from coupling to the environment and imprecision in the application of unitary operations \cite{Preskill_Fault_Tolerant}. Coupling to the environment produces decoherence, where the quantum state inside the computer becomes entangled with the environment and noise is introduced to the quantum state from the environment \cite{Pachos_TQC_Book,Zurek2003a}. Decoherence is a major problem in quantum computers, and can restrict the lifetime of a state and thus limit how much computation can be performed with a quantum computer \cite{Anyon_Computing_Nayak}.

Quantum error correction codes can be used to correct for the effects of decoherence, but they add significant overhead to any computation and require some maximum error rate for computation, typically around $10^{-4}$ or less, to be effective \cite{Anyon_Computing_Freedman,Quantum_Textbook}.

Even if all undesired coupling to the environment can be removed, there could remain random errors that could occur due to imprecision in the implementation of unitary operations, such as if a particle is rotated by 90.01$^\circ$ instead of 90$^\circ$ \cite{Anyon_Computing_Nayak}. A quantum computer with a low error rate is necessary for quantum computation to be successful or efficient.

\section{Topological Quantum Computing} \label{sec:quantum_computing}

\subsection{Anyons} \label{sec:anyons}

In three-dimensional space, particles can be classified as bosons or fermions. When one particle in 3D space is moved around another and returned to its original position, this path is topologically equivalent to not moving the particle at all, because the path can be deformed over the stationary particle into an arbitrarily small loop. This constraint makes the statistics involved in exchanging fermions and bosons very simple, producing a phase change of $\pi$ or $2\pi$ only \cite{Pachos_TQC_Book}.

When particles are constrained to two-dimensional space, it is possible to have more exotic exchange statistics, because a path where one particle is moved around another is no longer topologically trivial. These particles, which can have any phase change or even unitary operations, not just the $\pi$ and $2\pi$ phase shifts of fermions and bosons, are called anyons \cite{Wilczek1982a}. Specifically, anyons which result in a phase change when exchanging the positions of two particles are classified as Abelian anyons, because the operations produced by exchanging the anyons all commute with each other. However, it is possible for exchanging anyons to result in unitary operations beyond simple phase changes, and such particles are classified as non-Abelian anyons, because the operations produced by exchanging the anyons in general do not commute \cite{Anyon_Computing_Brennen,Anyon_Computing_Nayak,Anyon_Intro_Trebst,Anyons_Hall_Effect,Preskill_Lecture,Burton2016a}. It is non-Abelian anyons which are of interest to quantum computation.

The state of a system of anyons is defined by the anyons produced by fusing those anyons together, with each possible set of fusion outcomes representing one basis state in the Hilbert space of the quantum system of anyons. Each model of anyons contains rules regarding the possible outcomes of the fusion of two anyons. Abelian anyons have only a single possible fusion outcome for each fusion pair. When two non-Abelian anyons are fused, however, there are multiple possible fusion outcomes \cite{Preskill_Lecture}.

Adding more anyons to the system typically increases the number of possible states, and the factor by which this number increases is the quantum dimension. When an Abelian anyon is fused with another anyon, there is only one possible outcome, so Abelian anyons have a quantum dimension of 1. When two non-Abelian anyons are fused, then there are multiple possible outcomes, determined probabilistically, so non-Abelian anyons have a quantum dimension greater than 1 \cite{Anyon_Intro_Trebst,Pachos_TQC_Book,Anyon_Computing_Nayak}. 

The simplest model of non-Abelian anyons is the Fibonacci model. A Fibonacci anyon may fuse with another Fibonacci anyon to either annihilate or to produce another Fibonacci anyon. The number of possible fusion outcomes grows by the Fibonacci sequence when more anyons are added (hence the name), giving Fibonacci anyons a quantum dimension of the golden ratio, $\frac{1+\sqrt{5}}{2}$ \cite{Anyon_Computing_Nayak}. Furthermore, it has been demonstrated that the operations performed by exchanging Fibonacci anyons can reproduce any unitary operation to arbitrary accuracy (up to a global phase factor) \cite{Modular_Functor}, which makes Fibonacci anyons universal for quantum computation.

\begin{figure}
	\includegraphics[width=0.8\linewidth]{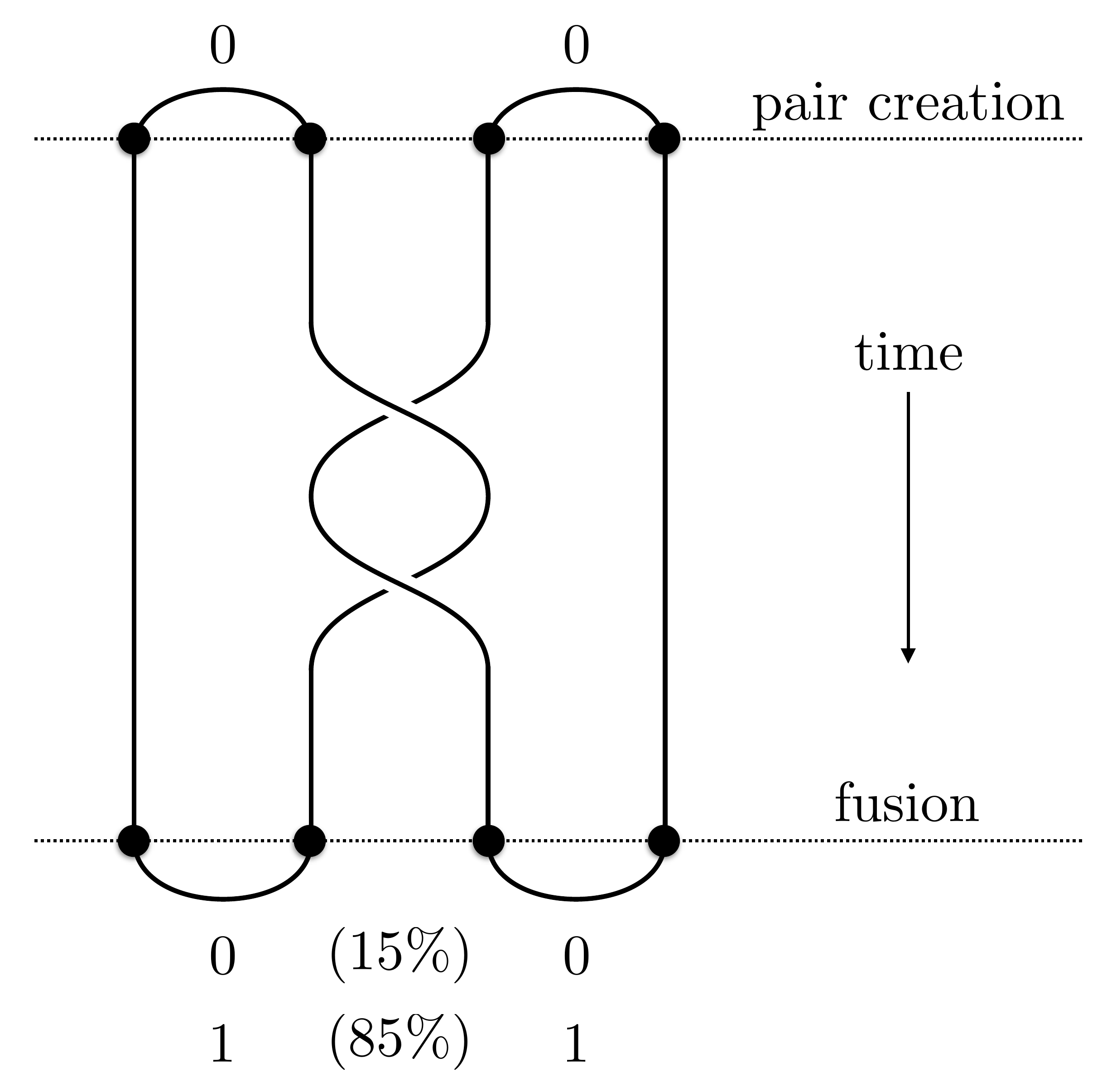}
	\caption{Four Fibonacci anyons in a row in $2+1$ dimensional space, where time travels downwards in the diagram. Two anyon pairs are created from the vacuum. The second and third anyons are exchanged twice, braiding the worldlines. Then the anyon pairs are fused again. There is an approximately $85\%$ probability that the anyons will not annihilate and instead produce another anyon.}
	\label{Fig:fib_braid}
\end{figure}

When exchanging anyons, their quantum state is manipulated, which changes the probabilities of the fusion outcomes. The operations performed by exchanging anyons are intrinsically topological \cite{Anyon_Computing_Nayak,Preskill_Lecture}. Consider $2+1$ dimensional space, where the anyons are spatially confined to a plane and the third dimension is time. As the anyons move in the plane, they trace worldlines in this $2+1$ dimensional space. As shown in \Figref{Fig:fib_braid}, exchanging two anyons results in braiding the worldlines. If two of such braids are topologically equivalent, and the particles have been kept sufficiently distant to minimise direct interactions between the anyons, they perform the same operation on the anyons \cite{Pachos_TQC_Book}.

Anyons, and groups of anyons, carry a charge- or spin-like quantity. For an individual particle, this simply denotes the particle type. For groups of anyons, this denotes the type of the resultant particle if all those anyons are fused together. The overall `charge' of a system of anyons is conserved, provided that it does not braid with other groups of anyons. Braiding within a group of anyons cannot change the `charge' of that group \cite{Preskill_Lecture}. Typically, anyons are created as anyon-antianyon pairs from the vacuum, meaning the pairs will each annihilate to the vacuum if they do not braid with any other anyons \cite{Anyon_Computing_Nayak,Carlos_Anyon_Computing}. If these anyons do perform braiding, there will in general be a non-zero probability that the anyon pairs will not fuse to vacuum.

\subsubsection{Fibonacci Anyons}

One of the simplest models of non-Abelian anyon is the Fibonacci anyon. The Fibonacci model appears in the ${\rm SU(2)}_3$ Witten--Cherns--Simmons topological quantum field theory \cite{Modular_Functor,Anyon_Computing_Freedman,Anyon_Computing_Nayak,Preskill_Lecture}. The Fibonacci model contains two particle types: the vacuum (with `charge' 0) here denoted by {\bf 0}, and the non-trivial anyon (with `charge' 1) here denoted by $\tau$. The vacuum is the absence of a particle. Explicitly, the fusion rules are

\begin{equation}
\begin{aligned}
\tau \otimes \tau  &= {\bf 0}\oplus \tau \\
{\bf 0} \otimes {\bf 0} &= {\bf 0} \\
{\bf 0} \otimes \tau  &= \tau  \\
\tau  \otimes {\bf 0} &= \tau ,
\end{aligned}
\label{eqn:fib_fuse}
\end{equation}
where $\otimes$ in this context denotes the fusion (merging) of two particles and $\oplus$ denotes multiple possible outcomes. In the following, we refer to the anyons ${\bf 0}$ and ${\bf \tau}$ of the Fibonacci model simply by their charges 0 and 1, respectively. Fusion of two Fibonacci anyons may result in either annihilation or creation of a new anyon. This makes the Fibonacci anyon its own anti-particle \cite{Pachos_TQC_Book,Preskill_Lecture}. From the last three rules, fusion with the vacuum does nothing. As is shown later, performing braiding can change the probabilities of the outcomes of this fusion.

Consider \Figref{Fig:fib_braid}. Since the pairs are created from the vacuum, each pair must have a net `charge' of 0. If the braid was not present, then the two pairs would individually fuse to vacuum with 100\% probability. However, by performing the braiding then fusing the particles, there is a non-zero probability that the fusion could give 1 instead of 0. The net `charge' of the whole system is still 0, though, so if the remaining two particles are fused, they must give the vacuum. From \eqref{eqn:fib_fuse} this is only possible if either both particles are 0 or both particles are 1. This means that, for this system, the outcome of the fusion of one of the pairs of anyons determines the fusion outcome of the other pair.

For Fibonacci anyons, the number of possible fusion outcomes, and thus the dimension of the Hilbert space, grows according to the Fibonacci sequence as more anyons are added. This gives Fibonacci anyons a quantum dimension $d_\tau$ of the golden ratio, $d_\tau=\phi = \frac{1+\sqrt{5}}{2}$, and is where Fibonacci anyons get their name \cite{Pachos_TQC_Book,Preskill_Lecture}. Generically, the quantum dimensions of the anyons satisfy $d_\alpha d_\beta=\sum_\gamma N_{\alpha\beta}^\gamma d_\gamma$, where the integer $N_{\alpha\beta}^\gamma$ is the number of distinguishable ways the anyons $\alpha$ and $\beta$ may be fused to yield an anyon $\gamma$. The total quantum dimension $\mathcal{D}$ of the anyon model is determined by the relation $\mathcal{D} =\sqrt{\sum_\alpha d^2_\alpha}$ \cite{Anyon_Computing_Nayak,Preskill_Lecture}. Abelian anyons have a quantum dimension equal to one, where as for non-Abelian anyons their quantum dimension is greater than one. Non-Abelian anyons, such as Fibonacci anyons, are thought to be capable of universal quantum computation by braiding alone if the square of their quantum dimension is not integer. The quantum dimension has also been linked to the passage of time by showing that a relational time for universal anyonic systems, such as the Fibonacci anyon model, is continuous where as for non-universal systems, such as the Ising anyon model, discrete time would emerge \cite{Nikolova2018a}.

There are several candidate systems which may exhibit the behaviours of Fibonacci anyons \cite{Alicea2015a,Sarma2015a}. One candidate is the fractional Hall effect at $v=12/5$, which can exhibit quasiparticle excitations which are predicted to follow the behaviour of Fibonacci anyons \cite{Anyon_Computing_Nayak,Anyon_Intro_Trebst,Anyon_Computing_Brennen,Anyons_Hall_Effect,Sarma_TQC,Fibonacci_Anyons_Hall1,RezayiRead1,Fibonacci_Wavefunctions,Anyon_Interferometry,Fibonacci_Hall_Size}. Another candidate is to construct networks of spin lattices which mimic the desired anyon behaviours \cite{Anyon_Computing_Brennen}. Other candidates such as rotating Bose-Einstein condensates \cite{BEC_Anyons}, dipolar boson lattices \cite{Fibonacci_Anyons_1D} and magnetic systems \cite{Fibonacci_Net} have also been proposed.

This work does not concern itself with any specific physical model of Fibonacci anyon and instead focuses on the macroscopic properties of the Fibonacci anyons with regards to braiding and fusion outcomes. In ignoring more specific physical implementations, we assume that it is possible to reliably create anyon pairs, move those anyons around each other, and determine the outcome of anyon fusion. In general, these tasks may be non-trivial in a physical system, but such challenges are beyond the scope of this work. Notwithstanding, the reader may choose to refer to the non-Abelian vortex anyon models mentioned in Sec.~\ref{sec:overview}.

\subsection{Braiding} \label{sec:fib_braid}

\subsubsection{The F Move}

\begin{figure}
	\includegraphics[width=0.8\linewidth]{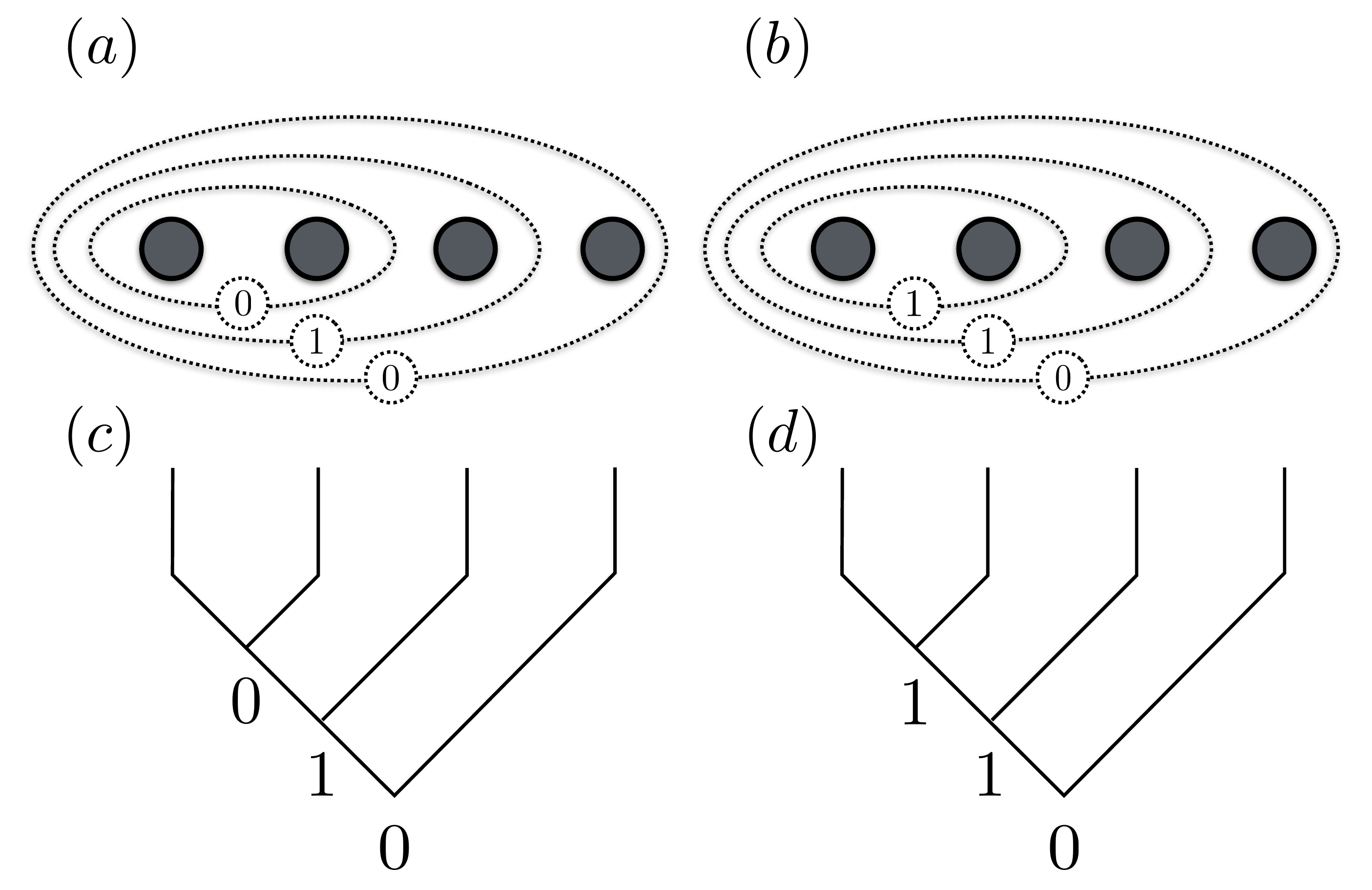}
	\caption{Diagrams for the two possible fusion outcomes for a set of four Fibonacci anyons with zero net overall `charge', where the anyons are fused from left to right. (a) and (b) describe the anyons with circle notation. (c) and (d) describe the anyons with tree diagrams, where time points downwards. (a) and (c) refer to the same anyons, as do (b) and (d).}
	\label{Fig:b4_fusion_tree}
\end{figure}

The fusion outcomes, and thus the basis states of the vector space, are readily visualised by the circle (or ellipse) notation in \Figref{Fig:b4_fusion_tree}(a) and (b), or the fusion tree diagrams, as in \Figref{Fig:b4_fusion_tree}(c) and (d).

With the circle notation, anyons are enclosed by circles, and these circles have marked the net `charge' of the anyons enclosed. It is implied that when fusion is performed, anyons within a circle will be fused before fusion with anyons outside that circle.

In the tree notation, the worldlines of the anyons are marked, with fusion occurring at the vertices. The outcome of each fusion is marked at each vertex.

\begin{figure}
	\includegraphics[width=0.8\linewidth]{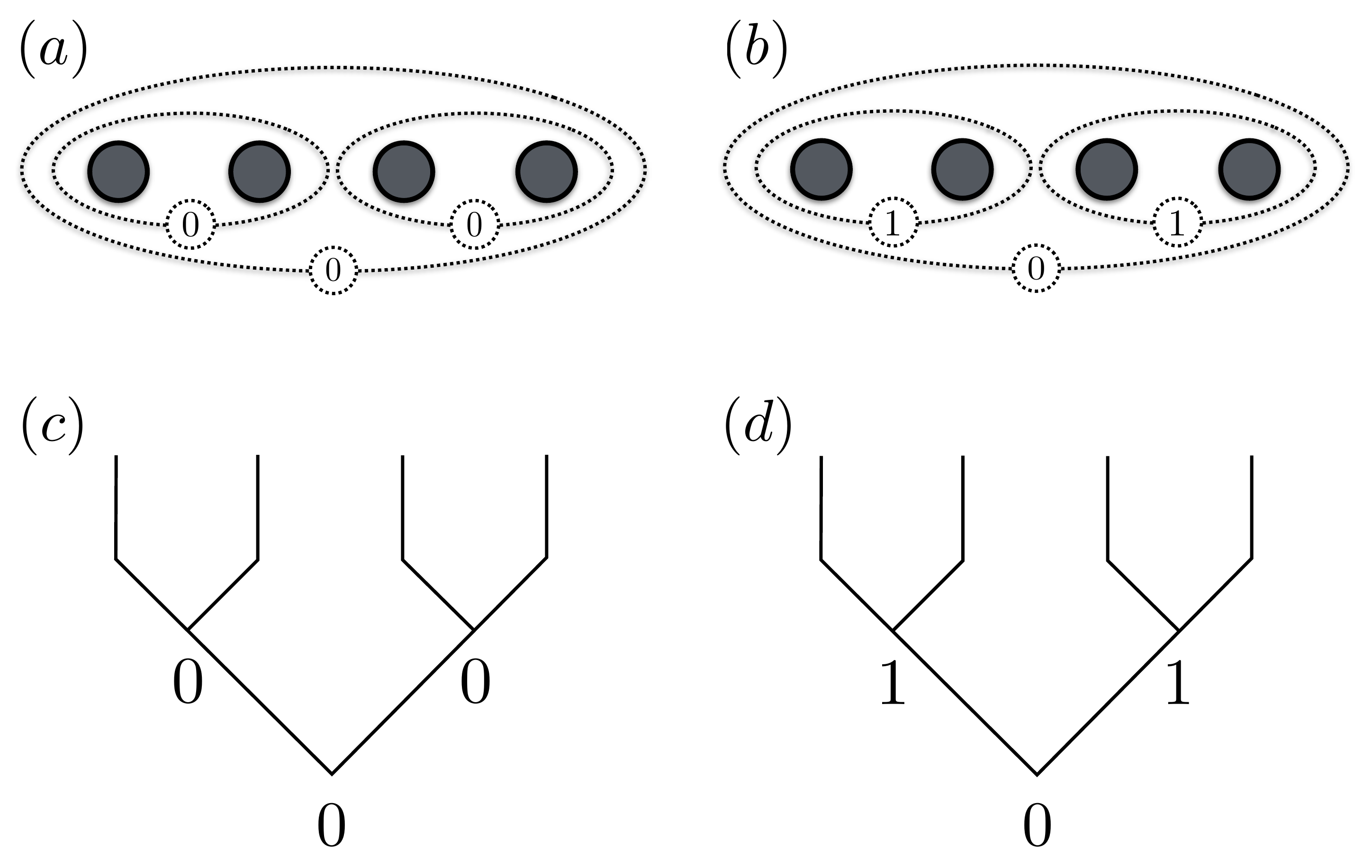}
	\caption{Diagrams for the two possible fusion outcomes for a set of four Fibonacci anyons with zero net overall `charge', where the anyons are fused in pairs. (a) and (b) describe the anyons with circle notation. (c) and (d) describe the anyons with tree diagrams, where time points downwards. (a) and (c) refer to the same anyons, as do (b) and (d).}
	\label{Fig:b4_fusion_tree_2}
\end{figure}

However, the choice of order in which the anyons are fused is somewhat arbitrary. \Figref{Fig:b4_fusion_tree} shows the basis when fusion is performed from left to right, while \Figref{Fig:b4_fusion_tree_2} shows the basis when fusion is performed in pairs. Both are valid choices for the basis states of the system. Changing between these bases, and any other set of bases, is done using an F move,

\begin{equation}
\vcenter{\hbox{\includegraphics[height=10ex]{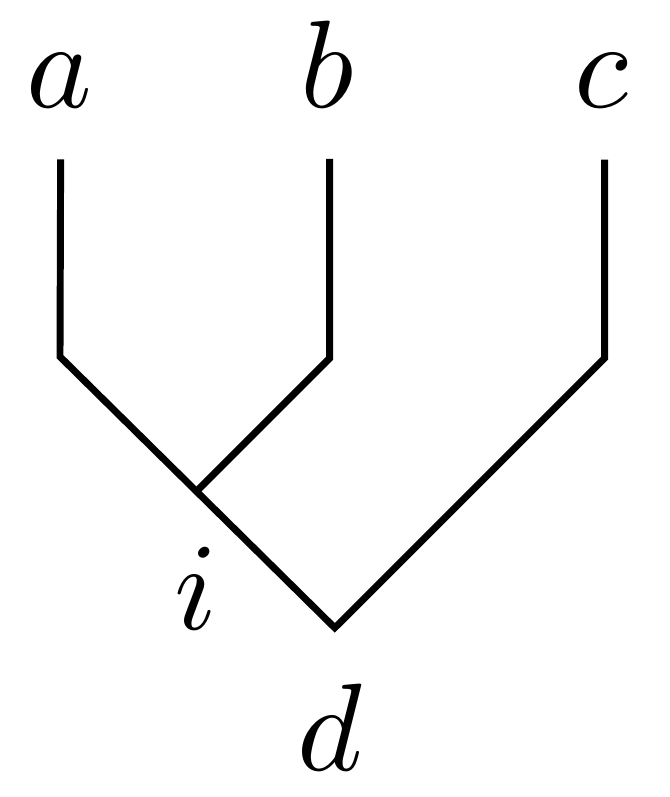}}} = \sum_{j} F(abcd)^i_j \vcenter{\hbox{\includegraphics[height=10ex]{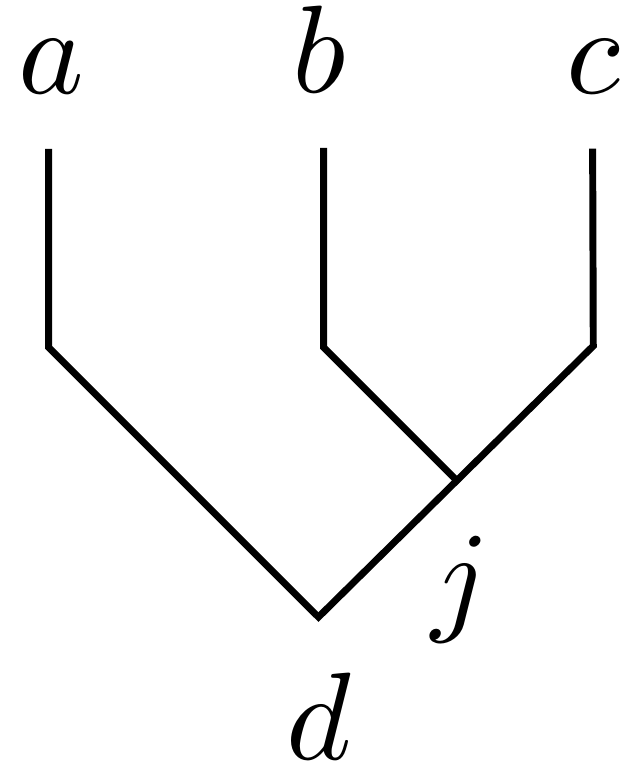}}}
\label{eqn:Fmove}
\end{equation}
where the sum over $j$ sums over each possible particle type, and $F(abcd)^i_j$ is an F coefficient. The F move is taken to act locally at any such segment of a fusion tree diagram. The F move is also applicable for where the tree diagrams with the corresponding labels are mirrored; the F move is its own inverse.

For Fibonacci anyons, the sum in \eqref{eqn:Fmove} may be explicitly expressed as
\begin{equation}
\vcenter{\hbox{\includegraphics[height=10ex]{"F-Move_Left"}}} = F(abcd)^i_0 \vcenter{\hbox{\includegraphics[height=10ex]{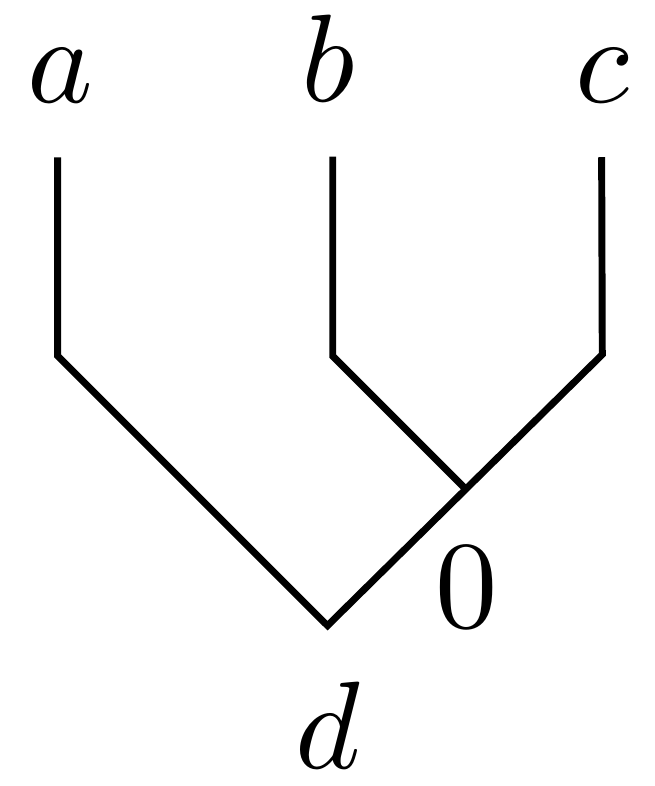}}}
+F(abcd)^i_1 \vcenter{\hbox{\includegraphics[height=10ex]{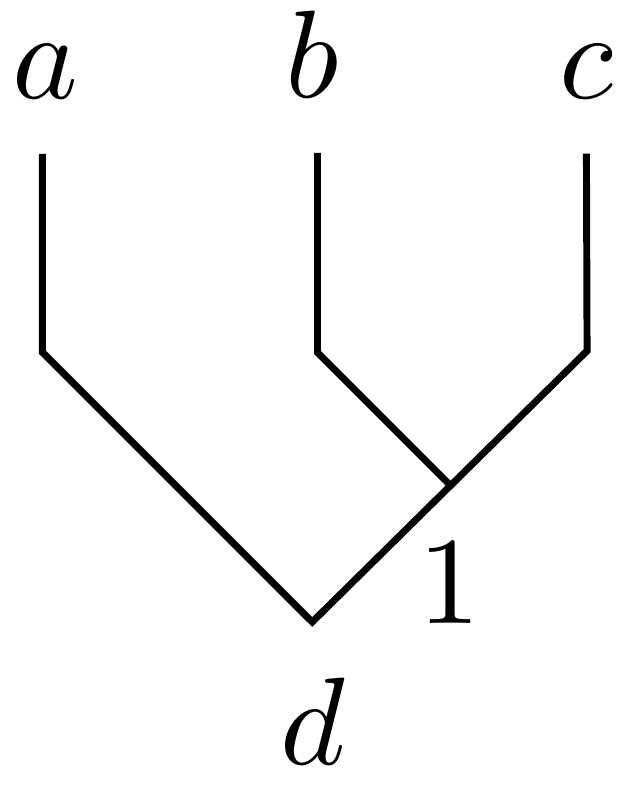}}}
\label{eqn:Fmove_fib}
\end{equation}
and $a$, $b$, $c$, $d$ and $i$ may have values of either 0 or 1.

\begin{figure}
	\includegraphics[width=\linewidth]{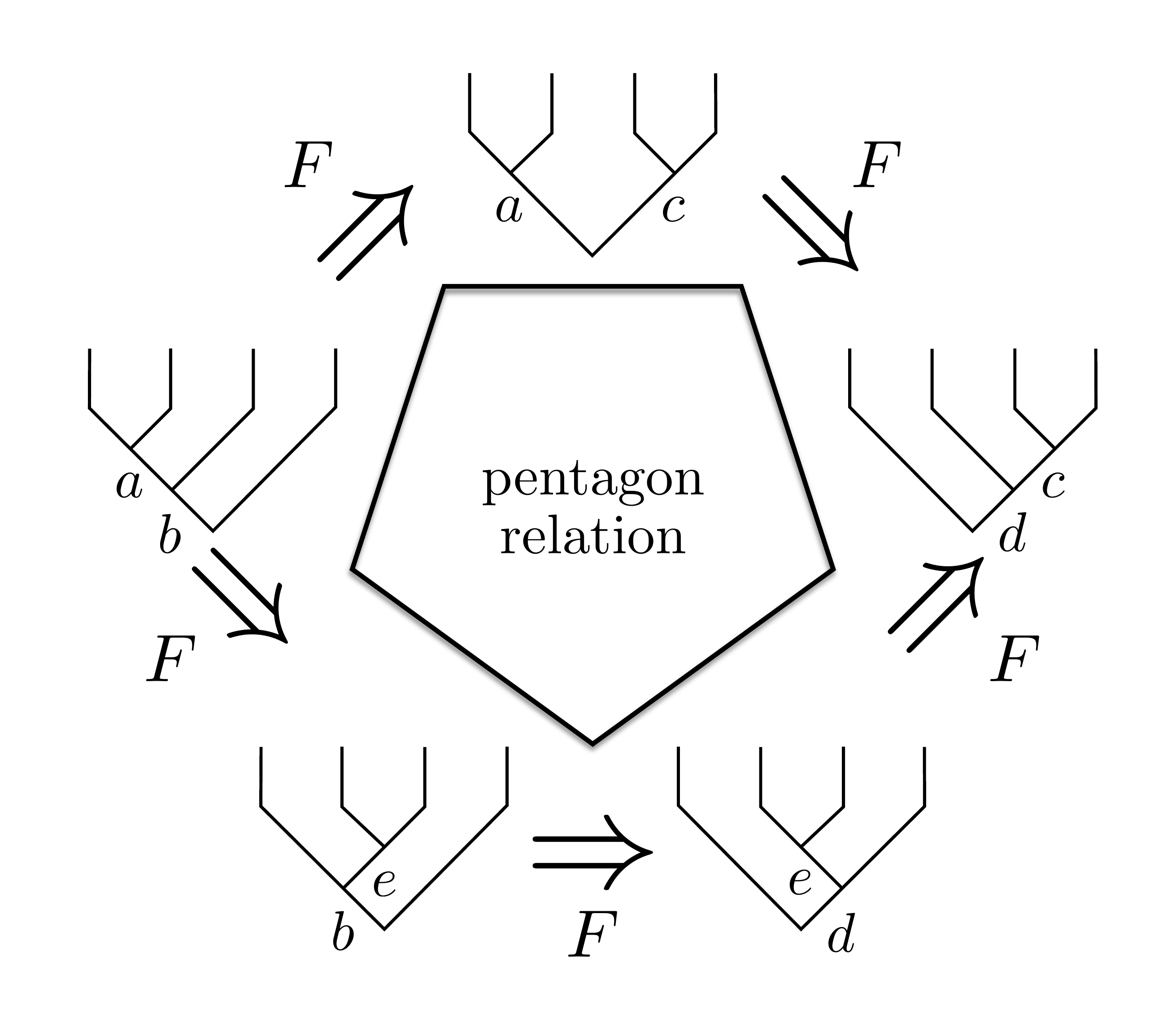}
	\caption{Diagrammatic representation of the pentagon relationship, where the F moves act on parts of the fusion tree.}
	\label{Fig:pentagon}
\end{figure}

To apply the F moves, it is necessary to know the F coefficients. First, we constrain the F move to be a unitary operation. Some of the coefficients can be calculated trivially. If the fusion diagram disobeys the fusion rules \eqref{eqn:fib_fuse}, then the corresponding F coefficient is zero. If this reduces \eqref{eqn:Fmove_fib} to having only a single term on the right hand side, then the remaining F coefficient must be equal to one. To find the values of any non-trivial coefficients requires solving a consistency relationship known as the pentagon relationship, illustrated in \Figref{Fig:pentagon}. The pentagon relationship shows two different combinations of F moves to go from one particular basis to another \cite{Preskill_Lecture}.

For the Fibonacci model, the only F coefficients which can not be solved trivially are those corresponding to the cases where $(abcd)=(1111)$. The pentagon relationship is sufficient to solve for these F coefficients.

Mathematically, the pentagon relationship for Fibonacci anyons gives the equation
\begin{equation}
F(11c1)^d_a F(a111)^c_b = \sum_{e=\{0,1\}} F(111d)^c_e F(1e11)^d_b F(111b)^e_a,
\label{eqn:pentagon}
\end{equation}
where the indices $a$, $b$, $c$, $d$, and $e$ correspond to those in \Figref{Fig:pentagon}. Using this, the trivial results, and the constraint that the F move should be a unitary operation, the remaining F coefficients for Fibonacci anyons, up to an arbitrary phase, are
\begin{equation}
\begin{gathered}
F(1111)^0_0 = 1/\phi, \\
F(1111)^0_1 = F(1111)^1_0 = 1/\sqrt{\phi}, \\
F(1111)^1_1 = -1/\phi ,
\end{gathered}
\label{eqn:F_coeff}
\end{equation}
where $\phi=(1+\sqrt{5})/2$ is the golden ratio \cite{Anyon_Intro_Trebst}. These four coefficients may also be expressed in matrix form \cite{Pachos_TQC_Book,Preskill_Lecture},
\begin{equation}
F(1111) = \begin{pmatrix}
1/\phi & 1/\sqrt{\phi} \\
1/\sqrt{\phi} & -1/\phi
\end{pmatrix}.
\label{eqn:F_coeff_matrix}
\end{equation}

While the matrix form is useful for the three anyon case, when considering more than three anyons it is more practical to work with the coefficients.

\subsubsection{The R Move}

To find the effect of braiding, it is also necessary to consider the effect of exchanging two particles. This is quantified using the R move,
\begin{equation}
\vcenter{\hbox{\includegraphics[height=10ex]{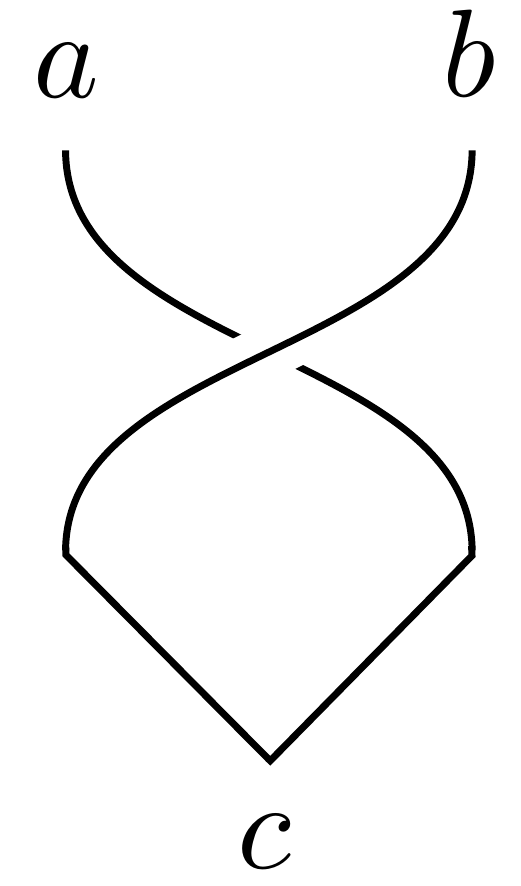}}} = R^{ab}_c \vcenter{\hbox{\includegraphics[height=10ex]{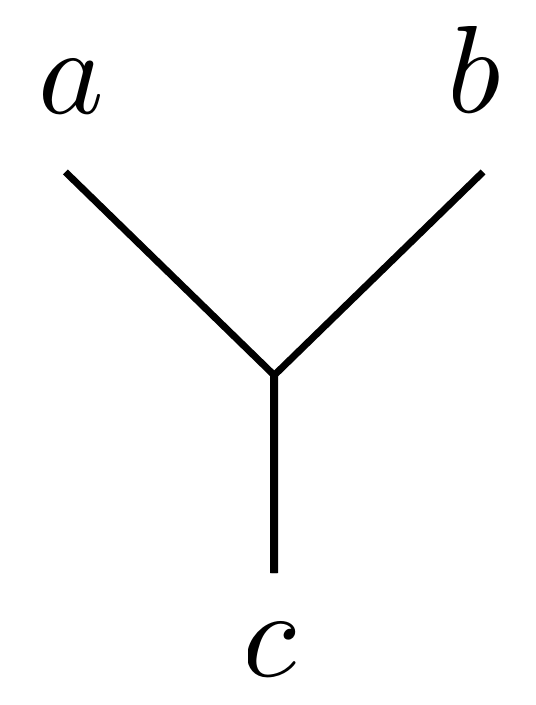}}},
\label{eqn:Rmove}
\end{equation}
which acts upon a braid immediately preceding a fusion, where $R^{ab}_c$ is an R coefficient. In our convention, time travels downwards. If the twist in \eqref{eqn:Rmove} was in the opposite direction, then the inverse of the R move would be applied. Because the R moves are unitary, the inverse R coefficients are the reciprocal or the complex conjugate of the regular R coefficients (both are equivalent in this case).

\begin{figure}
	\includegraphics[width=\linewidth]{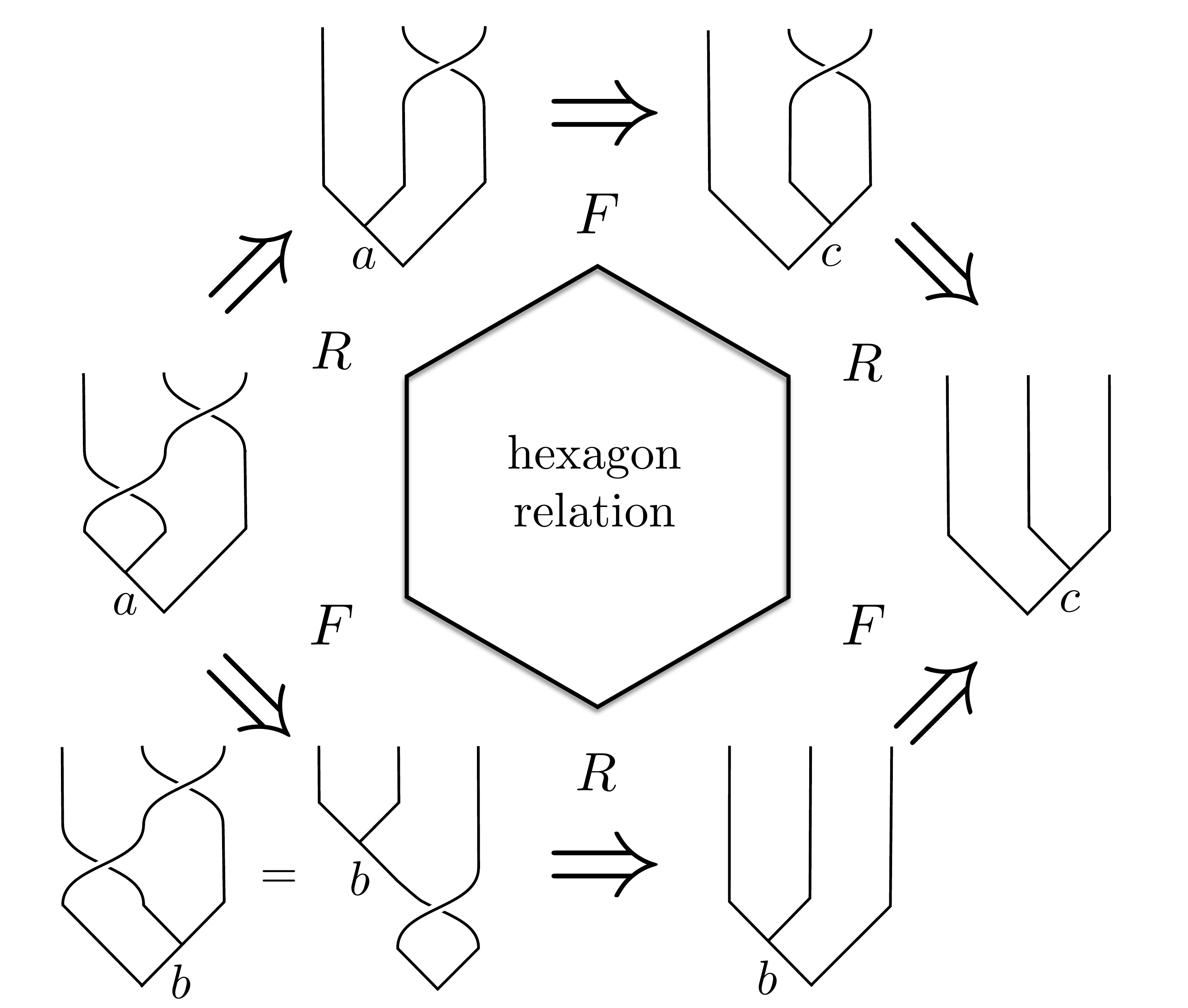}
	\caption{Diagrammatic representation of the hexagon relationship, where F and R moves act on parts of the fusion tree to yield the same outcome by two different paths.}
	\label{Fig:hexagon}
\end{figure}

The R coefficients for the cases where $a$ and $b$ are anything other than non-Abelian anyons (such as the vacuum) are trivially determined by the fusion rules, as for the F coefficients. The remaining R coefficients can be found using a consistency relation known as the hexagon relationship illustrated in \Figref{Fig:hexagon} \cite{Preskill_Lecture}.

For Fibonacci anyons, only the case $ab=11$ cannot be trivially solved. The values of these two R coefficients can be found by topological quantum field theory and the hexagon relationship \cite{Anyon_Computing_Nayak}. The values for these R coefficients are

\begin{equation}
\begin{aligned}
R^{11}_0 &= e^{-4\pi i/5}, \\
R^{11}_1 &= e^{3\pi i /5}.
\end{aligned}
\label{eqn:R_coeff}
\end{equation}

Using the same basis as for \eqref{eqn:F_coeff_matrix}, these R coefficients can also be expressed in matrix form \cite{Pachos_TQC_Book},
\begin{equation}
R^{11} = \begin{pmatrix}
e^{-4\pi i /5} & 0 \\
0 & e^{3 \pi i /5}
\end{pmatrix}.
\label{eqn:R_coeff_matrix}
\end{equation}

For non-Abelian anyons, the consistency conditions imposed by the pentagon and hexagon relations are complete, encapsulating all required topological consistencies and not requiring any other consistency relations \cite{Preskill_Lecture}. Together with the fusion rules, they are sufficient to derive the F and R coefficients \cite{Pachos_TQC_Book,Preskill_Lecture}.

\subsubsection{Braid Matrices}

Explicit matrix representations for the effect of braiding can be constructed by performing F and R moves acting on an appropriate basis state \cite{Preskill_Lecture}. The methodology here can be used for any anyon model once the F and R coefficients have been determined. For concreteness, we will apply this method to Fibonacci anyons.

The first step is to choose the basis to work in. We will consider the basis where the anyons are fused sequentially from left to right. For the purpose of demonstration we will consider the case of three Fibonacci anyons, although the method can readily be expanded to more.

\begin{figure}
	\includegraphics[width=\linewidth]{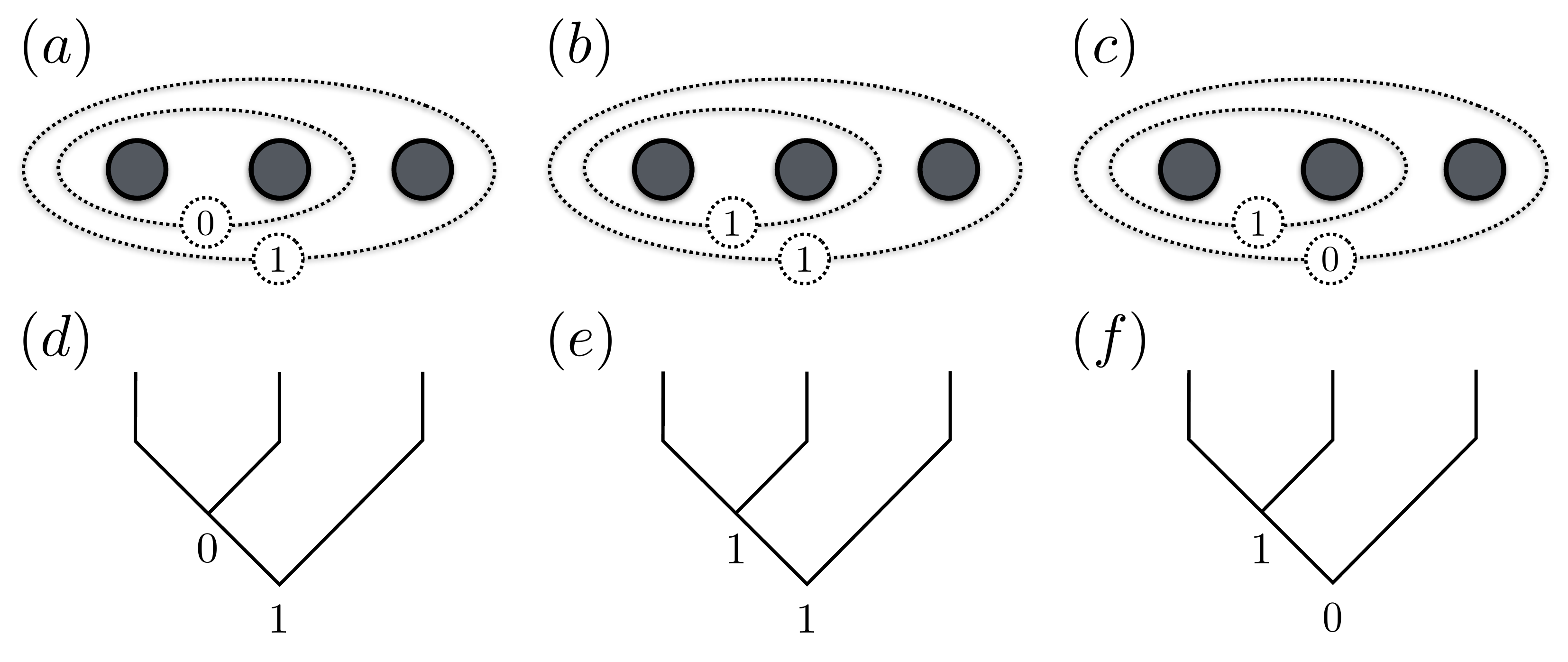}
	\caption{Representations of the basis states of the 3 anyon system, where (a) and (d) are the same anyons, as are (b) and (e), and (c) and (f). As bitstrings, these states can be represented as `01', `11' and `10' respectively. The labels for these states are $\ket{0}$, $\ket{1}$ and $\ket{N}$ respectively.}
	\label{Fig:3anyon_bases}
\end{figure}

The next step is to enumerate over the possible basis states. There are several possible representations. One representation is with circle notation, as in \Figref{Fig:3anyon_bases}(a)-(c), or fusion trees, as in \Figref{Fig:3anyon_bases}(d)-(f). The tree representation explicitly demonstrates the action of the F and R moves, although for computation it is often more useful to represent the bases as bitstrings, where each component of the bitstring relates to a particular fusion outcome. The action of an F move in this representation is to change one of the bits, and it is implied that the underlying basis has also changed. Another representation is to label the basis states with unique names. For reasons that are explained in Section \ref{sec:fib_compute}, we label the states in \Figref{Fig:3anyon_bases} as $\ket{0}$, $\ket{1}$ and $\ket{N}$. These states form an orthonormal set of basis vectors in $\mathbb{C}^3$.

The next step is to consider the braid that we wish to apply, then apply an appropriate sequence of F and R moves in order to convert that braid into a linear combination of our chosen basis states. We must do this for the braid acting on each of the possible basis states. Thus, by considering the action of the operation upon each of the basis vectors, the matrix specifying that operation can be constructed. We shall call the exchange of the first and second anyons $\sigma_1$, and the exchange of the second and third anyons $\sigma_2$. The braid matrix $\sigma_1$ can thus be constructed element by element as
\begin{gather}
\begin{split}
\sigma_1\ket{0} = \eqimage{"3_Anyon_01_b1_Left"}{8ex}=R^{11}_0 \eqimage{"3_Anyon_01_Left"}{8ex} = e^{-4\pi i /5}\ket{0},\\
\sigma_1\ket{1} = \eqimage{"3_Anyon_11_b1_Left"}{8ex}=R^{11}_1 \eqimage{"3_Anyon_11_Left"}{8ex} = e^{3\pi i /5}\ket{1},\\
\sigma_1\ket{N} = \eqimage{"3_Anyon_10_b1_Left"}{8ex}=R^{11}_1 \eqimage{"3_Anyon_10_Left"}{8ex} = e^{3\pi i /5}\ket{N},
\end{split}
\displaybreak[0] \notag 
\end{gather}
resulting in the diagonal braid matrix
\begin{equation}
\sigma_1 = \begin{pmatrix}
e^{-4\pi i /5} & 0 & 0 \\
0 & e^{3\pi i /5} & 0 \\
0 & 0 & e^{3\pi i /5}
\label{eqn:matrix_b1_3anyon}
\end{pmatrix}.
\end{equation}
Similarly, the braid matrix $\sigma_2$ can be constructed as
\begin{widetext}
\begin{gather*}
\sigma_2\ket{0} = \eqimage{"3_Anyon_01_b2_Left"}{8ex} = F(1111)^0_0\eqimage{"3_Anyon_01_b2_Right"}{8ex} + F(1111)^0_1\eqimage{"3_Anyon_11_b2_Right"}{8ex}
= F(1111)^0_0 R^{11}_0 \eqimage{"3_Anyon_01_Right"}{8ex} + F(1111)^0_1 R^{11}_1\eqimage{"3_Anyon_11_Right"}{8ex}\\
= F(1111)^0_0 R^{11}_0 F(1111)^0_0 \eqimage{"3_Anyon_01_Left"}{8ex} + F(1111)^0_0 R^{11}_0 F(1111)^0_1 \eqimage{"3_Anyon_11_Left"}{8ex}\\
+ F(1111)^0_1 R^{11}_1 F(1111)^1_0 \eqimage{"3_Anyon_01_Left"}{8ex} + F(1111)^0_1 R^{11}_1 F(1111)^1_1 \eqimage{"3_Anyon_11_Left"}{8ex} \\
= (\phi^{-2}e^{-4\pi i/5} + \phi^{-1}e^{3\pi i/5})\ket{0} + (\phi^{-3/2}e^{-4\pi i/5} - \phi^{-3/2}e^{3\pi i/5})\ket{1}
= \phi^{-1} e^{4\pi i/5} \ket{0} + \phi^{-1/2} e^{-3\pi i/5} \ket{1}, \displaybreak[0]\\
\sigma_2 \ket{1} = \eqimage{"3_Anyon_11_b2_Left"}{8ex} =  F(1111)^1_0\eqimage{"3_Anyon_01_b2_Right"}{8ex} + F(1111)^1_1\eqimage{"3_Anyon_11_b2_Right"}{8ex}
= F(1111)^1_0 R^{11}_0 \eqimage{"3_Anyon_01_Right"}{8ex} + F(1111)^1_1 R^{11}_1\eqimage{"3_Anyon_11_Right"}{8ex}\\
= F(1111)^1_0 R^{11}_0 F(1111)^0_0 \eqimage{"3_Anyon_01_Left"}{8ex} + F(1111)^1_0 R^{11}_0 F(1111)^0_1 \eqimage{"3_Anyon_11_Left"}{8ex}\\
+ F(1111)^1_1 R^{11}_1 F(1111)^1_0 \eqimage{"3_Anyon_01_Left"}{8ex} + F(1111)^1_1 R^{11}_1 F(1111)^1_1 \eqimage{"3_Anyon_11_Left"}{8ex} \\
= (\phi^{-3/2} e^{-4\pi i/5} - \phi^{-3/2} e^{3\pi i/5})\ket{0} + (\phi^{-1} e^{-4\pi i/5} + \phi^{-2} e^{3\pi i/5})\ket{1}
= \phi^{-1/2} e^{-3\pi i/5} \ket{0} - \phi^{-1} \ket{1},\displaybreak[0]\\
\sigma_2 \ket{N} = \eqimage{"3_Anyon_10_b2_Left"}{8ex} = F(1110)^1_1 \eqimage{"3_Anyon_10_b2_Right"}{8ex} = F(1110)^1_1 R^{11}_1 \eqimage{"3_Anyon_10_Right"}{8ex} = F(1110)^1_1 R^{11}_1 F(1110)^1_1 \eqimage{"3_Anyon_10_Left"}{8ex} = e^{3\pi i/5} \ket{N},\displaybreak[0]
\end{gather*}
\end{widetext}
resulting in the block diagonal braid matrix
\begin{equation}
\sigma_2 = \begin{pmatrix}
\phi^{-1} e^{4\pi i/5} & \phi^{-1/2} e^{-3\pi i/5} & 0 \\
\phi^{-1/2} e^{-3\pi i/5} & -\phi^{-1} & 0 \\
0 & 0 & e^{3\pi i/5}
\end{pmatrix}.
\refstepcounter{equation}\tag{\theequation}\label{eqn:matrix_b2_3anyon}
\end{equation}

Note that the matrices \eqref{eqn:matrix_b1_3anyon} and \eqref{eqn:matrix_b2_3anyon} are block diagonal, and the braiding matrices for larger numbers of anyons are also block diagonal. It is not possible for braiding to convert a $\ket{0}$ or $\ket{1}$ into an $\ket{N}$ or vice versa. This is because $\ket{0}$ and $\ket{1}$ have an overall `charge' of 1, while $\ket{N}$ has an overall `charge' of 0. As such, one may optionally simplify the basis by considering only the basis states with a particular overall `charge', where the final fusion result is identical.

By this method, it is straightforward to algorithmically construct the braiding matrices.

\begin{enumerate}
	\item Specify the basis states, indexing each fusion site.
	\item For each braiding operation, determine which F and R moves are necessary, noting which indices each of those moves act upon.
	\item Apply that operation to each basis vector, obtaining a linear combination of basis vectors with coefficients comprised of F and R coefficients.
	\item Substitute in the values of the F and R coefficients.
	\item Use the operation on each basis vector as the columns of the matrix.
\end{enumerate}

This method is applicable to any anyon braiding model for any arrangement of anyons, provided the values of the F and R coefficients can be determined.

\begin{figure}
	\includegraphics[width=\linewidth]{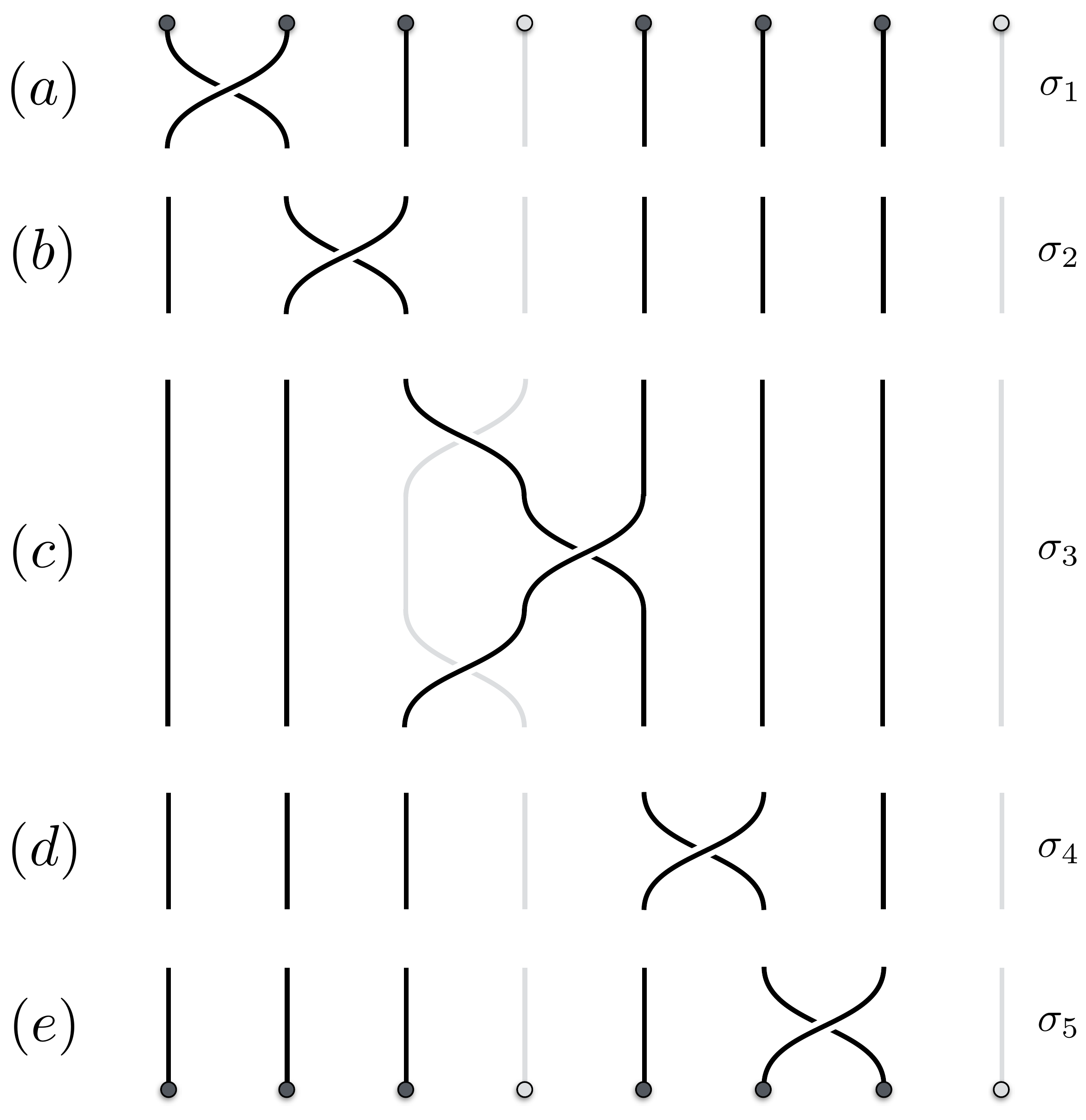}
	\caption{The elementary braiding operations which act on two qubits (not including their inverses). Time points downwards in these diagrams. The fourth and eighth strands are gray to signify that no braiding is done with them. Note that in $\sigma_3$, the braid occurs in front of the fourth strand, with the fourth strand not topologically involved in the braid. For consistency with the convention of  \cite{Bonesteel_Braid_Topologies}, the four anyons on the right is the first qubit and the four anyons on the left is the second qubit.}
	\label{Fig:2q_braids}
\end{figure}

\begin{figure}
	\includegraphics[width=0.7\linewidth]{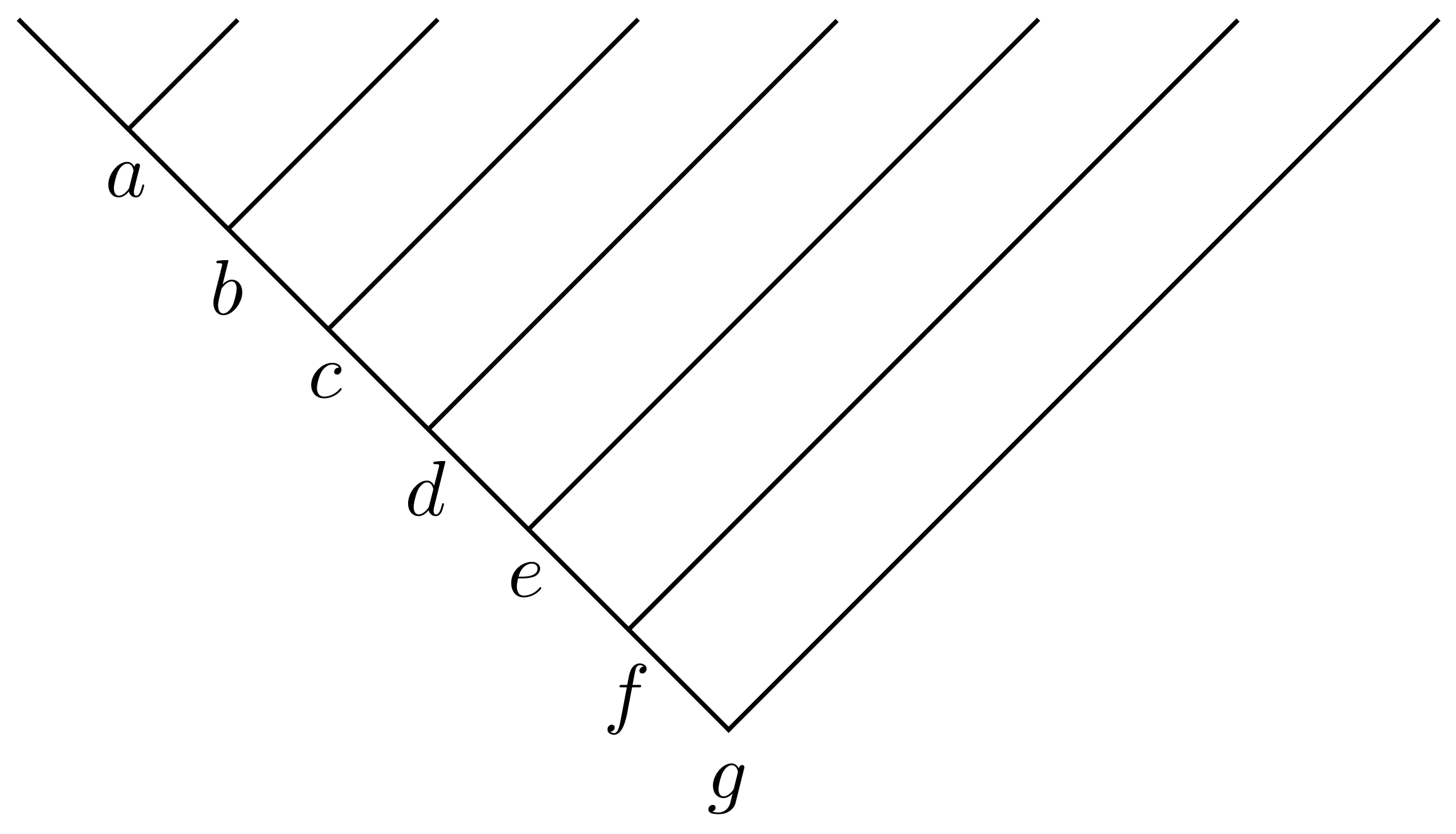}
	\caption{Fusion tree for eight anyons fused consecutively. Each of the fusion sites are indexed $a$ to $g$.}
	\label{Fig:eight_anyon_tree}
\end{figure}

As another example, we will consider the two-qubit eight anyon braiding operators presented in \Figref{Fig:2q_braids}. For computation, the system is grouped into qubits of four anyons, with the rightmost qubit being the first qubit for consistency with the notation of  \cite{Bonesteel_Braid_Topologies}. We will consider the fusion tree given by \Figref{Fig:eight_anyon_tree}.

The basis states of this system can be considered as bitstrings denoting the fusion outcomes, in the order `$abcdefg$'. We will consider just the states with an overall `charge' of zero.

The four states describing the qubits are $\ket{00}=`0101010$', $\ket{01}=`1101010$', $\ket{10}=`0101110$' and $\ket{11}=`1101110$'. The other nine basis states, which are nominally non-computational states so do not receive any special labels, are `1011010', `1011110', `1010110', `0111010', `0111110', `0110110', `1111010', `1111110', and `1110110'.

Next, we find the sequences of F and R moves necessary to apply each of the operators used in \Figref{Fig:2q_braids}.

To apply $\sigma_{1}$, as in \Figref{Fig:2q_braids}(a), we need to apply an R move, where the fusion outcome is that at `$a$', $R^{11}_a$.

To apply $\sigma_2$, as in \Figref{Fig:2q_braids}(b), we need to apply an F move involving `$a$' and `$b$', moving the location of `$a$', then apply an R move involving the newly moved `$a$', then another F move to return the fusion tree to its original arrangement. $F(111b)^a \rightarrow R^{11}_a \rightarrow F(111b)^a$. In computation, it is necessary to remember that, when applying an F move which modifies the site indexed `$a$', the value of `$a$' changes as well, and a superposition of states is often produced.

To apply $\sigma_4$, as in \Figref{Fig:2q_braids}(d), we need to apply the sequence of moves $F(c11e)^d \rightarrow R^{11}_d \rightarrow F(11ce)^d$.

To apply $\sigma_5$, as in \Figref{Fig:2q_braids}(e), we need to apply the sequence of moves $F(d11f)^e \rightarrow R^{11}_e \rightarrow F(11df)^e$.

To apply the slightly more complicated $\sigma_3$, as in \Figref{Fig:2q_braids}(c), we work upwards, undoing one twist at a time. $F(a11c)^b \rightarrow R^{11}_b \rightarrow F(11ac)^b \rightarrow F(b11d)^c \rightarrow R^{11}_c \rightarrow F(11bd)^c \rightarrow F(a11c)^b \rightarrow (R^{11}_b)^{-1} \rightarrow F(11ac)^b$.

\begin{figure}
	\includegraphics[width=\linewidth]{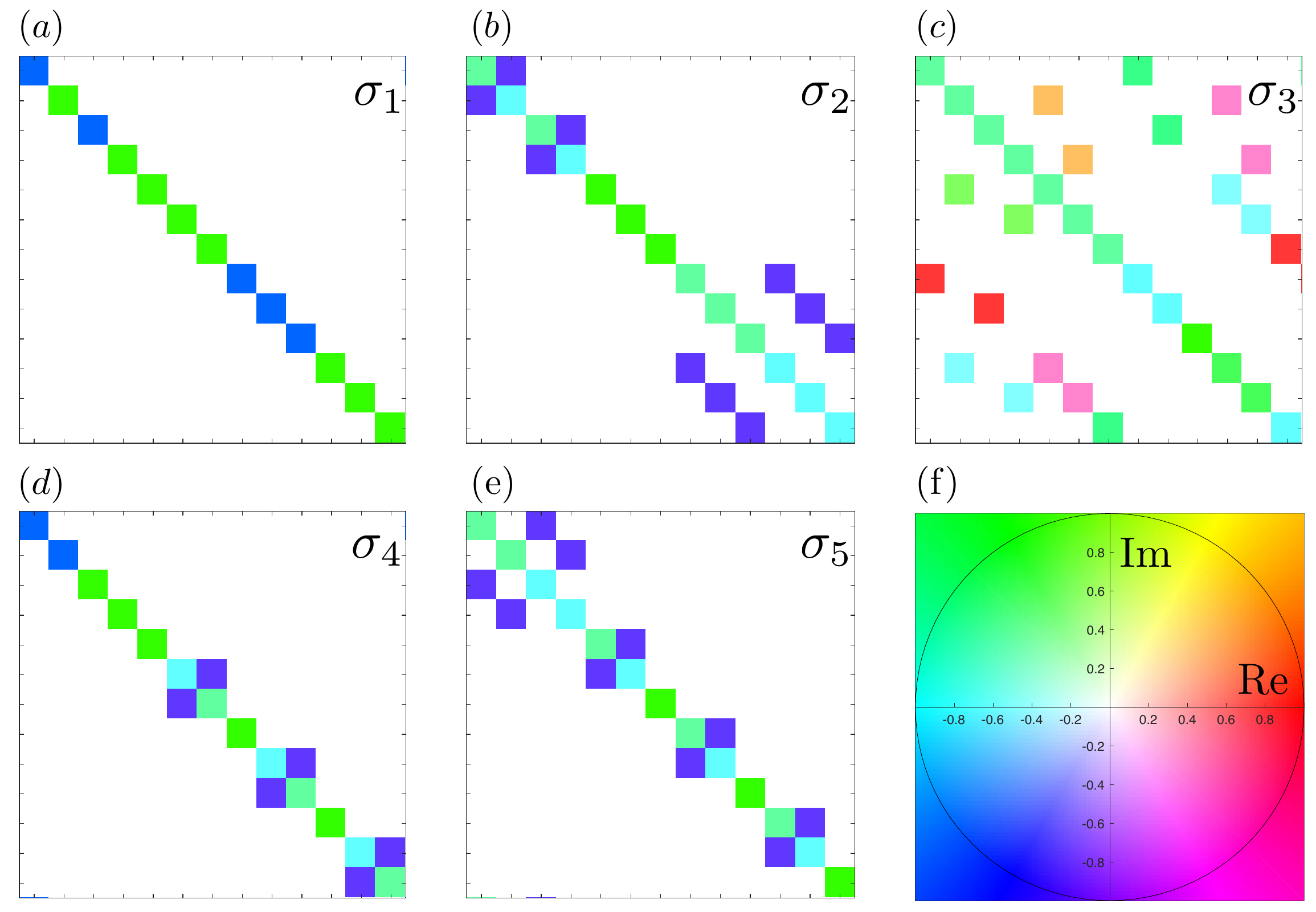}
	\caption{Representation of braiding matrices (a)-(e) for the two-qubit operations in \Figref{Fig:2q_braids}. The phase and magnitude of complex numbers are represented here with hue and saturation, respectively, as indicated by the colormap (f), with white squares being equal to zero. The first four states are the computational states, while the remaining nine states are non-computational states. Observe that only $\sigma_3$ has leakage between computational and non-computational states because it has non-zero off-diagonal elements causing transitions between the $4\times4$ computational block and the $9\times9$ non-computational block.}
	\label{Fig:2q_matrices}
\end{figure}

Given the basis states and the sequences of moves, it is then simply a matter of calculation to evaluate numerical values for these braid matrices, which in this case are $13\times 13$. The matrices are presented in \Figref{Fig:2q_matrices}.

The constructed braid matrices \eqref{eqn:matrix_b1_3anyon} and \eqref{eqn:matrix_b2_3anyon}, and all other braid matrices, obey the braiding identities \eqref{eqn:braid_identities}. However, we note that the convention for the direction of braiding and the direction in which braids are added from the braidword for anyons, as used by  \cite{Anyon_Computing_Nayak,Bonesteel_Braid_Topologies}, is the opposite to the convention for braids of a general topological nature as used by  \cite{Aharonov_Jones_Algorithm1,Knot_Majorana_Kauffman}. Anyon braids can be made to relate to the other convention by drawing the worldlines with time pointing upwards, rather than downwards, as in  \cite{Vadym_Compiling}, or alternatively the direction of general topological braids can be reversed to match the convention for anyonic braids, as in  \cite{Pachos_TQC_Book}. The choice of convention for the chirality of braiding is, to an extent, arbitrary, for two anyon models with opposite chirality are simply the complex conjugates of each other \cite{Preskill_Lecture}, although it is important that consistency is maintained once a convention is chosen.

\begin{figure}[!t]
	\includegraphics[width=0.8\linewidth]{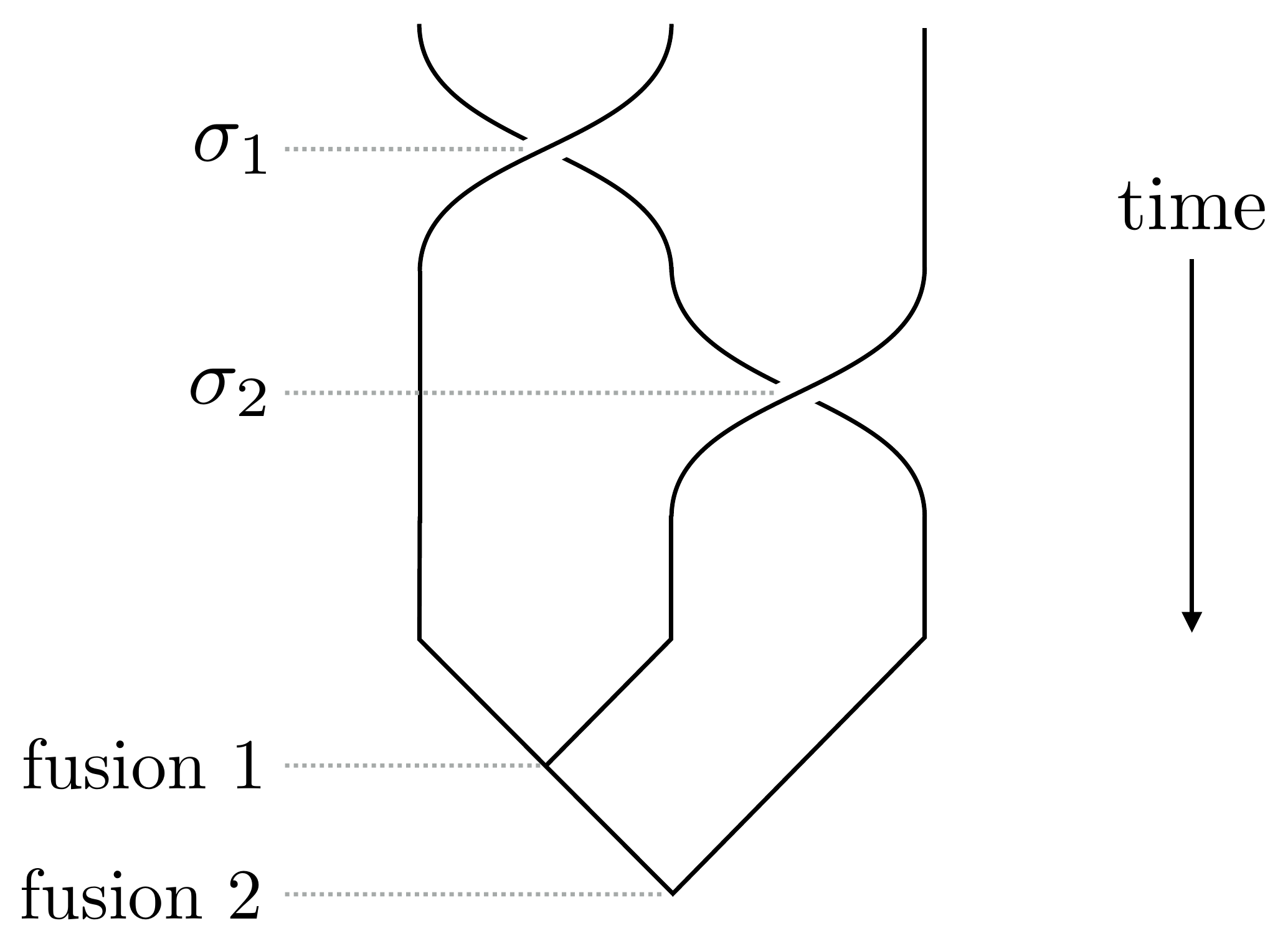}
	\caption{The elementary braid $\sigma_1$ chronologically followed by $\sigma_2$, followed by fusion. The braidword is thus $\sigma_1 \sigma_2$.}
	\label{Fig:testbraid_b1b2}
\end{figure}

Finally, we will make a note about the chronology of the braids and the order of matrix multiplication in this braiding formalism. Consider the braid in \Figref{Fig:testbraid_b1b2}, where the first and second anyons are exchanged ($\sigma_1$), then the second and third anyons are exchanged ($\sigma_2$), then the anyons are fused. This braiding operation acts on the state $\ket{\psi_i}$, which is the fusion outcome that would occur without this braiding. However, while time points downwards, the derivation of the braiding operations works upwards, starting at the fusion and unraveling the braid via R moves. One can derive the effects of this particular braid using the algorithm above, by applying an F move, then an R move, then an F move, then an R move. Alternatively, one can left-multiply the initial state by $\sigma_2$ to unravel the elementary braid closest to the fusion, then left-multiply by $\sigma_1$ to unravel the next elementary braid. Both methods are equivalent. The state of the system at fusion is $\ket{\psi_f}=\sigma_1 \sigma_2 \ket{\psi_i}$.

Note that matrices are left-multiplied onto the initial state in reverse chronological order. To apply the braiding operations in chronological order, the braiding matrices must be right-multiplied in chronological order, then finally the initial state must be right-multiplied to that matrix to find the final state. This is an important detail to remember when constructing these braids.

\subsection{Using Fibonacci Anyons for Computing} \label{sec:fib_compute}

Because braiding non-Abelian anyons performs a unitary transformation, it can be used for quantum computation. The advantage of anyonic systems for quantum computation over other quantum computers is that the quantum states and computations are topologically protected. The computations, performed by braiding the anyons, are highly resistant to local perturbations, because the operation performed depends only on the topology of the braid. The state of the system is encoded non-locally and can only be measured by actually fusing the anyons, and not by any local interactions, which makes the states resistant to decoherence. This is a topological quantum computer, and it has intrinsic fault tolerance which makes it potentially superior to other forms of quantum computers \cite{Anyon_Computing_Freedman,Anyon_Computing_Nayak,Kitaev_Anyon_Computing,Preskill_Lecture}, provided an effective physical implementation can be developed.

Topological quantum computers are not entirely immune to errors. The anyons must be kept sufficiently distant to prevent interactions and quantum tunneling between them \cite{Anyon_Computing_Freedman,Kitaev_Anyon_Computing}. The anyons must be moved sufficiently slowly that the system evolves adiabatically \cite{Simanek1992a,Virtanen2001a}, without causing excitations from the motion \cite{Pachos_TQC_Book}. Thermal fluctuations can create spurious anyon pairs \cite{Simula2006a,Hadzibabic2006a,Kosterlitz2017a}, which might braid non-trivially with the intentionally created anyons \cite{Anyon_Computing_Nayak,Kitaev_Anyon_Computing}. And multi-qubit operations can cause leakage into a non-computational state \cite{Carlos_Anyon_Computing,Qubit_Leakage}. Fortunately, there exists numerous error correction protocols that can suppress these kinds of errors in a reasonably efficient manner \cite{ErrorCorrectionThesis,Burton_Fibonacci_Error,Wootton_Error_Correction,Carlos_Anyon_Computing}. A physical implementation of a topological quantum computer will need to implement similar protocols to remove erroneous anyon pairs if they form, as well as dealing with any other errors. Here, however, we will focus on just the computation that can be performed by braiding the anyons.

Fibonacci anyons satisfy an important consideration for quantum computing, which is universality, a property not shared by all non-Abelian anyons \cite{Anyon_Computing_Nayak,Bonesteel_Fibonacci_General,Pachos_TQC_Book}. A universal quantum computer is capable of performing any quantum computation. By using braiding of Fibonacci anyons alone, it is possible to approximate any unitary matrix to arbitrary accuracy, up to an overall phase \cite{Anyon_Computing_Brennen,Anyon_Computing_Freedman,Anyon_Computing_Nayak,Bonesteel_Fibonacci_General}. Formally, the set of elementary braiding operations for 3 Fibonacci anyons forms a dense set in ${\rm SU(2)}$, and similar results apply for more anyons \cite{Modular_Functor}. In a quantum computer, universality requires a minimum set of quantum gates, such as the set of single qubit braids and a two qubit gate such as the controlled-NOT (CNOT) gate \cite{Quantum_Basics}. Single qubit gates are relatively easy to construct for Fibonacci anyons, and the CNOT gate has also been constructed \cite{Bonesteel_Braid_Topologies,Improved_Injection_Weave}, demonstrating that Fibonacci anyons are indeed universal for quantum computation.

Not all anyon models, such as Ising anyons, provide universality by braiding alone \cite{Sarma2015a}. Such computers need to be supplemented by non-topological operations in order to achieve universality \cite{Pachos_TQC_Book,Shor_TQC_Resources,Ising_Universal}. There even exist schemes for topological quantum computing which entirely replace braiding with different topological operations such as measurement \cite{Measurement_Only_TQC}. However, we shall only consider here a topological quantum computer constructed from Fibonacci anyons and using only braiding to perform computations.

\subsubsection{Topological Qubits}

In our computer, we define a qubit to be composed of four anyons in a row of zero net `charge'. This has the bases described by \Figref{Fig:b4_fusion_tree}. Being a two-state system, this is an appropriate choice for a qubit. The qubit is initialised by pair creation, which ensures that each pair has zero initial `charge', which makes the initial state have the `010' fusion outcome, which will be labeled $\ket{0}$. The `110' outcome, where the first pair of anyons fuse to 1, will be labeled $\ket{1}$. Multiple qubits are arranged in a row.

While the four anyons need to be created to form the qubit, it can be observed that a system of four anyons with zero net overall `charge' is equivalent, in terms of possible fusion outcomes, to a system of three anyons with net overall `charge' of 1. In line with the notation of  \cite{Anyon_Computing_Nayak,Bonesteel_Braid_Topologies}, we can simplify our system to consider braiding with just three anyons in each qubit for the purposes of computation, as in \Figref{Fig:3anyon_bases}. This is justifiable because, with four anyons with zero net `charge', braiding the first and second anyons performs an identical operation to braiding the third and fourth anyons, making the latter elementary braiding operation redundant.

Our basis states are $\ket{0}$ and $\ket{1}$, which are characterised by the first fusion outcome. The $\ket{N}$ state is not possible if the overall `charge' of the four-anyon qubit is zero, making it a non-computational state. It is not possible for braiding within a single qubit to cause leakage from the computational states $\ket{0}$ and $\ket{1}$ into a non-computational state $\ket{N}$, although leakage occurs during braiding between two qubits and must be carefully managed and minimised \cite{Qubit_Leakage,Improved_Injection_Weave}.

By omitting the state $\ket{N}$, our braid matrices for a single qubit become, cf. Eqns~(\ref{eqn:matrix_b1_3anyon}) and (\ref{eqn:matrix_b2_3anyon}),
\begin{gather}
\sigma_1 = \begin{pmatrix}
e^{-4\pi i /5} & 0 \\
0 & e^{3\pi i /5}
\end{pmatrix}, \\
\sigma_2 = \begin{pmatrix}
\phi^{-1} e^{4\pi i/5} & \phi^{-1/2} e^{-3\pi i/5}\\
\phi^{-1/2} e^{-3\pi i/5} & -\phi^{-1}
\end{pmatrix}.
\end{gather}

By simplifying the computational space to include only three anyons per qubit, we reduce the number of different elementary braiding operations, making it easier to compile braids, while still maintaining universality. The fourth anyon in the qubit still exists, for physical reasons, but no braiding is performed with it for computation.

\subsubsection{Computation by Braiding}

Once all the qubits have been initialised into the $\ket{0}$ state by pair creation, braiding may be performed on the anyons to perform the computation. How specific braids which perform specific computations (gate operations) are found is detailed in Section \ref{sec:compile}. If an initial state other than the $\ket{0}$ state is required, then a braid equivalent to the NOT gate can be performed to change the required $\ket{0}$ qubits into $\ket{1}$ qubits.

In two qubit braids with a total of 8 anyons to fuse, the vector space is 13 dimensional. Only four of these basis states are in our computational subspace. The remaining 9 dimensions contain non-computational states, where the overall `charge' of each of the qubits is not zero. In particular, the elementary braid $\sigma_3$ (in \Figref{Fig:2q_braids} and \Figref{Fig:2q_matrices}), which is the braid between the two qubits, results in leakage into non-computational states. As such it is necessary to carefully construct the braids in such a way that leakage into non-computational states is minimised.

Hypothetically, one could remove this constraint on two qubit braids by changing the computational space to be the entire fusion space of all the anyons, rather than be partitioned into qubits. However, this would come at the cost of the modularity and easy expandability that qubits provide, since the fusion space does not have a tensor product decomposition \cite{Preskill_Lecture}, and also make compatibility with qubit-based algorithms difficult. As such, we will continue to focus on qubits. In a practical implementation, the leakage errors that occur in two qubit braids could potentially be dealt with by quantum error correction schemes \cite{Carlos_Anyon_Computing}.

\begin{figure}
	\includegraphics[width=\linewidth]{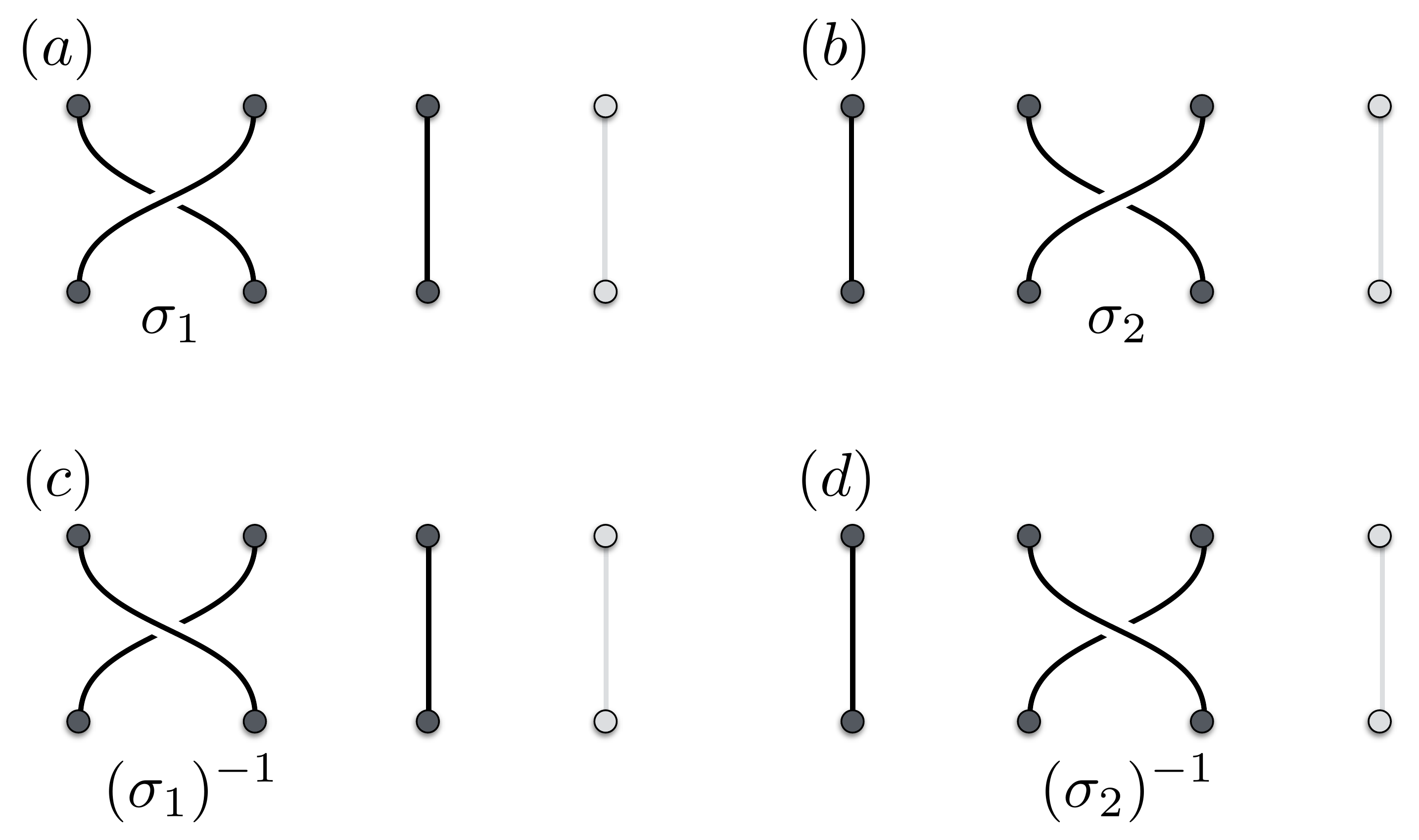}
	\caption{The elementary braiding operations $\sigma_1$ and $\sigma_2$, and their inverses $\sigma_1^{-1}$ and $\sigma_2^{-1}$, for a single qubit.  Time points downwards in these diagrams. The fourth strand is light gray to signify that no braiding is done with it. Here, the pegs corresponding to the anyons are also shown.}
	\label{Fig:1q_braids}
\end{figure}

The set of elementary single qubit braids and their inverses is in \Figref{Fig:1q_braids}. The set of elementary two qubit braids is in \Figref{Fig:2q_braids}. In our three anyon qubit convention, the fourth anyon in each qubit sits behind all the other anyons and is not involved in braiding.

\subsubsection{Measurement}

Once all the braiding has been performed, the state of the computer is measured by fusing anyons together sequentially from left to right. Each qubit should yield the result of either $\ket{0}$ or $\ket{1}$, as described in \Figref{Fig:b4_fusion_tree}. Any other fusion result indicates leakage into a non-computational state, and the computer would return an error. Otherwise, the computer would return a bitstring corresponding to the measured states of the qubits. This is the essential operating principle of a topological quantum computer.

One could measure the state of a single qubit by only fusing the first two anyons in that qubit, assuming the computer is in a computational state. However, if one wishes to perform further computation with that qubit after the measurement, unless a method to split an anyon into two anyons is present then that qubit would have been effectively destroyed, and that qubit would have to be freshly initialised. However, since braiding Fibonacci anyons is universal for quantum computation, it is not necessary to supplement braiding with intermediate measurements.

\begin{figure}
	\includegraphics[width=0.8\linewidth]{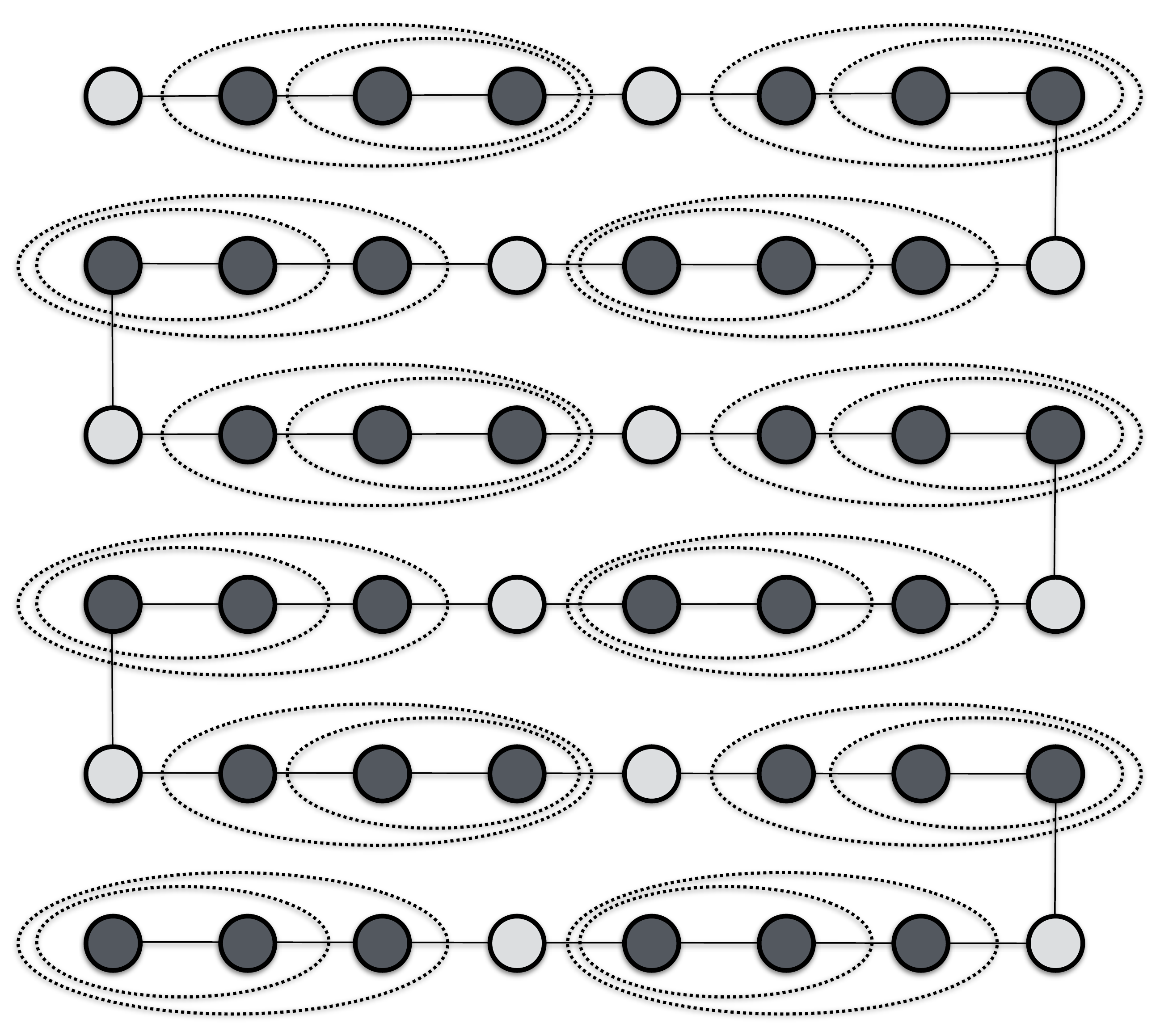}
	\caption{A square grid of anyons, which corresponds to a folded line of anyons.}
	\label{Fig:anyon_grid}
\end{figure}

Finally, we mention that some works which investigated systems of anyons considered two-dimensional arrays of anyons rather than linear chains \cite{ErrorCorrectionThesis,Burton_Fibonacci_Error,Wootton_Error_Correction,Kitaev_Anyon_Computing}. These works, however, either focused on the implementation of error correction codes or worked with anyons in a largely theoretical framework, rather than performing any quantum computation. For this work, we will focus on the simpler one-dimensional array of anyons, where all the qubits are arranged in a row and can only interact with their immediate neighbours. One can interconvert between a grid and a line by observing that a grid can be made from a folded line, as in \Figref{Fig:anyon_grid}. Similarly, the qubit density may be further increased straightforwardly by folding the two-dimensional sheet of anyons to extend the qubit manifold to the third spatial dimension.

\subsection{Compiling Braids for Computation}\label{sec:compile}

In order to perform any specific quantum algorithm in a specific quantum computer, it is necessary to find a sequence of physical operations within the quantum computer which approximates the required quantum gates for that algorithm to a desired accuracy. In classical computers, this involves deciding how many bits to be used to store each number, which determines the machine precision. In the case of anyons, this sequence of physical operations is a braid. A braid which exactly implements a given operation does not always exist, so approximations are usually necessary \cite{No_Exact_Braids}. While a braid approximating an arbitrary unitary matrix up to an overall phase to arbitrary accuracy theoretically exists, finding this braid is typically non-trivial.

\subsubsection{Error Metrics}

In searching for these braids, it is necessary to have a metric for how closely one matrix matches another. This can be done using the operator norm of the difference between those two matrices \cite{Bonesteel_Braid_Topologies,Compiling_Hashing,Improved_Injection_Weave}. The operator norm gives the largest value by which a matrix can change the length of a vector. If two matrices are equal, then their difference will give the zero matrix, which has an operator norm of zero. If two matrices are dissimilar, then their difference will be dissimilar to the zero matrix, and so have a larger operator norm. A smaller value for this error metric indicates a better approximation.

Explicitly, the operator norm can be evaluated using
\begin{equation}
\abs{\abs{A}}=\sqrt{\operatorname{maxEigenvalue}(A^{\dagger} A)}. \label{eqn:operator_norm}
\end{equation}

It is also important to consider that, while Fibonacci braiding forms a dense set in $\rm SU(2)$ \cite{Modular_Functor}, the group of $2\times 2$ unitary matrices with a determinant of 1, it does not form a dense set in $\rm U(2)$, the group of arbitrary $2\times 2$ unitary matrices. This can be shown by observing that the determinants of the elementary Fibonacci braiding matrices are all $e^{\pm \pi i/5}$. This means that the determinant of the matrix representing any braid is $e^{n\pi i/5}$, where $n$ is some integer. This means that it is not possible for braiding of Fibonacci anyons to approximate a unitary matrix which has another determinant. However, it is possible for braiding to approximate that matrix times a certain overall phase factor.

The determinant of an arbitrary unitary matrix $U$ is $\det(U)=e^{i\phi}$, and for demonstration purposes we shall assume that $U$ is a $2\times 2$ matrix. We can change the determinant of $U$ by multiplying it by a phase factor. Suppose we want to construct a certain unitary $U'$, which is a member of $\rm SU(2)$ and differs from $U$ only by an overall phase. Then $\det(U')=e^{-i\phi}\det(U)=\det(e^{-i\phi/2}U)$, so $U' = e^{-i\phi/2}U$. Therefore, it is possible to find braids which approximate any unitary matrix $U$ up to an overall phase factor. Since in quantum mechanics the overall phase of a system cannot be measured, this overall phase is generally of no consequence.

When comparing how closely two matrices match using the operator norm, we require a way to cancel out the phase difference between the two unless we have already guaranteed that both are within $\rm SU(2)$. The determinant of a $2\times2$ unitary matrix returns the square of the phase. This gives two possible solutions for the actual phase of each unitary matrix, but one will provide the phase difference between those two matrices, while the other will be off by a factor of $-1$. For using the operator norm, we can compare both possible phase differences and choose the one that results in the closest match.

While this provides the phase difference between the two matrices, it is relatively expensive computationally. A simpler method for comparing two $2\times2$ unitary matrices while ignoring their global phase is the global phase invariant distance defined by  \cite{Vadym_Compiling},
\begin{equation}
d(U,V) = \sqrt{1-\abs{\operatorname{\rm tr}(U V^{\dagger})}/2}, \label{eqn:phase_distance}
\end{equation}
where $\operatorname{\rm tr}$ is the standard matrix trace, the sum of diagonal elements. This distance $d(U,V)$ gives a measure of the difference between $U$ and $V$ while ignoring their overall phases. If $U$ and $V$ are equal and unitary, then $UV^{\dagger}=\mathbb{I}_2$. The trace of $\mathbb{I}_2$ is 2 (hence the division by 2, to normalise the trace). \eqref{eqn:phase_distance} would then give a value of zero. If $U=e^{i \phi} V$, then $UV^{\dagger}=e^{i\phi} \mathbb{I}_2$, so the trace is $2e^{i\phi}$, and the absolute value of the trace will be 2, which would also lead to a distance of zero. As such, \eqref{eqn:phase_distance} is insensitive to the global phase of the matrices.

However, it turns out that the global phase invariant distance and the operator norm after global phases have been removed are actually equivalent, up to a normalisation constant of $\sqrt{2}$.

Consider the unitary matrices $A,B\in \rm SU(2)$. Note that $-B$ is also a member of $ \rm SU(2)$. As such, when finding the operator norm of the difference between these two matrices while ignoring the global phase between them, we will want to take the minimum of $\abs{\abs{A-B}}$ and $\abs{\abs{A+B}}$, such that
\begin{equation}
\begin{gathered}
\abs{\abs{A\mp B}} = \sqrt{\operatorname{Eig}[(A\mp B)^{\dagger}(A\mp B)]} \\
= \sqrt{\operatorname{Eig}[A^{\dagger} A +B^{\dagger} B \mp A^{\dagger} B \mp B^{\dagger} A]}.
\end{gathered}
\label{eqn:opnorm_proof1}
\end{equation}
Let $C=A^{\dagger}B$. Note that $C\in \rm SU(2)$, so it is a matrix of the form
\begin{equation}
C=\begin{pmatrix}a & b \\ -b^* & a^* \end{pmatrix},\text{ where }\abs{a}^2+\abs{b}^2 = 1.
\label{eqn:opnorm_proof2}
\end{equation}
Hence,
\begin{equation}
\begin{gathered}
\abs{\abs{A\mp B}} = \sqrt{\operatorname{Eig}[2\mathbb{I} \mp (C + C^{\dagger})]} \\
= \sqrt{\operatorname{Eig}\begin{pmatrix}
	2 \mp (a + a^*) & b - b \\
	b^* - b^* & 2 \mp (a^* + a)
	\end{pmatrix}} \\
= \sqrt{\operatorname{Eig}\begin{pmatrix}
	2 \mp 2\Re(a) & 0 \\
	0 & 2 \mp 2\Re(a)
	\end{pmatrix}} \\
= \sqrt{2\mp 2\Re(a)}.
\end{gathered}
\label{eqn:opnorm_proof3}
\end{equation}
Taking the minimum of \eqref{eqn:opnorm_proof3}, we obtain
\begin{equation}
\abs{\abs{A\mp B}} = \sqrt{2 - 2\abs{\Re(a)}}. \label{eqn:opnorm_proof4}
\end{equation}
Now consider the global phase invariant distance.
\begin{equation}
\begin{gathered}
\sqrt{2} d(A,B) = \sqrt{2-\abs{\operatorname{\rm tr}(A B^{\dagger})}} \\
= \sqrt{2-\abs{\operatorname{\rm tr}(B^{\dagger} A)}} = \sqrt{2-\abs{\operatorname{\rm tr}(C^{\dagger})}} \\
= \sqrt{2 - \abs{a^* + a}} = \sqrt{2 - 2\abs{\Re(a)}}. \label{eqn:opnorm_proof5}
\end{gathered}
\end{equation}

Since \eqref{eqn:opnorm_proof4} and \eqref{eqn:opnorm_proof5} are the same, this proves that these two apparently different metrics are actually the same. The computational advantage of \eqref{eqn:phase_distance} is that it removes any global phase difference between the two matrices in a facile manner, but both will produce the same result up to a normalisation constant.

For consistency with  \cite{Bonesteel_Braid_Topologies}, we will report errors of braids with the normalisation inherent to the operator norm, which is $\sqrt{2}$ times more than \eqref{eqn:phase_distance}.

\subsubsection{Compiling Single Qubit Braids}

The simplest method for finding a braid which approximates a given unitary is by an exhaustive search, checking each possible braidword up to a certain length to see if it constructs the desired unitary matrix to within a desired accuracy. Exhaustive search was the method used in this work to find braids. The time taken for this brute force method grows exponentially as $\mathcal{O}(b^L)$, where $b$ is the number of elementary operations in the search and $L$ is the length of the braid.

There are several optimisations we made to make the exhaustive search more efficient. Most of these optimisations work on eliminating braidwords which are equivalent to previously investigated braidwords from the search. If an elementary braid and its inverse were adjacent to each other, the braid was rejected. By noting that, for Fibonacci anyons, $\sigma_i^6 = \sigma_i^{-4}$ (or, equivalently, $\sigma_i^{10}=\mathbb{I}$), any braidword with six or more consecutive identical elements was rejected. Due to topology and braiding identities, it was noted that certain triplets of elementary braids were equivalent to other triplets (eg. $\sigma_1 \sigma_2 \sigma_1 = \sigma_2 \sigma_1 \sigma_2$, $\sigma_2 \sigma_1 \sigma_2^{-1} = \sigma_1^{-1} \sigma_2 \sigma_1$), so from each pair only one pattern was kept while the other rejected. In certain cases, applying these equivalences would result in cancellation from adjacent inverses, which led to a selection of four element patterns which were also rejected. By these optimisations, we reduced the number of braidwords of length 18 or less to 33,527,163, which is a tiny fraction of the potential 91,625,968,980 braidwords.

There are two more optimisations that were made, pertaining to trading memory for speed. Generating braidwords of length $n$ can be done by appending elements to the end of the valid braidwords of length $n-1$. This avoids generating many braidwords which contain patterns which would lead to them being rejected for redundancy, at the cost of using more memory since the braids have to be recorded. However, using the other optimisations, the memory requirements were not onerous. Finally, the generated valid braidwords could be saved to file and retrieved later, saving on computation at later times.

\begin{figure}
	\includegraphics[width=\linewidth]{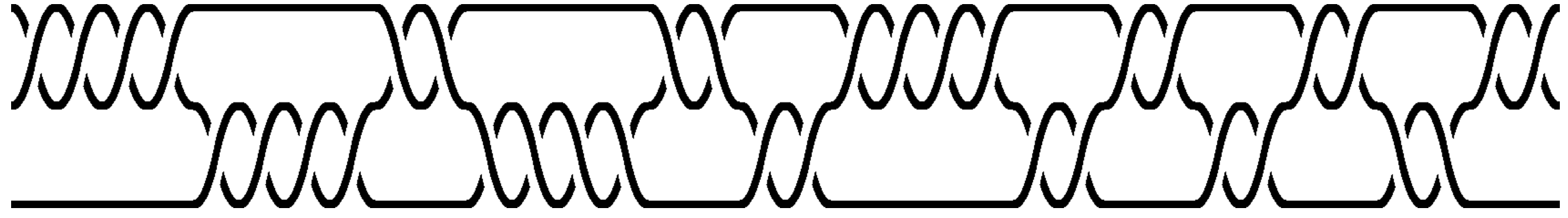}
	\caption{A weave approximating the Hadamard gate, $\tfrac{1}{\sqrt{2}}\left(\begin{smallmatrix}1&1\\1&-1\end{smallmatrix}\right)$, with an error of 0.003. Time points to the right in this diagram.\raggedright The braid consists of 34 elementary braiding operations, and has the braidword $\sigma_{2}^{-4} \sigma_{1}^{-4} \sigma_{2}^{2} \sigma_{1}^{4} \sigma_{2}^{2} \sigma_{1}^{-2} \sigma_{2}^{-4} \sigma_{1}^{-2} \sigma_{2}^{-2} \sigma_{1}^{-2} \sigma_{2}^{-2} \sigma_{1}^{2} \sigma_{2}^{-2}$, which corresponds to the matrix $e^{3.4558i}\tfrac{1}{\sqrt{2}}\left(\begin{smallmatrix}0.9997 + 0.0017i &  1.0003 - 0.0039i\\
		1.0003 + 0.0039i & -0.9997 + 0.0017i\end{smallmatrix}\right)$.\raggedright}
	\label{Fig:hadamard_weave}
\end{figure}

A special sub-class of braids are weaves, where one strand, the warp strand, moves around other stationary strands, the weft strands \cite{Bonesteel_Weave}. For braiding anyons, this can be achieved by moving one anyon while keeping the other anyons stationary. It has been shown that any operation which can be performed with a braid can also be performed with a weave \cite{Bonesteel_Weave}. Furthermore, it has also been shown that, in the three anyon case, weaves with the warp starting and finishing at the middle position (with generators $\sigma_1^{\pm2}$ and $\sigma_2^{\pm2}$) also forms a dense set in $\rm SU(2)$ \cite{Bonesteel_Fibonacci_General}, meaning such weaves can approximate any arbitrary unitary operation up to an overall phase. An example of a weave approximating a matrix is given in \Figref{Fig:hadamard_weave}.

Using the elementary weaving operations $\sigma_1^{\pm2}$ and $\sigma_2^{\pm2}$, an exhaustive search for single qubit weaves is similar to the exhaustive search for single qubit braids. Inverses could be canceled as before. There were no simple topologically equivalent patterns as for braids. However, $(\sigma_i^2)^3 = (\sigma_i^{-2})^2$, meaning that any braidword with three or more consecutive identical elementary weaves could be rejected. For braidwords with up to 18 elementary weaving operations (which corresponds to 36 elementary braiding operations), these optimisations reduced the number of braidwords from 91,625,968,980 to 178,918,056.

Given the braidword approximating an operation, the inverse can be found by simply reversing the order and inverting the signs of the powers of the braidword.

\subsubsection{Convergence of Single Qubit Braids} \label{sec:braid_convergence}

When searching for braids which approximate a target operation, it is important that these braids converge quickly. If increasing the desired accuracy requires increasing the length of the braid by an exponential amount, then computation using braiding would quickly become impractical.

\begin{figure}
	\includegraphics[width=\linewidth]{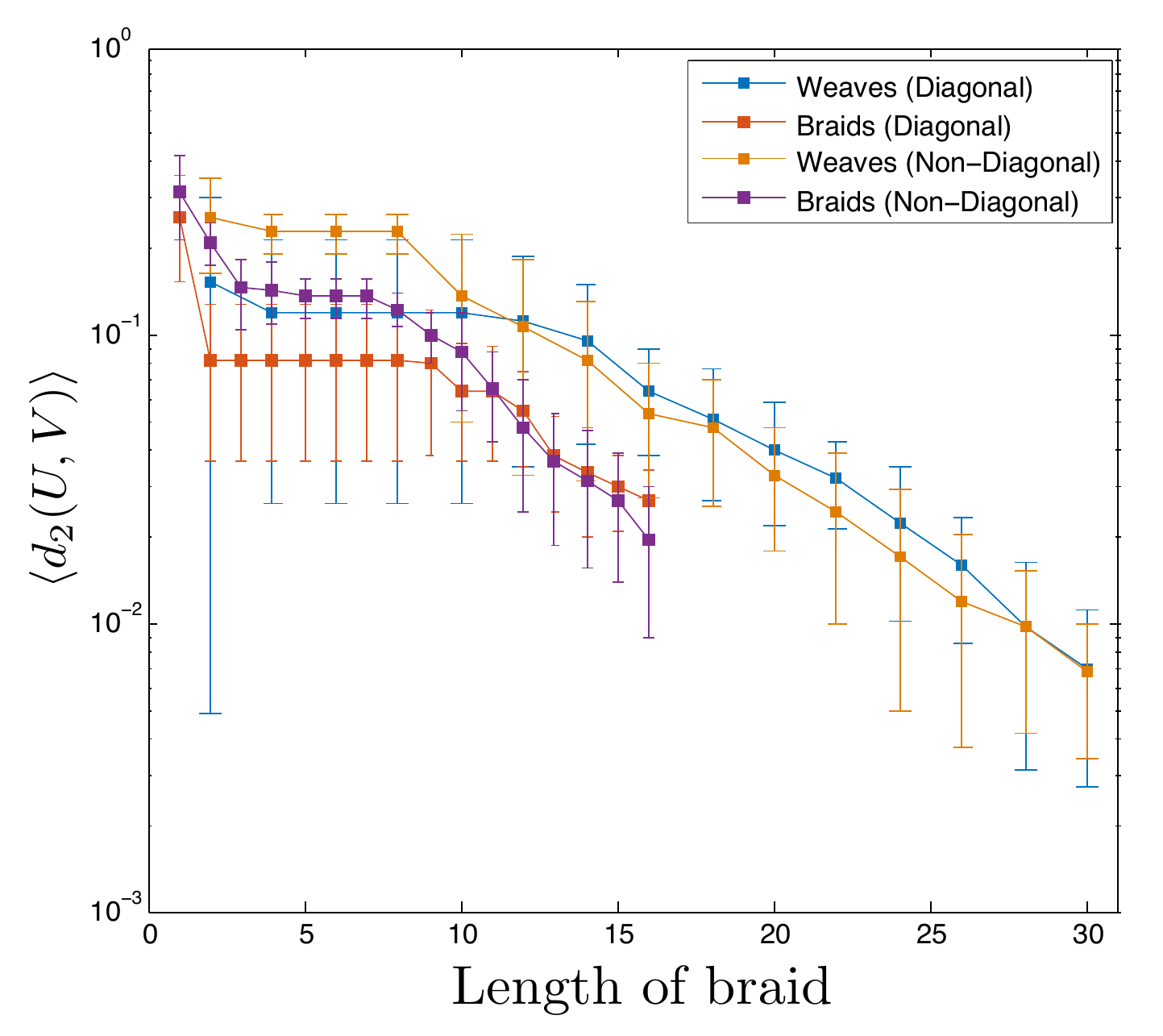}
	\caption{The average error $d_2(U,V)=\sqrt{2}{d(U,V)}$ achieved for braids and weaves of specified lengths found by exhaustive search. Error bars indicate the standard deviation across the sample of target matrices tested. Length of braid is measured in terms of elementary braiding operations, where one elementary weaving operation is equivalent to two elementary braiding operations. The convergence for diagonal and non-diagonal target matrices are plotted separately.}
	\label{Fig:braid_convergence}
\end{figure}

We have numerically tested the rate of convergence of braids found via an exhaustive search. For our sample of target matrices, we have used the matrices for the Aharonov-Jones-Landau algorithm described in Section \ref{sec:algorithm_derive}, because they are relevant to the application in this work. We chose two sets of matrices. The first set had $n=2$, and were all diagonal matrices. The second set was the second matrix for $n=3$, considering just the $2\times2$ block, which was a non-diagonal matrix. For both sets, we used values of $k$ between 4 and 13. The $k=5$ and $(n=2,k=10)$ cases were omitted for braids and the $(n=2,k=10)$ case omitted for weaves, because those cases had exact solutions. Braidwords containing up to 16 elementary braiding operations for braids and 15 elementary weaving operations for weaves were investigated. The results of this search, showing the average rate of convergence with respect to braid length, are in \Figref{Fig:braid_convergence}.

For braids and weaves longer than 10 elementary braiding operations, we can observe that the length of the braid is proportional to $\log(1/\epsilon)$, as indicated by the approximately linear slope when the error axis is plotted with a logarithmic scale. This matches the lower bound predicted by  \cite{Compiling_Proof} and observed by  \cite{Vadym_Compiling}. It means that if an error that is an order of magnitude smaller is desired, the length of the braid only needs to increase by a few elementary braids. This demonstrates that it is possible to use braiding of Fibonacci anyons to efficiently approximate any $2\times2$ unitary operation up to a global phase.

For braids less than 10 elementary braiding operations long (or 5 elementary weaving operations), they tend to converge more slowly than longer braids. This suggests that when approximating operations using this method, there is a minimum length of braid required before the braids begin converging to the desired operation.

We compared the rates of convergence for diagonal and non-diagonal target matrices. Below 10 elementary braiding operations, the diagonal matrices have a smaller error, because at that length using diagonal generators provides a better approximation than other braids of that length. As the length increases, diagonal and non-diagonal target matrices converge at the same rate. Diagonal matrices no longer have the advantage, because in order to improve upon their approximations it becomes necessary to use generators with off-diagonal elements. This suggests that above a minimum length, the form of the matrix has minimal effect on the rate of convergence, although further tests with a wider variety of matrices would be required to confirm this.

We also compared the rates of convergence of braids and weaves. For the same number of elementary braiding operations, braids converge more rapidly than weaves. This is to be expected, because weaves are a sub-set of braids, so weaves would form a less dense set in $\rm SU(2)$ than braids. However, because each elementary weaving operation consists of two elementary braiding operations, the weaves that can be searched within a fixed amount of time are almost twice as long as braids. Because the weaves are longer, in terms of elementary braiding operations, the error for the longest weaves we can search is lower than the error for the longest braids we can search. Thus, for finding accurate approximations to an operation within a fixed amount of time, weaves are better than braids, allowing us to readily approximate operations to within 1\% using an exhaustive search.

\subsubsection{Alternatives to Exhaustive Search}

While exhaustive search methods provide the shortest possible braid approximating a given operation, because they investigate every possible braid, exhaustive search is not efficient in terms of time. Performing an exhaustive search on all braidwords up to length 18, given the previously specified optimisations, takes on the order of one hour on a standard desktop computer. There exist methods which can make exhaustive searches more efficient, such as by decomposing the target matrix into other matrices \cite{Compiling_Decompose_Single_Qubit}, but these do not change the fundamental scaling of brute force. The time taken to find braids grows exponentially with the length of the braid, which quickly makes improving upon the accuracy of braids by exhaustive search impractical.

\begin{figure}
	\includegraphics[width=\linewidth]{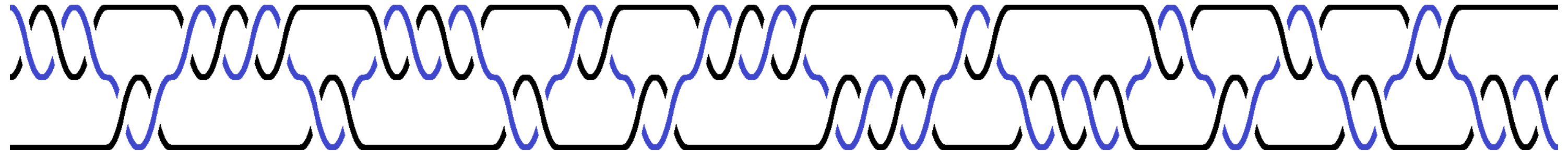}
	\caption{The injection weave presented in  \cite{Bonesteel_Braid_Topologies}. The warp strand is coloured blue. Time points to the right in this diagram. This weave approximates the identity with an error of 0.0015, and consists of 48 elementary braiding operations. The braidword is $\sigma_{2}^{3} \sigma_{1}^{-2} \sigma_{2}^{-4} \sigma_{1}^{2} \sigma_{2}^{4} \sigma_{1}^{2} \sigma_{2}^{-2} \sigma_{1}^{-2} \sigma_{2}^{-4} \sigma_{1}^{-4} \sigma_{2}^{-2} \sigma_{1}^{4} \sigma_{2}^{2} \sigma_{1}^{-2} \sigma_{2}^{2} \sigma_{1}^{2} \sigma_{2}^{-2} \sigma_{1}^{3} $.}
	\label{Fig:inject_weave}
\end{figure}

For constructing more accurate braids, the Solovay-Kitaev theorem is particularly useful \cite{Solovay_Kitaev_Explained,Kitaev_Theorem}. The Solovay-Kitaev theorem provides a method which can efficiently find, from a fixed set of quantum gates, a sequence of quantum gates which approximates the desired unitary operation to within an arbitrary accuracy.

The Solovay-Kitaev theorem can be implemented with a recursive algorithm, where the base cases are seeded by words found by exhaustive search up to a maximum length, and are combined to form increasingly more accurate approximations \cite{Solovay_Kitaev_Explained}. For a target error $\epsilon$, the time taken to run the algorithm and the length of the resulting braidword is polynomial in $\log(1/\epsilon)$ \cite{Solovay_Kitaev_Explained}.

The Solovay-Kitaev theorem is a very generalised theorem, applying to any set of invertible gates, although it is poorer than the asymptotic lower bound of being linear in $\log(1/\epsilon)$ \cite{Compiling_Proof}. There exists other algorithms which improve upon the time and length efficiency of the Solovay-Kitaev theorem. An algorithm by  \cite{Compiling_Hashing,Burrello2010a} can find target matrices in $\rm SU(2)$ with any set of generators, with a length scaling as $\mathcal{O}((\log(1/\epsilon))^2)$. Another algorithm by  \cite{Compiling_Geometric} can find braids of Fibonacci anyons in a single qubit approximating a target matrix, although its scaling was not specified. An algorithm presented by \cite{Vadym_Compiling} is specific to braiding of Fibonacci anyons in a single qubit, and efficiently finds braids which have a length which scales linearly with $\log(1/\epsilon)$. Evolutionary algorithms are also being developed which provide a generically applicable method to search for braids with the ability to trade between length and accuracy \cite{McDonald2013a,Santana2014a}.

More recently, Ross and Selinger developed a fast new probabilistic algorithm for approximating arbitrary single-qubit phase gates optimally with an expected runtime of $\mathcal{O}(\rm{polylog}(1/\epsilon$)) \cite{Ross2014a}. Their algorithm requires a factoring oracle such as a quantum computer but still achieves near-optimal performance in the absence of a factoring oracle.

While algorithms such as these will be important for constructing arbitrarily accurate braids or for constructing braids quickly and on demand, they are beyond the scope of this work. The use of exhaustive search is adequate for compiling first-order approximations to single qubit operations, at least for the purposes of demonstration.

\begin{figure}
	\includegraphics[width=\linewidth]{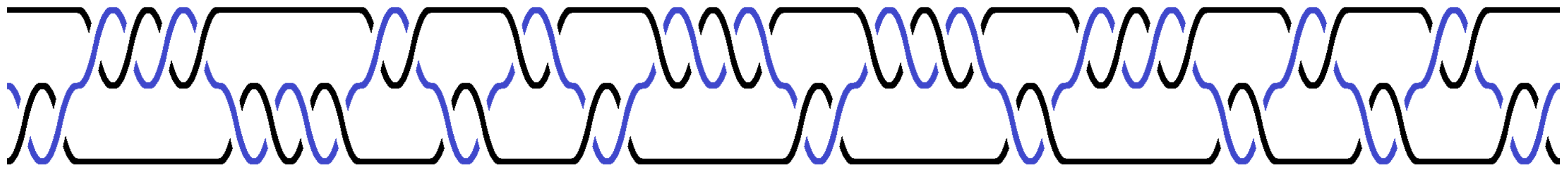}
	\caption{The NOT weave (with an overall phase factor of $i$) presented in  \cite{Bonesteel_Braid_Topologies}. The warp strand is coloured blue. Time points to the right in this diagram. This weave approximates the NOT gate with an error of 0.00086, and consists of 44 elementary braiding operations. The braidword is $\sigma_{1}^{-2} \sigma_{2}^{-4} \sigma_{1}^{4} \sigma_{2}^{-2} \sigma_{1}^{2} \sigma_{2}^{2} \sigma_{1}^{-2} \sigma_{2}^{4} \sigma_{1}^{-2} \sigma_{2}^{4} \sigma_{1}^{2} \sigma_{2}^{-4} \sigma_{1}^{2} \sigma_{2}^{-2} \sigma_{1}^{2} \sigma_{2}^{-2} \sigma_{1}^{-2} $.}
	\label{Fig:NOT_weave}
\end{figure}

\subsubsection{Compiling Two Qubit Braids} \label{sec:two_qubit_braids}

While single qubit operations have only four elementary operations, two qubit operations in our formalism have ten elementary operations. Thus, for a reasonable braidword length of 16, the number of possible braidwords increases by a factor of approximately two million from one qubit to two qubits. This factor would likely be reduced by a couple of orders of magnitude by accounting for the far-commutativity of braiding operations, but the problem of finding two qubit operations from the elementary operators is still far larger than the equivalent problem for single qubit operations.

Fortunately, the work of  \cite{Bonesteel_Braid_Topologies} has provided a simpler and more intuitive method to construct controlled two qubit operations with negligible leakage into non-computational states, including a relative of the CNOT gate. For this construction, we will consider the subclass of braids known as weaves.

Observe that if the warp strand is removed from a weave, then the weave will become the identity.

An important weave is the injection weave, which approximates the identity but moves the warp strand over by two positions, illustrated in \Figref{Fig:inject_weave}. This weave is useful for constructing other weaves because it allows the warp strand to be moved without affecting the state of the system.

Next,  \cite{Bonesteel_Braid_Topologies} found a weave corresponding to the NOT gate (with an overall phase of $i$), where the warp strand began and finished in the second position, as in \Figref{Fig:NOT_weave}.

\begin{figure}
	\includegraphics[width=\linewidth]{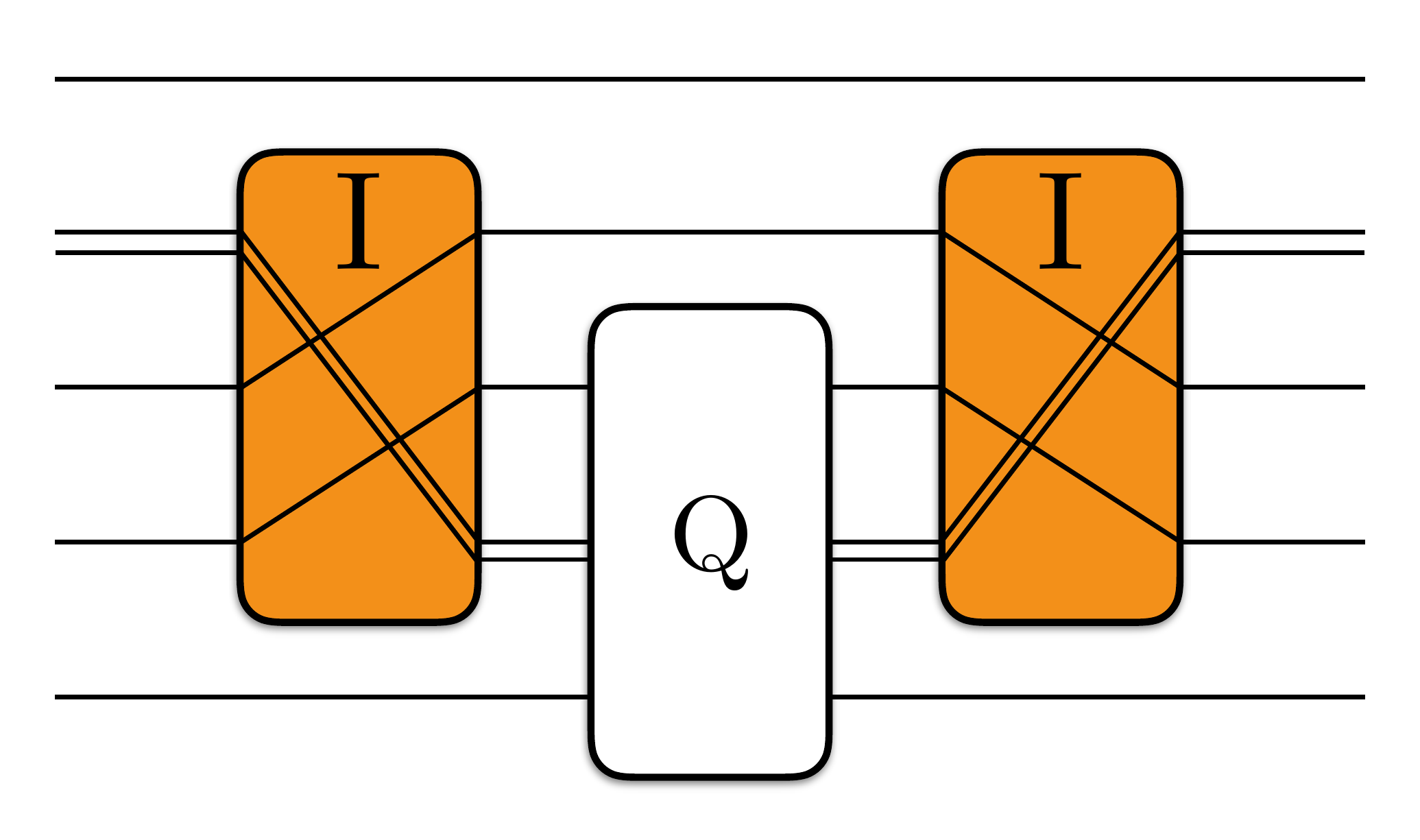}
	\caption{Schematic representation of the method used by  \cite{Bonesteel_Braid_Topologies} to produce a controlled operation. Note that the warp strand is a pair of anyons rather than a single anyon. $Q$ is the desired controlled operation, with the warp starting and ending in the middle. $I$ is an injection weave.}
	\label{Fig:CWeave_standard}
\end{figure}

\begin{figure*}
	\includegraphics[width=\linewidth]{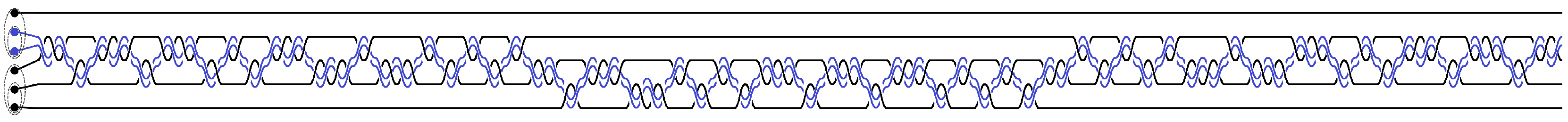}
	\caption{The weave producing the relative of the CNOT gate presented in  \cite{Bonesteel_Braid_Topologies}. Specifically, it implements the controlled version of the NOT gate multiplied by $i$, $\left(\begin{smallmatrix}0 & i \\ i & 0\end{smallmatrix} \right)$, with an error of 0.0007 and leakage of $6\times 10^{-6}$. The blue pair of strands are the warp strands. Time points to the right in this diagram. Counting the pair of anyons in the warp strands separately, this weave consists of 280 elementary braiding operations.}
	\label{Fig:CNOT_weave}
\end{figure*}

Then, weaving with a pair of anyons was performed. Specifically, the pair of anyons on the left hand side of the first qubit was selected as the warp strand. This pair has `charge' 0 if the qubit is in the $\ket{0}$ state and `charge' 1 if the qubit is in the $\ket{1}$ state. This means that any weave performed with this pair of qubits performs an operation other than the identity if and only if the first qubit is in the $\ket{1}$ state. This satisfies the controlled part of any controlled operation. To perform the CNOT gate,  \cite{Bonesteel_Braid_Topologies} used an injection weave to move the warp anyons into the second qubit, performed the NOT weave, then used the inverse injection weave to return the warp anyons to their original positions, as illustrated in \Figref{Fig:CWeave_standard} and \Figref{Fig:CNOT_weave}. If the first qubit is in the $\ket{1}$ state, the NOT weave is performed on the second qubit (with an additional phase shift of $i$). If the first qubit is in the $\ket{0}$ state, the identity is applied. By using the injection weaves to perform braiding between the two qubits, negligible leakage into non-computational states occurs. If a tighter threshold for leakage errors is required, then a more accurate injection weave can be compiled.

This method used for constructing the CNOT gate could be used for constructing any arbitrary controlled operation. The general method, illustrated in \Figref{Fig:CWeave_standard}, uses an injection weave to insert the pair of anyons from the control qubit into the middle of the target qubit, performs a weave $Q$ in the target qubit, with the warp strand beginning and ending at the central position, and then performs the inverse injection weave to move the pair of anyons back to their original position in the control qubit. If the pair of anyons has charge `0', with the first qubit in the $\ket{0}$ state, then no change occurs. If the pair of anyons has charge `1', corresponding to the $\ket{1}$ state, then the operation $Q$ is applied to the target qubit.

While the set of weaves $Q$ form a dense set in $\rm SU(2)$, the situation is complicated slightly if the desired operation is an arbitary unitary matrix, in which case the weave $Q$ will approximate the desired operation up to an overall phase. For instance, the CNOT weave produced by  \cite{Bonesteel_Braid_Topologies} is not actually the CNOT gate, but the CNOT gate composed with a $\pi/2$ phase gate on the first qubit (which changes the phase of the $\ket{1}$ state by $\pi/2$ relative to the $\ket{0}$ state). We can correct for this unintended phase shift by applying another phase gate to the first qubit, in this case a $-\pi/2$ phase gate. In general, this phase gate may have an additional overall phase, but since this overall phase is applied to the whole system it is inconsequential.

Alternatives and variants to this method are presented in  \cite{Improved_Injection_Weave,Hormozi_Compiling,Carnahan_Diagonal_Weaves,Two_Qubit_Generic_Anyon_Braids}, but the principle of divide and conquer and dimension reduction to produce two qubit operations remains.

A special two qubit operation is the SWAP gate, where the states of two qubits are interchanged. In a topological quantum computer, where the qubits comprise of anyons with a net `charge' of zero, the SWAP gate can be exactly achieved by physically swapping the two qubits. Because each qubit has zero `charge', moving one whole qubit around another performs no operation other than changing the locations of the qubits.

Operations across larger numbers of qubits, such as controlled-controlled operations \cite{Xu2011a}, can be composed from single qubit and two qubit operations \cite{Quantum_Basics,Quantum_Textbook}.

\subsection{Simulating Generic Quantum Algorithms}\label{sec:conventionalQC}

\begin{figure*}
	\includegraphics[width=0.9\linewidth]{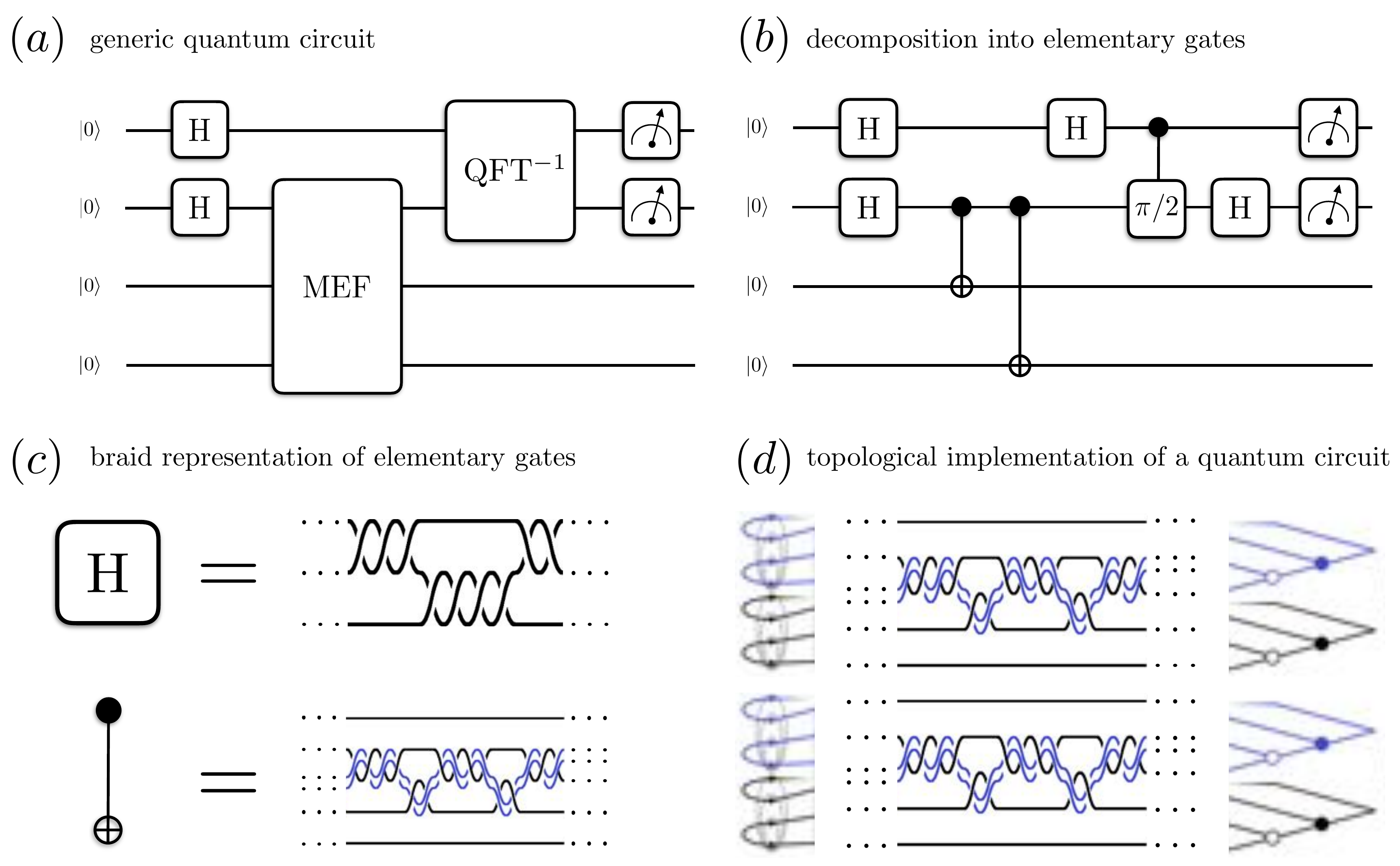}
	\caption{Modular approach to topological quantum computation. (a) Generic quantum circuit implementing a quantum algorithm that itself may be topological or conventional and may or may not involve error correcting modules. This particular instance corresponds to Shor's factorisation algorithm to factorise 15. (b) Decomposition of the generic circuit diagram in (a) into a circuit involving elementary single qubit and two-cubit controlled gates only \cite{Lu2007a}. (c) Mapping of elementary single qubit and two-qubit controlled gates such as the Hadamard and CNOT gates, onto the corresponding anyon braids and weaves. (d) Reconstruction of the full quantum circuit in (a) in terms of braiding and fusion of anyons.}
	\label{Fig:ModularTQC}
\end{figure*}

When using topological quantum computing to implement an arbitrary quantum algorithm, a modular or a direct approach could be chosen. In the modular method, which is the most readily generalisable method, the following steps are taken:

\begin{enumerate}
	\item Express the quantum algorithm as a generic quantum circuit as in \Figref{Fig:ModularTQC}(a).
	\item Decompose the quantum circuit into a set of one qubit and controlled two qubit gates as in \Figref{Fig:ModularTQC}(b), or some other appropriate set of elementary operations.
	\item Compile braidwords corresponding to each of the required elementary gate operations to desired accuracy within the chosen anyon model as in \Figref{Fig:ModularTQC}(c).
	\item Concatenate the braidwords corresponding to these elementary gates as they appear in the quantum circuit diagram as in \Figref{Fig:ModularTQC}(d) (in reverse order, as per the discussion at the end of Sec.~\ref{sec:fib_braid}).
\end{enumerate}

Note that steps 1 and 2 are generic to most forms of quantum computing, regardless of whether they are topological or conventional. It is steps 3 and 4 which are specific to topological quantum computation.

In contrast to this modular approach, in the direct method one would directly search for the optimal multi-qubit braidword corresponding to the full unitary operator of the quantum algorithm, thus avoiding compounding errors from each elementary gate operation. Although this direct approach would in principle yield the most accurate computation, finding such braids (without an access to a quantum computer) is prohibitively costly in general and the direct method is therefore restricted to a small number of specialized applications, such as that explored in Sec.~\ref{sec:exact_ajl}.

A universal topological quantum computer is capable of performing the action of any unitary quantum algorithm including Shor factoring \cite{Lu2007a}, Grover search, and Deutsch-Jozsa quantum algorithms \cite{Dicarlo2009a,Watson2018a}. In what follows we have chosen to use the AJL quantum algorithm, which itself happens to be inherently topological, to demonstrate the operation of a topological quantum computer. However, from the operational point of view, once the complete braidword corresponding to the chosen algorithm has been compiled, there is no difference whether the underlying quantum algorithm is based on topology or not.

Simulating the operation of quantum computers and algorithms using classical digital computers and optimising the performance of such simulations is an important and growing field of research in its own right \cite{Raedt2007a,Pednault2017a}. The two main differences pertinent to classical simulations of conventional and topological quantum computers are how error correction and gate decomposition are dealt with.

First, if one wishes to simulate a conventional quantum computer at the scale of physical operations (as we will do for a topological quantum computer), then this simulation should include error correcting codes. In principle one could set the error rate in the simulation to zero, but since the error rate of conventional quantum computers tends to be quite high, assuming that the error can be set to zero is a poor assumption and would lead to results dissimilar to the actual set of physical operations which the conventional quantum computer would have to perform.

In contrast, a topological quantum computer is thought to have a significantly lower rate of spontaneous errors than a conventional quantum computer. This means that, if one wishes to simulate the sequence of physical operations (such as braiding) needed to perform a computation in a topological quantum computer, ignoring the occurrence of spontaneous errors should be a far better approximation than for a conventional quantum computer. On the other hand, the topological protection is not absolute and some combination of spontaneous and systematic errors such as quasiparticle poisoning \cite{Sarma2015a} may still affect the operation of a topological quantum computer \cite{divincenzo2013quantum} and still warrant the usage of error correcting protocols \cite{ErrorCorrectionThesis,Burton_Fibonacci_Error,Wootton_Error_Correction,Carlos_Anyon_Computing}. Nevertheless, error correction should still play a lesser role in topological quantum computers than in conventional quantum computers.

Second, and perhaps most significantly, the set of elementary physical operations within a topological quantum computer is different to those within other quantum computers. While some quantum computers are capable of performing a continuous set of operations (such as evolving a particle at a higher energy for some time to change its phase), topological quantum computers have a strictly discrete set of elementary physical operations, with a countable set of elementary braiding operations. To apply the desired unitary gate operation, it is necessary to search through the exponentially large set of possibilities of braids and weaves to find an optimal or near optimal approximation to the gate. To do this efficiently is an area of research in its own right, as mentioned in Sec.~\ref{sec:compile}, and is a major part of topological quantum computation. These pre-compiled braids are specific to each anyon model, so must be compiled separately for each anyon model. Determining the correspondence between a braid and a gate in a topological quantum computer is equivalent to determining how to physically realise the gate operations in a conventional quantum computer.

\section{Topological Quantum Algorithm}\label{sec:algorithm}

A quantum computer is of little use without an algorithm to run on it. An algorithm which is closely related to topology, and which has a relatively simple quantum component, is the algorithm developed by Aharonov, Jones and Landau (AJL) to find the value of the Jones polynomial at the roots of unity, $t=e^{2\pi i/k}$ \cite{Aharonov_Jones_Algorithm1}.

In brief, the algorithm was derived by considering the knot as the trace closure of a braid. The skein relation \eqref{eqn:Bracket_Rule_1} was applied to the braid, forming a linear superposition of disjoint loops. To find the Jones polynomial, a function which effectively counts those loops is necessary, and the defining properties of such a function were noted. It was then observed that the diagrams of loops was a representation of what is known as a Temperley Lieb algebra. A matrix representation of the Temperley Lieb algebra was presented, leading to a translation between each elementary braid and a unitary matrix, such that a braid would be represented by a unitary matrix which was the product of certain elementary unitary matrices. The function for counting the loops, named a Markov trace, was found to be a weighted sum over the diagonal matrix elements. If the knot is the plat closure of a braid instead of a trace closure, this can be related to a trace closure, and the algorithm simplifies to determining a single element of the matrix. The diagonal matrix elements could be computed using a simple quantum algorithm known as the Hadamard Test \cite{Aharonov_Jones_Algorithm1,Lomonaco_Jones_Algorithm}. The classical details of the algorithm and the pertinent aspects of its derivation are reproduced in Sec.~\ref{sec:algorithm_derive}. The quantum part of the algorithm is explained in Sec.~\ref{sec:hadamard}

\subsection{The AJL Algorithm}\label{sec:algorithm_derive}

The first step is to convert the link for which the Jones polynomial is to be calculated for into the trace closure of a braid. This braid of $n$ strands is a member of the braid group $B_n$, and is given by a braidword or the product of certain elementary braids $b_j$.

Recall that the computationally difficult part of calculating the Jones polynomial is the computation of the Kauffman bracket polynomial, which follows the rules

\begin{gather}
\picbracket{"Bracket_Crossing"} = A \picbracket{"Bracket_Horizontal_Smooth"} + A^{-1} \picbracket{"Bracket_Vertical_Smooth"} \label{eqn:Bracket_Rule_1_2} \\
\anglebrackets{K \sqcup O} = -(A^2+A^{-2})\:\anglebrackets{K} = d\:\anglebrackets{K} \label{eqn:Bracket_Rule_3_2} \\
\anglebrackets{O}=1. \label{eqn:Bracket_Rule_2_2}
\end{gather}

The skein relation \eqref{eqn:Bracket_Rule_1_2} can be used to construct an alternative representation of the braid group,

\begin{equation}
\begin{aligned}
\rho_A: B_n \rightarrow {\rm TL}_n(d), \: \rho_A(b_j) &= A E_j + A^{-1}\mathbb{I}, \\
\rho_A(b_j^{-1}) &= A \mathbb{I} + A^{-1} E_j
\end{aligned}
\label{eqn:jones_rep}
\end{equation}
where, graphically,
\begin{equation}
b_j = \eqimage{"elementary_braid"}{6ex},\; E_j = \eqimage{"capcup"}{6ex},\text{ and } \mathbb{I} = \eqimage{"elementary_identity"}{6ex}.
\label{eqn:ajl_elem_components}
\end{equation}

When $\rho_A$ is applied to a product of elementary braids, then it is applied individually to each component of that product.

\begin{figure}
	\includegraphics[width=\linewidth]{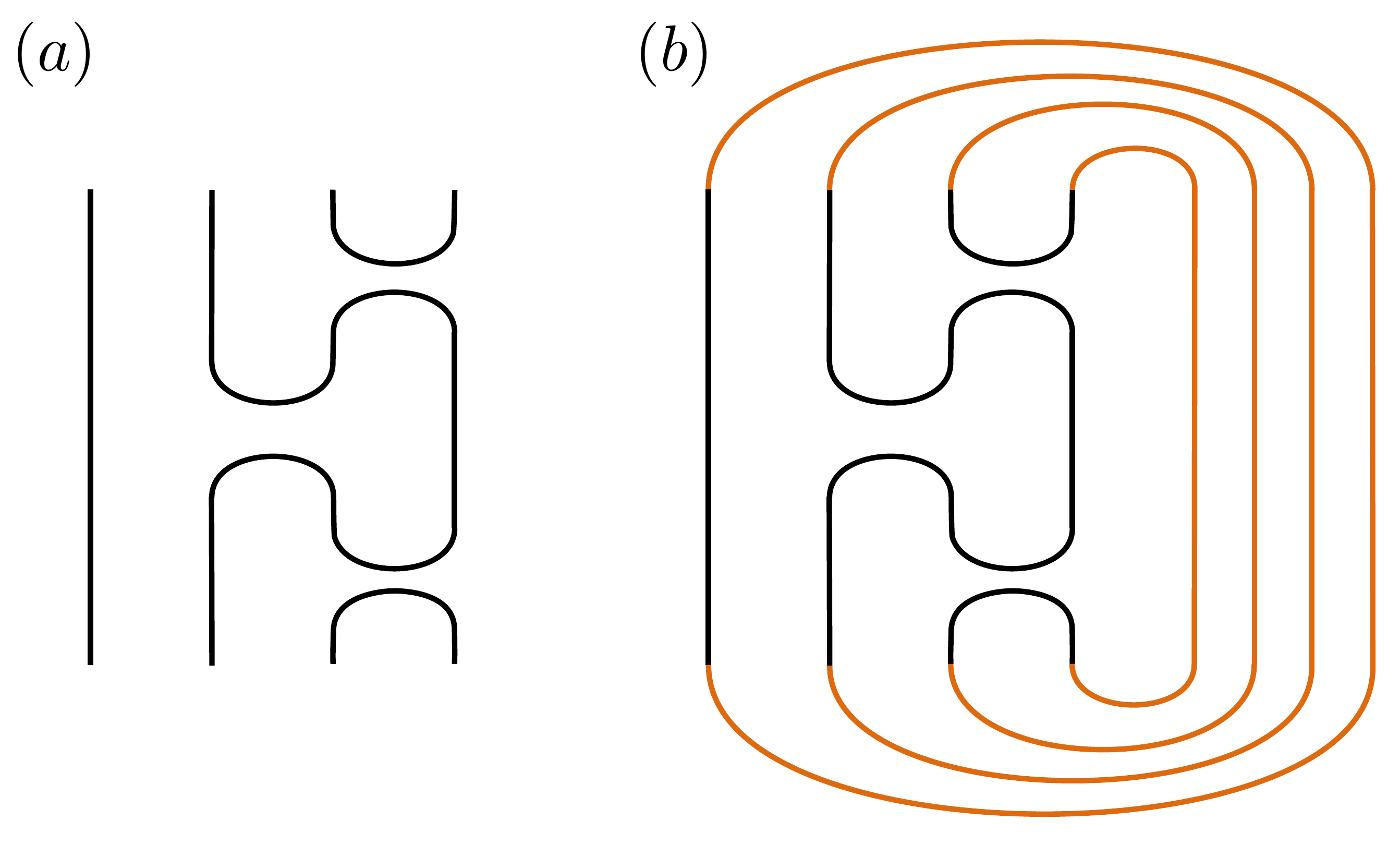}
	\caption{(a) A graphical depiction of $E_3 E_2 E_3$, which is a member of ${\rm TL}_4$. Note that this arrangement is topologically equivalent to $E_3$. (b) The trace closure of (a). This diagram contains 3 loops.}
	\label{Fig:TL_sample}
\end{figure}

Where the braids $b_j$ (with $1\le j < n$) are the generators of the braid group $B_n$, the cap-cups $E_j$ (with $1\le j < n$) are the generators of ${\rm TL}_n(d)$. Graphically, this forms a diagram similar to \Figref{Fig:TL_sample}(a), with the diagram built from bottom to top when read from left to right. ${\rm TL}_n(d)$ is a Temperley Lieb algebra, which obeys the properties

\begin{align}
E_i E_j &= E_j E_i,\; \abs{i-j} \ge 2, \label{eqn:TL_rule_1}\\
E_j E_{j\pm1} E_j & = E_j, \label{eqn:TL_rule_2}\\
E_j^2 &= d E_j. \label{eqn:TL_rule_3}
\end{align}

All these properties can be demonstrated graphically. \eqref{eqn:TL_rule_1} indicates the commutativity of distant elements. \eqref{eqn:TL_rule_2} indicates a bend which is topologically trivial, illustrated in \Figref{Fig:TL_sample}(a). \eqref{eqn:TL_rule_3} indicates that a disjoint loop can be removed by replacing it with a constant $d=-A^2-A^{-2}$, as per \eqref{eqn:Bracket_Rule_3_2}.

We now need a linear function which acts on an object in ${\rm TL}_n(d)$ to implement the remaining rules \eqref{eqn:Bracket_Rule_3_2} and \eqref{eqn:Bracket_Rule_2_2} to find the Kauffman bracket polynomial. Such a function effectively counts the loops in the closure of an object in ${\rm TL}_n(d)$, an example of which is in \Figref{Fig:TL_sample}(b). Let this function, acting on a space with $n$ strands, be $f_n: {\rm TL}_n(d) \rightarrow \mathbb{C}$, defined such that the bracket polynomial of the trace closure of braid $B$ is

\begin{equation}
\anglebrackets{B^{\rm tr}} = f_n ( \rho_A ( B) ). \label{eqn:AJL_kauffman_1}
\end{equation}

In accordance with \eqref{eqn:Bracket_Rule_3_2} and \eqref{eqn:Bracket_Rule_2_2}, this function can be defined as

\begin{equation}
f_n(X) = d^{a-1},\label{eqn:AJL_kauffman_2}
\end{equation}
where $a$ is the number of loops in the closure of the diagram. By considering the topology of these diagrams, it can be shown that this function has the properties
\begin{align}
f_1(\mathbb{I}) &= 1, \label{eqn:f_rule_1}\\
f_{n+1}(X) &= d f_n(X),\text{ for } X\in {\rm TL}_n(d), \label{eqn:f_rule_2} \\
f_n(XY) &= f_n(YX), \label{eqn:f_rule_3}\\
f_{n+1}(X E_{n}) &= f_{n}(X), \text{ for } X\in {\rm TL}_{n}(d). \label{eqn:f_rule_4}
\end{align}

However, this function is not normalised, because from \eqref{eqn:f_rule_1} and \eqref{eqn:f_rule_2} it follows that $f_n(\mathbb{I})=d^{n-1}$. Ideally, our function should have a value of 1 when acting on the identity. As such, let us define a new function $\widetilde{\rm Tr}$, the Markov trace, which is a normalised version of $f_n$. It is given by
\begin{equation}
\widetilde{\rm Tr}(X) = d^{1-n} f_n(X) = d^{a-n}, \label{eqn:trace_def_1}
\end{equation}
such that the Kauffman bracket polynomial is given by
\begin{equation}
\anglebrackets{B^{\rm tr}} = d^{n-1} \widetilde{\rm Tr}(\rho_A(B)). \label{eqn:AJL_kauffman_3}
\end{equation}

Following from the properties of $f_n$, the Markov trace has the properties
\begin{equation}
\begin{aligned}
\widetilde{\rm Tr}(\mathbb{I})&=1, \\
\widetilde{\rm Tr}(XY) &= \widetilde{\rm Tr}(YX), \\
\widetilde{\rm Tr}(X E_n) &= d^{-1} \widetilde{\rm Tr}(X), \text{ for } X\in {\rm TL}_n (d).
\end{aligned}
\label{eqn:trace_properties}
\end{equation}

These properties are sufficient to uniquely define the Markov trace for any representation of the Temperley Lieb algebra.

\subsubsection{Unitary Representation of the Braid Group}

Next, a unitary representation of the braid group within ${\rm TL}_n(d)$ must be constructed, and at this point it is necessary to choose an integer value of $k\ge 3$. This value of $k$ determines the value of the other constants
\begin{equation}
t = e^{2\pi i/k},\; A = i e^{-\pi i / 2k},\; d = 2 \cos(\pi/k), \label{eqn:AJL_constants}
\end{equation}
including the point $t$ at which the polynomials are evaluated.

The representation of the Temperley Lieb algebra that will be considered here is the path model representation. This will involve defining a function $\Phi: {\rm TL}_n(d) \rightarrow \mathbb{C} U(\mathcal{H}_{n,k})$, which takes the Temperley Lieb algebra and represents it as a matrix with dimensions parameterised by $n$ and $k$.

\begin{figure}
	\includegraphics[width=\linewidth]{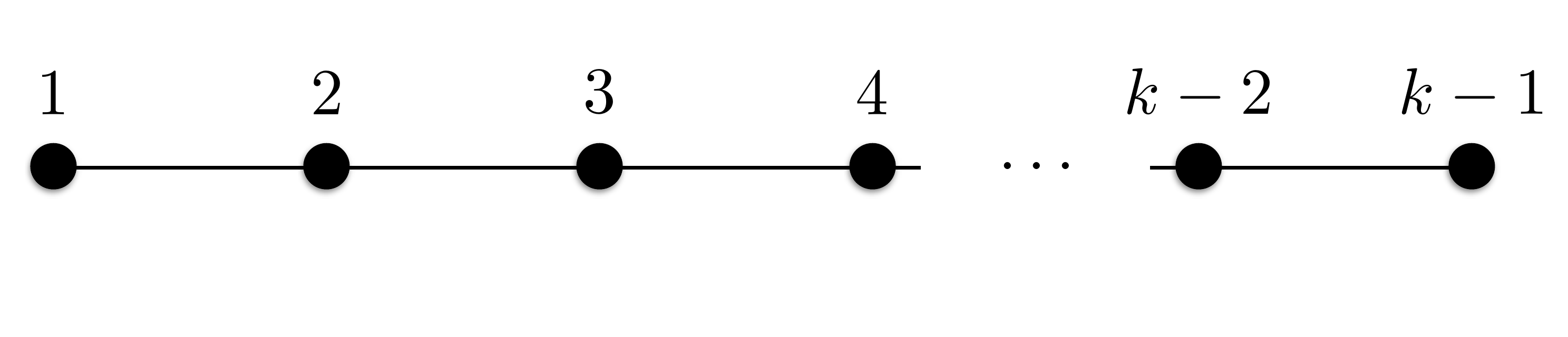}
	\caption{A schematic representation of $G_k$, a graph of $k-1$ vertices and $k-2$ edges.}
	\label{Fig:path_graph}
\end{figure}

For this function, consider a graph $G_k$ containing $k-1$ vertices and $k-2$ edges. These vertices can be arranged in a row, numbered from left to right from 1 to $k-1$, as in \Figref{Fig:path_graph}. The adjacency matrix of this graph has an eigenvalue $d=2\cos (\pi/k)$, and the components of the corresponding eigenvector are given by
\begin{equation}
\lambda_l = \sin(\pi l/k),\; 1\le l < k. \label{eqn:lambda}
\end{equation}

Now consider the set of paths of length $n$ starting at vertex 1 in $G_k$. Call this set $\mathcal{P}_{n,k}$. Each path $p$ can be represented as a bitstring of length $n$, where `0' is a step to the left and `1' is a step to the right. The set of these paths will be used to define a set of orthonormal basis vectors in the Hilbert space $\mathcal{H}_{n,k}$, $\{\ket{p}:p\in \mathcal{P}_{n,k}\}$.

Let $p^{j-1 \lrcorner}$ be the subpath from 1 to $j-1$ (that is, where the bitstring $p$ has been truncated to the first $j-1$ elements), $p^{\llcorner j \dots j+1 \lrcorner}$ be the subpath from $j$ to $j+1$, and $p^{\llcorner j+2}$ be the subpath from $j+2$ to $n$. Define $l$ in this context to be the endpoint of $p^{j-1\lrcorner}$.

Just as ${\rm TL}_n(d)$ has the generators $E_j$ ($1\le j < n$), the matrix representation given by $\Phi$ has the corresponding generators $\Phi_j=\Phi(E_j)$. As for any matrix, $\Phi_j$ can be constructed by considering its action on each of the basis states $\ket{p}$. Using \eqref{eqn:lambda} and the endpoint $l$ of subpath $p^{j-1\lrcorner}$, $\Phi_j$ can be defined as

\begin{widetext}
\begin{equation}
\Phi_j \ket{p} =
	\begin{cases}
	0 & \text{if } p^{\llcorner j \dots j+1 \lrcorner} = `00' \\
	\frac{\lambda_{l-1}}{\lambda_l} \ket{p} + \frac{\sqrt{\lambda_{l-1}\lambda_{l+1}}}{\lambda_l} \ket{p^{j-1\lrcorner} 10 p^{\llcorner j+2}} & \text{if } p^{\llcorner j \dots j+1 \lrcorner} = `01' \\
	\frac{\sqrt{\lambda_{l-1}\lambda_{l+1}}}{\lambda_l} \ket{p^{j-1\lrcorner} 01 p^{\llcorner j+2}} + \frac{\lambda_{l+1}}{\lambda_l}\ket{p} & \text{if } p^{\llcorner j \dots j+1 \lrcorner} = `10' \\
	0 & \text{if } p^{\llcorner j \dots j+1 \lrcorner} = `11'.
	\end{cases}
\label{eqn:phi}
\end{equation}
\end{widetext}

Note that $\lambda_0=\lambda_k=0$. Note also that $\Phi_j$ only transforms between paths $p$ with the same endpoint, such that the representation $\Phi$ is block diagonal when the bases $\ket{p}$ are arranged in order of endpoint.

A braidword $B$ can thus be expressed as a matrix $\Phi(\rho_A(B))$ which is a product of the elementary unitary matrices $\Phi(\rho_A(b_j))$, defined by
\begin{equation}
\begin{gathered}
\Phi(\rho_A(b_j)) = A \Phi_j + A^{-1} \mathbb{I}.
\end{gathered}
\label{eqn:braid2matrix}
\end{equation}
To denote the elementary AJL matrix $\Phi(\rho_A(b_j))$ for a given $k$, $n$, and an elementary braid $b_j$, we shall use the symbol
	\begin{equation}
	\Theta_j(n,k).
	\label{eqn:ajl_label}
	\end{equation}

\subsubsection{The Markov Trace}

The Markov trace 
\begin{equation}
\widetilde{\rm Tr}(X) = \frac{1}{N} \sum_{p\in P_{n,k}} \lambda_l \bra{p} X \ket{p},
\label{eqn:trace_def_2}
\end{equation}
where $l$ is the endpoint of path $p$ and $N$ is a normalisation constant, 
\begin{equation}
N=\sum_{p\in P_{n,k}} \lambda_l,
\label{eqn:trace_normalisation}
\end{equation}
can be defined to act in the image of $\Phi$.

Because of the uniqueness of the properties of the Markov trace, \eqref{eqn:trace_properties}, and because this function \eqref{eqn:trace_def_2} satisfies those properties, \eqref{eqn:trace_def_2} defines our unique Markov trace function. Note that the Markov trace is merely a weighted version of the standard matrix trace, with the diagonal entries weighted according to the corresponding endpoint and normalised such that $\widetilde{\rm Tr}(\mathbb{I}) = 1$.

An equivalent representation, which may be more convenient for larger $k$ and $n$ (where the size of the matrix is larger), is to note that $\Phi_j$ and thus any matrix in the image of $\Phi$ is block diagonal, with each block corresponding to a different value of $l$. The matrix $X$ in \eqref{eqn:trace_def_2} may thus be broken into smaller matrices $X\vert_l$, with the subspaces restricted to paths with the endpoint $l$, and then have \eqref{eqn:trace_def_2} act on $X\vert_l$ rather than the entire $X$ for each component of the sum. This representation will be more convenient for quantum computation, where compiling unitary operations corresponding to large matrices is difficult.

\begin{figure}
	\includegraphics[width=\linewidth]{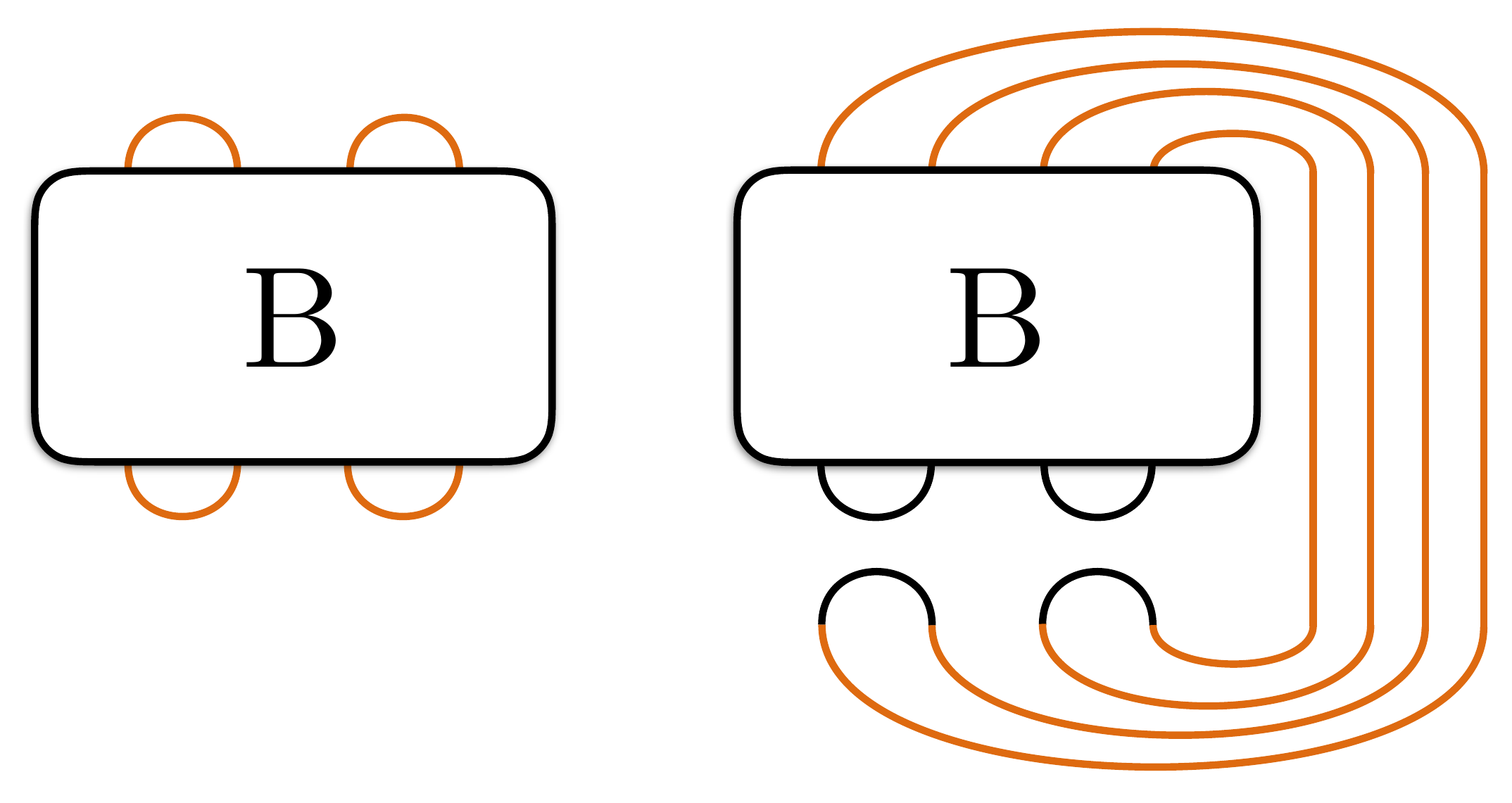}
	\caption{On the left is the plat closure of a braid $B$. On the right is a topologically equivalent object, expressed as the trace closure of the braid $B$ with the appended cap-cups $E_1$ and $E_3$.}
	\label{Fig:plat2trace}
\end{figure}

The Markov trace may be used to compute the Kauffman bracket polynomial by
\begin{equation}
\anglebrackets{B^{\rm tr}} = d^{n-1} \widetilde{\rm Tr}(\Phi(\rho_A(B))), \label{eqn:AJL_kauffman_4}
\end{equation}
and the Jones polynomial may be calculated from \eqref{eqn:Jones2}. The difficulty of calculating the Jones polynomial of the trace closure of a braid is thus reduced to multiplying together certain matrices (which grow exponentially in dimension with increasing $n$) and determining a weighted sum of their diagonal elements.

\subsubsection{Plat Closures}

This result may be extended to include plat closures \cite{Aharonov_Jones_Algorithm1}. As illustrated in \Figref{Fig:plat2trace}, the plat closure of a braid $B$ is topologically equivalent to the trace closure of the object $C=B E_1 E_3\dots E_{n-1}$. This means that

\begin{equation}
\begin{split}
\anglebrackets{B^{\rm pl}}=\anglebrackets{C^{\rm tr}}=d^{n-1}\widetilde{\rm Tr}(\Phi(C))\\
=d^{n-1}\widetilde{\rm Tr}(\Phi(\rho_A(B))\Phi_1\Phi_3\dots\Phi_{n-1}).
\end{split}
\label{eqn:plat1}
\end{equation}
 
Let us consider the action of $\Phi_1\Phi_3\dots\Phi_{n-1}$ on the path vectors. By \eqref{eqn:TL_rule_1}, $\Phi_j$'s commute if their indices differ by more than one. By the definition of $\Phi_j$ \eqref{eqn:phi}, $\Phi_1 \ket{p}$ is only non-zero if the first two bits in the path are `10'. Note that the path starting with `01' is invalid, because it would require stepping from vertex 1 to vertex 0 in $G_k$, but vertex 0 does not exist. The value of this non-trivial result is

\begin{equation}
\begin{split}
	\Phi_1\ket{`10\dots\text{'}}=\frac{\lambda_2}{\lambda_1}\ket{`10\dots\text{'}}=\frac{\sin(2\pi/k)}{\sin(\pi/k)}\ket{`10\dots\text{'}}\\
	=2\cos(\pi/k)\ket{`10\dots\text{'}}=d\ket{`10\dots\text{'}}.
\end{split}
\label{eqn:phi_plat_1}
\end{equation}

By similar logic, we can infer that the next two bits must also be `10' and the operation is also a multiplication by $d$. Thus, by induction, we can state that
\begin{equation}
\begin{aligned}
\Phi_1 \Phi_3 \dots \Phi_{n-1} = d^{n/2} \ket{\alpha}\bra{\alpha},
\end{aligned}
\label{eqn:phi_plat_2}
\end{equation}
where $\ket{\alpha} = \ket{`1010\dots 10\text{'}}$. This means that, by the definition of the Markov trace \eqref{eqn:trace_def_2},
\begin{equation}
\begin{aligned}
\widetilde{\rm Tr}(\Phi(C)) = d^{n/2} \widetilde{\rm Tr}(\Phi(\rho_A(B))\ket{\alpha}\bra{\alpha})
\\= \frac{d^{n/2}\lambda_1}{N}\bra{\alpha}\Phi(\rho_A(B))\ket{\alpha}.
\end{aligned}
\label{eqn:plat_trace_1}
\end{equation}

To generalise the value of the constant terms, consider the case where $B$ is the identity. This means the closure of the braid has $n/2$ loops. The fundamental function of the Markov trace \eqref{eqn:trace_def_1} is to count loops, meaning that $\widetilde{\rm Tr}(C) = d^{n/2-n} = d^{-n/2}$. Comparing this to \eqref{eqn:plat_trace_1}, noting that $\bra{\alpha}\Phi(\rho_A(\mathbb{I}))\ket{\alpha}=1$, the constant terms are $\lambda_1/N = d^{-n}$, which means that the Markov trace becomes

\begin{equation}
\widetilde{\rm Tr}(\Phi(C)) = d^{-n/2} \bra{\alpha}\Phi(\rho_A(B))\ket{\alpha},
\label{eqn:plat_trace_2}
\end{equation}
and from \eqref{eqn:plat1} we can state that the Kauffman bracket polynomial of the plat closure of a braid $B$ is
\begin{equation}
\anglebrackets{B^{\rm pl}} = d^{\tfrac{n}{2} - 1} \bra{\alpha}\Phi(\rho_A(B))\ket{\alpha}.
\label{eqn:plat2}
\end{equation}

As such, the difficulty of calculating the Jones polynomial for the plat closure of a braid is reduced to multiplying together certain matrices and finding a particular diagonal matrix element.

\subsubsection{An Example}

We conclude this explanation of the algorithm with an explicit demonstration of deriving the matrices $\Phi_j$ and their application to the Hopf link.

\begin{figure}
	\includegraphics[width=0.5\linewidth]{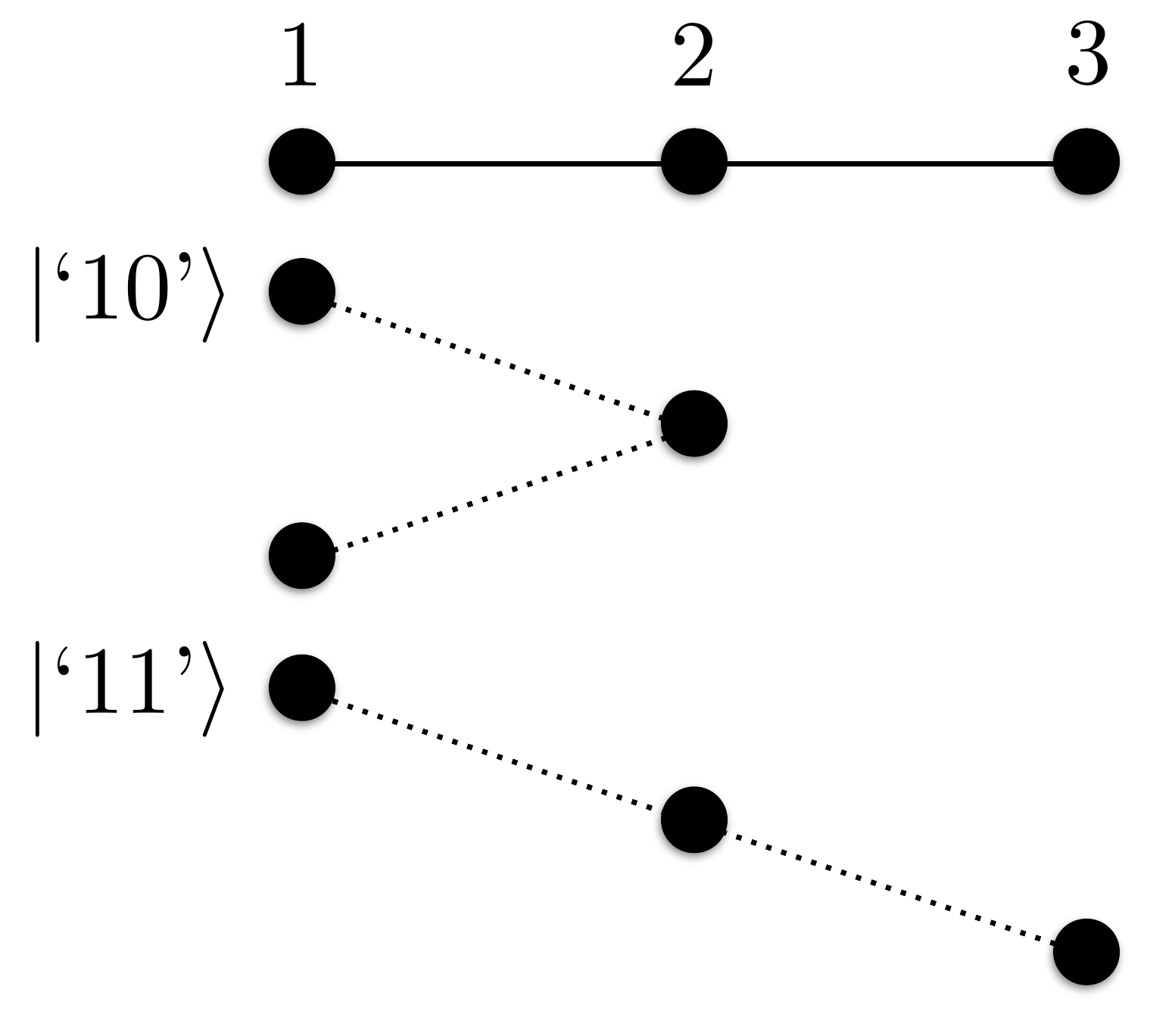}
	\caption{Illustration of the two possible paths when $n=2$ and $k=4$. The graph $G_4$ has three vertices. The path `10' ends at vertex 1, while the path `11' ends at vertex 3.}
	\label{Fig:paths_n2k4}
\end{figure}

Consider $n=2$ and $k = 4$. The constants are $A=ie^{-\pi i/8}$ and $d=2\cos(\pi/4)=\sqrt{2}$. The only possible value for $j$ is 1. The graph $G_4$ has three vertices, and the possible paths are $\mathcal{P}_{2,4} = \{ `10\text{'},\; `11\text{'} \}$, illustrated in \Figref{Fig:paths_n2k4}.  These will form our basis vectors, $\ket{`10\text{'}}=\left(\begin{smallmatrix}1 \\ 0\end{smallmatrix}\right)$ and $\ket{`11\text{'}}=\left(\begin{smallmatrix}0 \\ 1\end{smallmatrix}\right)$.

In this case, $j=1$, so the endpoint of $p^{j-1\lrcorner}=p^{0\lrcorner}$ is the endpoint of the path with no steps, which is 1. Using \eqref{eqn:phi}, we evaluate $\Phi_1$ to be
\begin{equation}
\begin{aligned}
\Phi_1 \ket{`10\text{'}} &= \frac{\sqrt{\lambda_0\lambda_2}}{\lambda_1}\ket{`01\text{'}} + \frac{\lambda_2}{\lambda_1} \ket{`10\text{'}}\\
&= \frac{\lambda_2}{\lambda_1} \ket{`10\text{'}} = d \ket{`10\text{'}} = \sqrt{2} \ket{`10\text{'}}, \\
\Phi_1 \ket{`11\text{'}} &= 0,
\end{aligned}
\label{eqn:phi_example1}
\end{equation}
such that
\begin{equation}
\Phi_1 = \begin{pmatrix}
\sqrt{2} & 0 \\
0 & 0
\end{pmatrix}.
\label{eqn:phi_example2}
\end{equation}

The corresponding unitary matrix representation of the elementary braiding operator in $B_2$ for $k=4$ is thus
\begin{equation}
\begin{aligned}
\Theta_1(2,4) &= A\Phi_1 + A^{-1} \mathbb{I}_2 \\
&= i e^{-\pi i/8} \begin{pmatrix}
\sqrt{2} & 0 \\
0 & 0
\end{pmatrix} - i e^{\pi i/8} \begin{pmatrix}
1 & 0 \\ 0 & 1
\end{pmatrix} \\
&= i e^{\pi i/8} \begin{pmatrix}
\sqrt{2} e^{-\pi i/4} - 1 & 0 \\
0 & -1
\end{pmatrix}\\
& = e^{\pi i/8} \begin{pmatrix}
1 & 0 \\
0 & -i
\end{pmatrix}.
\end{aligned}
\label{eqn:AJL_braid_example}
\end{equation}

Now consider the trace closure of the Hopf link, as in \Figref{Fig:hopf_braids}(b), which has the braidword $b_1^2$. The matrix which corresponds to this braidword is

\begin{equation}
X = -e^{\pi i/4} \begin{pmatrix}
2e^{-\pi i/2} - 2\sqrt{2}e^{-\pi i/4} + 1 & 0\\0 & 1
\end{pmatrix}.
\label{eqn:AJL_hopf_example_mat}
\end{equation}

Noting that `10' has an endpoint of 1 and `11' has an endpoint of 3, the Markov trace of the matrix \eqref{eqn:AJL_hopf_example_mat} is

\begin{equation}
\widetilde{\rm Tr}(X) = \frac{(-e^{\pi i/4})}{\lambda_1+\lambda_3} \left( \lambda_1 (2e^{-\pi i/2} - 2\sqrt{2}e^{-\pi i/4} + 1) + \lambda_3 \right),
\end{equation}
where
\begin{equation}
\lambda_1 = \sin(\pi/4) = \frac{\sqrt{2}}{2},\; \lambda_3 = \sin(3\pi/4) = \frac{\sqrt{2}}{2}. 
\end{equation}
Hence,
\begin{align}
\widetilde{\rm Tr}(X) &= \frac{1}{\sqrt{2}} \frac{\sqrt{2}}{2} \left(-2e^{-\pi i/4} + 2\sqrt{2} - 2e^{\pi i/4} \right)\notag \\
&= \sqrt{2} - 2\cos\left(\frac{\pi}{4}\right) = 0.
\label{eqn:AJL_hopf_example_trace}
\end{align}

\begin{figure}
	\includegraphics[width=\linewidth]{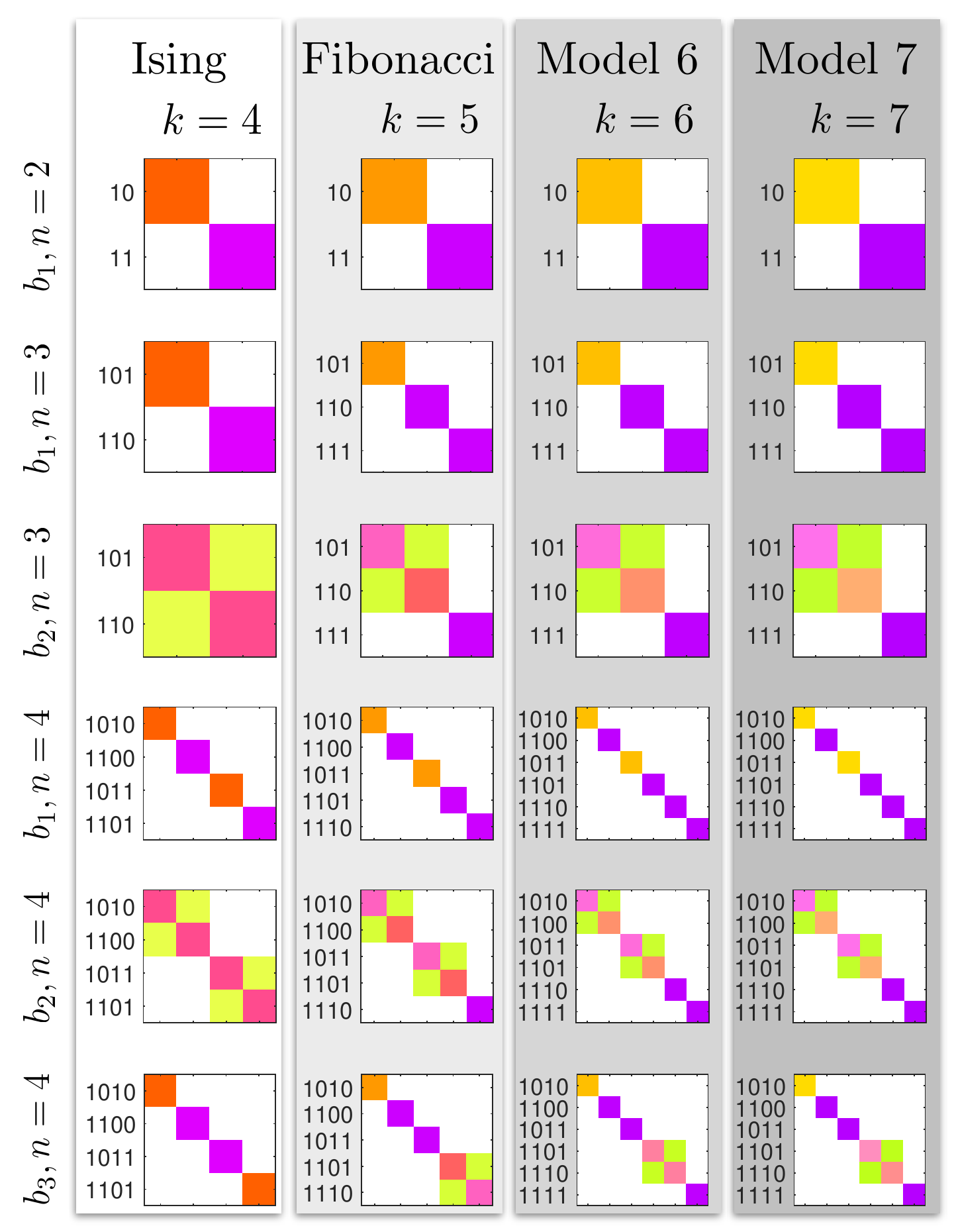}
	\caption{A selection of the elementary AJL matrices, $\Theta_j(n,k)$, with $k$ between 4 and 7 and $n$ between 2 and 4. Each coloured square represents a complex number, where the hue and saturation correspond to the phase and magnitude respectively, as indicated by the legend in Fig.~\ref{Fig:2q_matrices}. White squares are matrix elements with a value of zero. The squares in the matrices are indexed according to the bitstrings corresponding to their paths, sorted in order of path endpoint. The columns corresponding to different values of $k$ are grouped with the anyon models to which the matrices are related to (see Secs~I and \ref{sec:exact_ajl}), with $k=4$ and 5 related to Ising and Fibonacci anyons, respectively, and $k=6$ and 7 related to models yet to be identified.}
	\label{Fig:ajl_matrices}
\end{figure}

From this, the Kauffman bracket polynomial at $A=ie^{-\pi i/8}$ evaluates to
\begin{equation}
\anglebrackets{K}=d^{2-1} \widetilde{\rm Tr}(X) = \sqrt{2} (0) = 0.
\label{eqn:AJL_hopf_example_bracket}
\end{equation}

Using the conventional algorithm, we had previously found that the Kauffman bracket polynomial of the Hopf link was $-A^4-A^{-4}$ in \eqref{eqn:hopf_bracket_derive}. For $A=ie^{-\pi i/8}$, this evaluates to $-e^{-\pi i/2} - e^{\pi i/2} = i - i = 0$, which agrees with \eqref{eqn:AJL_hopf_example_bracket}. The calculation for other values of $k$ and $n$ and for different braids is similar.

\subsubsection{AJL Matrices}

We shall make a few observations about the form of the braiding matrices, $\Theta_j(n,k)$, derived for the AJL algorithm. For reference, representations of some of these matrices are provided in \Figref{Fig:ajl_matrices}.

For $k=3$, all the matrices are scalar and equal to 1. Further evaluation reveals that the Jones polynomial at $t=e^{2\pi i/3}$ evaluates to exactly 1, regardless of the knot being evaluated. This makes $k=3$ a trivial case.

For $n=2$, the matrices are all $2\times2$ and diagonal.

For $n=3$, the matrices are up to 3 dimensional, but are block diagonal with a $2\times2$ and a $1\times1$ block.

For $n=4$, the matrices are up to 6 dimensional, but the first block (corresponding to paths ending at vertex 1) is $2\times2$, with the next blocks being $3\times3$ and $1\times1$.

For larger $n$ (and $k>3$), the matrices are larger and the first block is larger than $2\times2$.

\begin{figure}
	\includegraphics[width=\linewidth]{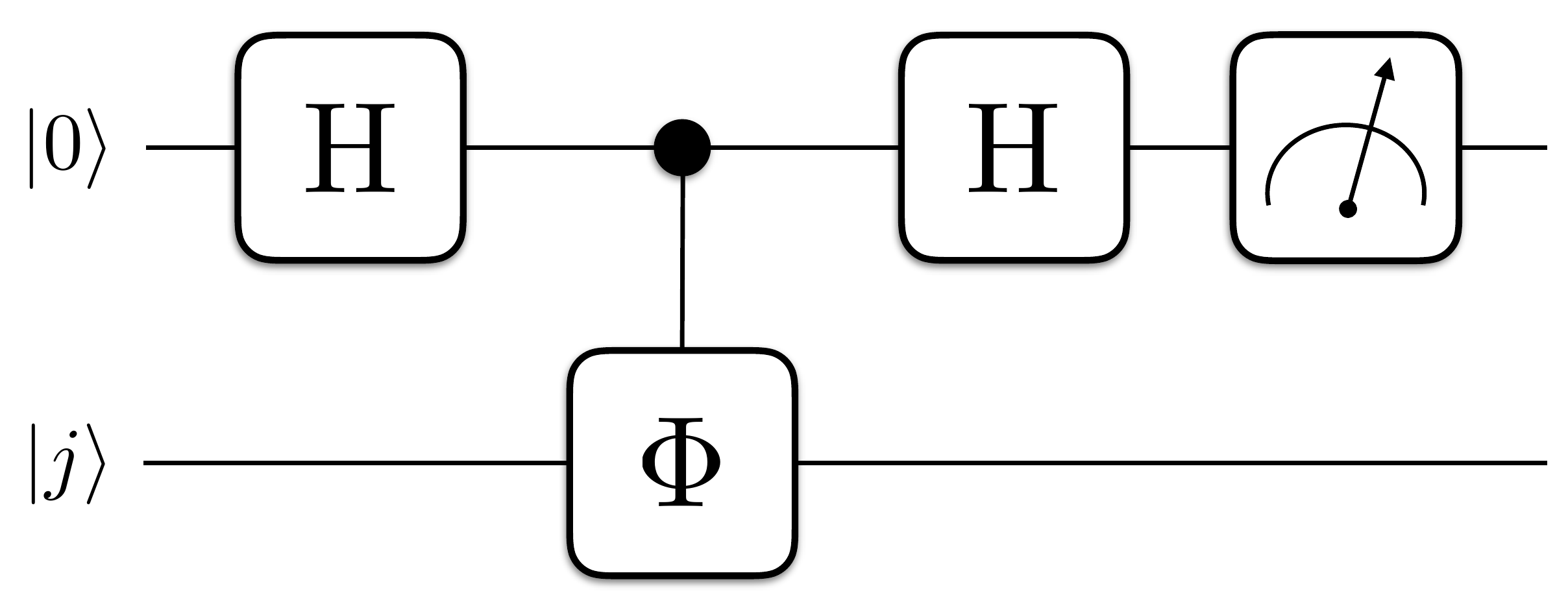}
	\caption{Quantum circuit diagram for the Hadamard test for evaluating the real component of a matrix element. $H$ is the Hadamard gate. The first/top qubit will end in the state $\ket{0}$ with probability of $\tfrac{1}{2}(1+\Re \bra{j}\Phi\ket{j})$. The gates are left-multiplied onto the initial qubits on the left in order from left to right, so mathematically this circuit is ${(H\otimes \mathbb{I}) (\ket{0}\bra{0}\otimes \mathbb{I} + \ket{1}\bra{1}\otimes \Phi)(H\otimes \mathbb{I})(\ket{0}\otimes\ket{j})}$.}
	\label{Fig:hadamard_real}
\end{figure}

If $k\le n+1$, then the graph $G_k$ is shorter than the length of some paths of length $n$. This has the effect of truncating the matrices to exclude some of the basis states which would be found at larger $k$. Additionally, for $k > n+1$, the dimension of the resulting matrices is independent of $k$, because the graph $G_k$ is longer than any possible path of length $n$.

For a given $k$ and $b_j$, there is some duplication of the structure of the matrices between different $n$. The first $2\times2$ block of $\Theta_1(3,k)$ and $\Theta_1(4,k)$ are identical to $\Theta_1(2,k)$. The first $2\times2$ block of $\Theta_2(4,k)$ is the same as that for $\Theta_2(3,k)$. The first $2\times2$ block of $\Theta_3(4,k)$ is identical to that for $\Theta_1(4,k)$. Since one braid can correspond to multiple different AJL matrices, this degree of redundancy is used to simplify the search for braids corresponding to these matrices.

From here on, for simplicity, $\Phi$ will be used to denote any product of AJL matrices $\Theta_j(n,k)$, which are in the image of the function $\Phi$.

\subsection{Hadamard Test}\label{sec:hadamard}

The AJL algorithm, as presented in the previous section, can be computed classically and exactly. However, the dimensions of the matrices $\Phi$ grows exponentially with the number of strands $n$, and is also greater for larger $k$ (up to $k=n+1$), which means that for arbitrary $n$ the AJL algorithm cannot be efficiently computed on a classical computer, although for fixed $n$ the complexity only grows linearly with the number of crossings in the knot.

The part of the AJL algorithm which can be improved by use of a quantum computer is the evaluation of the diagonal matrix elements for the Markov trace, and this is performed with a simple quantum circuit called the Hadamard test \cite{Aharonov_Jones_Algorithm1,Lomonaco_Jones_Algorithm}.

\begin{figure}
	\includegraphics[width=\linewidth]{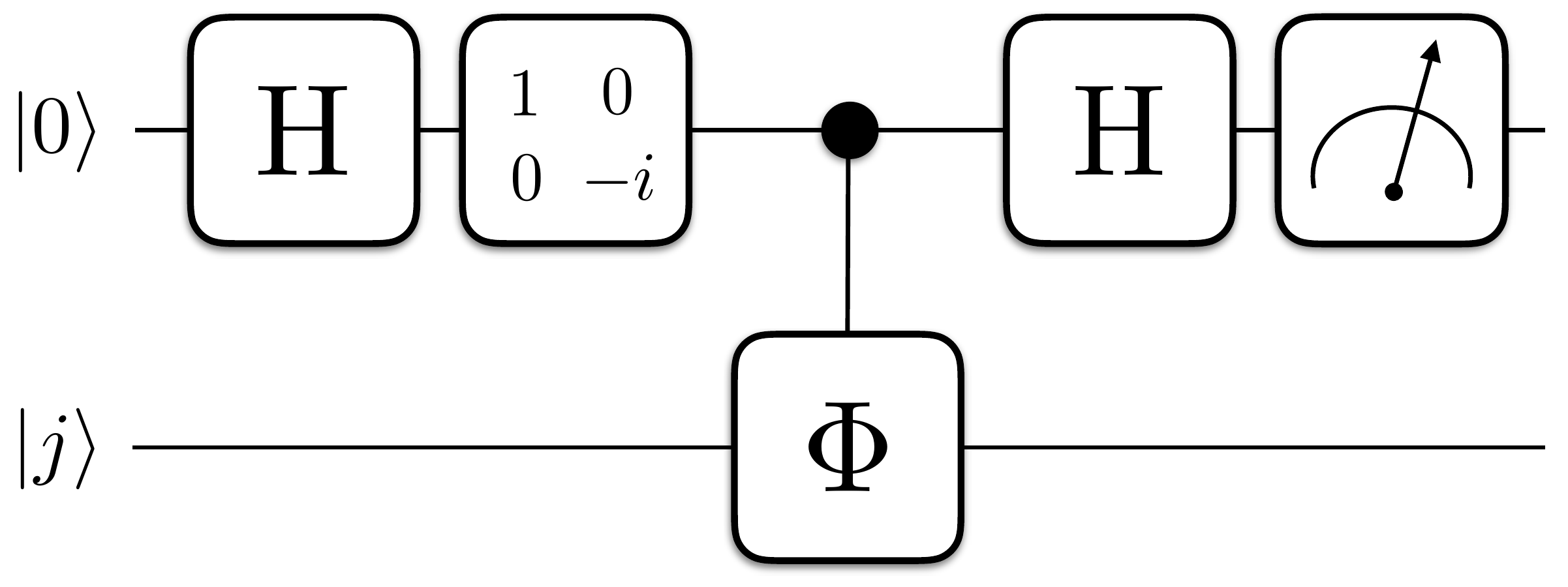}
	\caption{Quantum circuit diagram for the Hadamard test for evaluating the imaginary component of a matrix element. $H$ is the Hadamard gate. The top/first qubit will end in the state $\ket{0}$ with probability of $\frac{1}{2}(1+\Im \bra{j}\Phi\ket{j})$. The gates are left-multiplied onto the initial qubits on the left in order from left to right, so mathematically the circuit is ${(H\otimes \mathbb{I}) (\ket{0}\bra{0}\otimes \mathbb{I} + \ket{1}\bra{1}\otimes \Phi) (\left(\begin{smallmatrix}1&0\\0&-i\end{smallmatrix}\right)\otimes\mathbb{I})(H\otimes \mathbb{I})(\ket{0}\otimes\ket{j})}$.}
	\label{Fig:hadamard_imaginary}
\end{figure}

\subsubsection{Quantum Circuit}

The Hadamard test is described using quantum circuit diagrams in \Figref{Fig:hadamard_real} and \Figref{Fig:hadamard_imaginary}. A qubit register with the qubit $\ket{0}$ and the qubit(s) $\ket{j}$ is prepared. A Hadamard gate 
\begin{equation}
H=\frac{1}{\sqrt{2}} \begin{pmatrix}
1 & 1 \\
1 & -1
\end{pmatrix}
\label{eqn:hadamard_matrix}
\end{equation}
is applied to the first qubit. A controlled operation is performed, with the first qubit being the control qubit and $\ket{j}$ being the target qubit, performing the operation $\Phi$. Another Hadamard gate is applied to the first qubit, then the first qubit is measured. If the first qubit is in the state $\ket{0}$, then the computer returns the number 1. If it is in the state $\ket{1}$, then the computer returns the number $-1$. The average value will be $\Re\bra{j}\Phi\ket{j}$.

The imaginary part can be obtained by applying a $-\pi/2$ phase gate to the first qubit between the Hadamard gates; that is, an operation which induces a $-\pi/2$ phase shift between the $\ket{0}$ and $\ket{1}$ states. By repeated application of the Hadamard test over different basis vectors $\ket{j}$, the diagonal matrix elements of $\Phi$ can be determined.

For the controlled operation, if the desired operation $\Phi$ is a product of several matrices, then the controlled operation can be decomposed into a product of controlled operations. Symbolically, $\text{Controlled-}(AB)=(\text{Controlled-}A)(\text{Controlled-}B)$, although following the construction in Sec.~\ref{sec:two_qubit_braids} it is simpler for a topological quantum computer to concatenate the weaves corresponding to $A$ and $B$.

\begin{figure}
	\includegraphics[width=\linewidth]{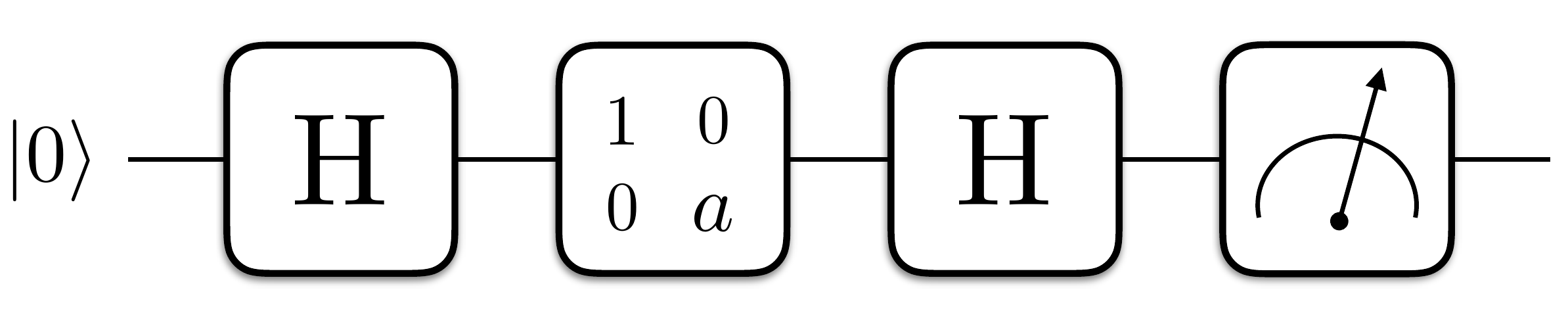}
	\caption{Quantum circuit diagram for evaluating the real component of a scalar $a$. The qubit will end in state $\ket{0}$ with probability of $\tfrac{1}{2}(1+\Re(a))$.}
	\label{Fig:hadamard_scalar}
\end{figure}

If $\Phi$ is one-dimensional, this corresponds to simple scalar multiplication by a complex phase. In practice, it is probably more effective to perform this scalar multiplication on a classical computer. However, it can still be performed on a quantum computer using a slight modification of the Hadamard test. The controlled operation can be replaced with a single qubit phase gate, where the phase shift is the same as the desired scalar, as in \Figref{Fig:hadamard_scalar}. As before, if this scalar is a product of scalars, then a product of phase gates can be used.

\subsubsection{Hadamard Test in the AJL Algorithm}

For the AJL algorithm, our $\Phi$ is composed of the product of certain matrices $\Theta_j(n,k)$ corresponding to elementary braiding operations. As such, to implement the Hadamard test for the AJL algorithm, it is only necessary to compile controlled operations corresponding to each elementary braiding operation for each value of $n$ and $k$, subverting the need to compile braids corresponding to every unique knot. It is also at no point necessary to explicitly compute the value of the matrix product $\Phi$, which would save considerable computation for large matrices.

A key step is to compile braids corresponding to controlled versions of the elementary matrices $\Theta_j(n,k)$. It is only necessary to compile each operation once, since it can be recorded and used again. When these matrices are $2\times2$ in size, the controlled operations can be constructed using the weaving method of  \cite{Bonesteel_Braid_Topologies}, possibly after concatenating the single qubit weaves into a single weave corresponding to the knot being investigated.

If these matrices are larger, then an alternative method is needed. They could be constructed by composing single qubit braids and select two qubit braids (such as CNOT) across more than two qubits. Hypothetically, these operations could also be constructed as a weave across a single qudit, composed of more than four anyons and thus having a fusion space with dimension greater than two. Additionally, the dimension of the matrices are not always the same as the dimensions of the computational space. When there is a mismatch in matrix size, a block diagonal matrix within a larger computational space can be computed, where one of the blocks corresponds to the target matrix.

Alternatively, the bitstrings in the path model representation of the matrices can be treated as defining the qubits directly, and a quantum circuit designed to apply the rules in \eqref{eqn:phi} and \eqref{eqn:braid2matrix} can be applied rather than a unique gate constructed for each $\Phi$. This would have the advantage of being readily generalisable to large $n$, and that not even the elementary matrices will have to be explicitly computed. However, while such a quantum circuit can theoretically be implemented efficiently \cite{Aharonov_Jones_Algorithm1}, it is beyond the scope of this work to design such a circuit.

Regardless of the method chosen, compiling these larger matrices will be computationally expensive due to the larger number of generators and substantially more complicated than the simple $2\times2$ case. In this work we will focus on just $2\times2$ matrices, or those which can be decomposed into $2\times2$ matrices.

The Hadamard test can be used to determine individual diagonal matrix elements, but for the AJL algorithm applied to trace closures the Markov trace needs to be calculated.

A straightforward implementation, as described in  \cite{Lomonaco_Jones_Algorithm}, is to iterate over each basis state, applying the Hadamard test to each state a sufficient number of times to obtain each diagonal matrix element to the desired accuracy. However, the time complexity of this approach scales with the number of matrix elements, and since the number of matrix elements grows exponentially with $n$, this would compromise the efficiency of the algorithm.

Another implementation, as described in  \cite{Aharonov_Jones_Algorithm1}, is to randomly select a basis state for each iteration of the Hadamard test, with each state chosen with probability proportional to $\lambda_l$. The average result will converge to the Markov trace, without having to explicitly compute each of the diagonal matrix elements.

\subsubsection{Convergence of the Hadamard Test}\label{sec:hadamard_convergence}

The Hadamard test is a stochastic method, requiring many iterations to obtain an average value. It is useful to know how quickly the Hadamard test converges, and how many iterations are needed to obtain a desired accuracy.

To test the rate of convergence of the Hadamard test, we used classical code which mimics the output of the Hadamard test, and tested it on randomly selected AJL matrices. The parameter $n$ was selected as a number between 2 and 8. The parameter $k$ was selected as a number between 4 and 13. Then a random braid matrix from that set of parameters, including inverses, was selected.

For each trial, we ran the Hadamard test for a particular number of iterations for each of the real and imaginary components and averaged those outputs to obtain an estimate for that matrix. For each matrix and number of iterations, we performed 1000 trials in order to obtain statistical behaviour. For each trial, we measured the distance between the result of the Hadamard test and the exact value, then took the percentiles of that data.

\begin{figure}
\includegraphics[width=\linewidth]{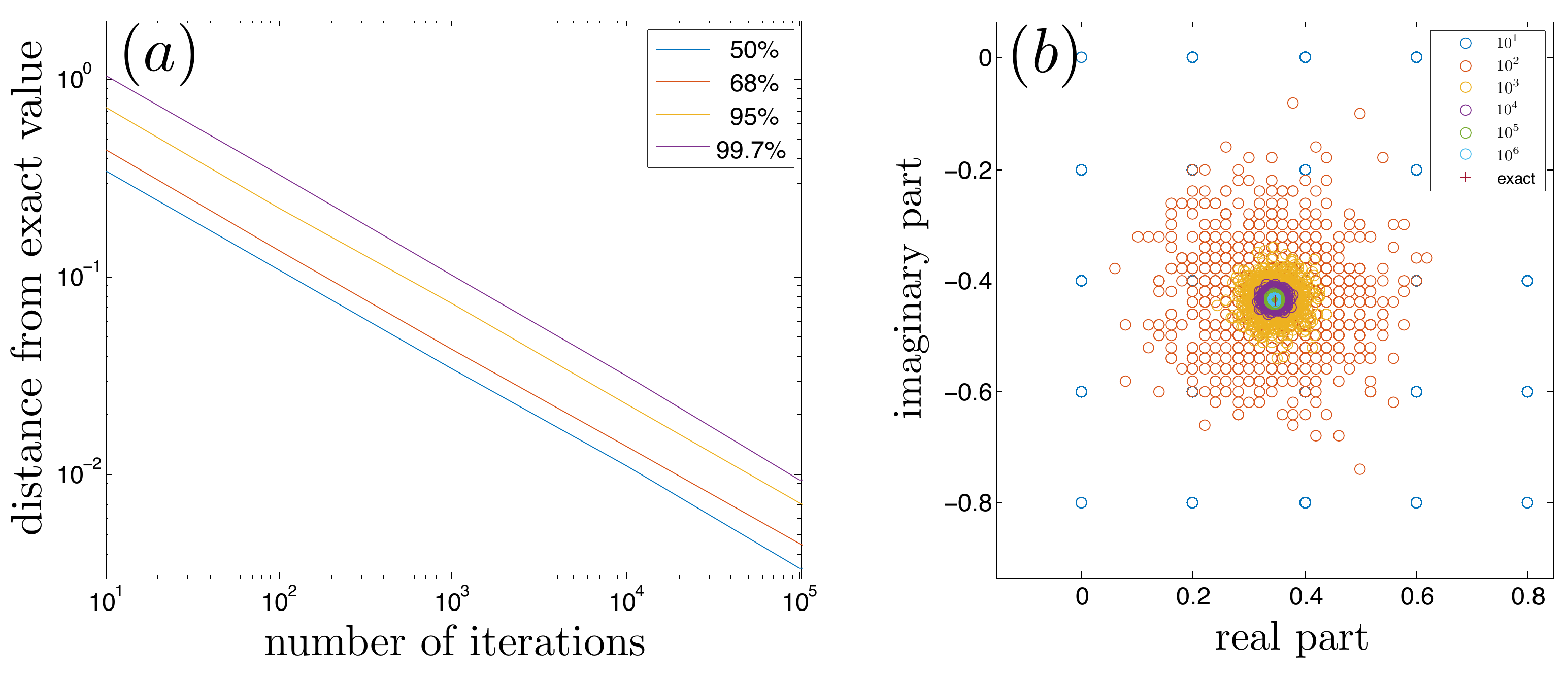}
	\caption{Convergence of the Hadamard test for finding a single matrix element for a particular matrix. (a) The distance from the exact solution within which a given percentage of the results lay as functions of the number of iterations on a log-log scale. (b) The outputs of individual trials are compared against the exact solution.}
	\label{Fig:hadamard_convergence_single}
\end{figure}

In \Figref{Fig:hadamard_convergence_single} are representative results for the convergence of the Hadamard test when testing for a single matrix element. We found that the convergence follows a power law, where the error scales proportionate to $\frac{1}{\sqrt{N}}$ with the number of iterations of the Hadamard test.

\begin{figure}
	\includegraphics[width=\linewidth]{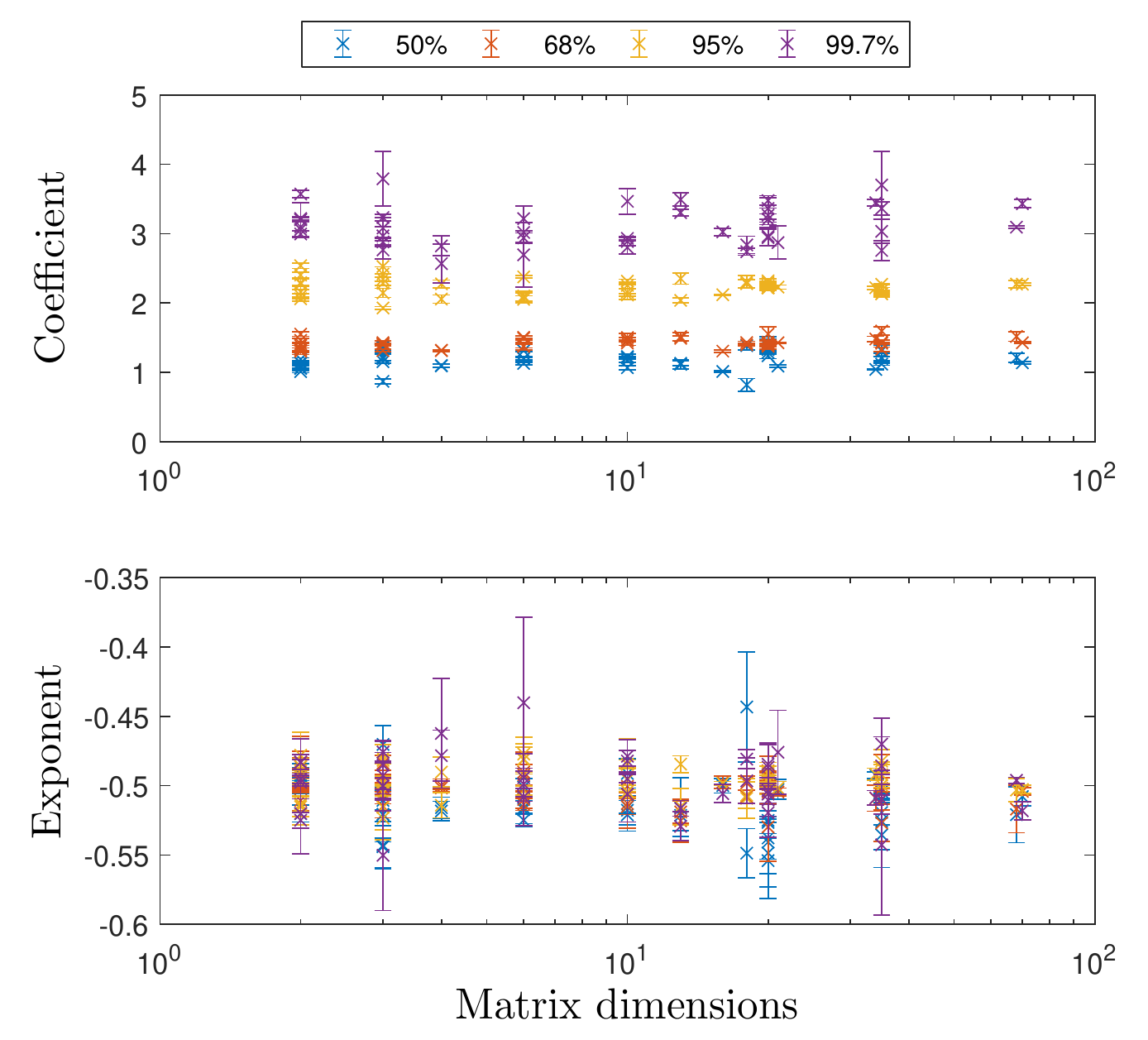}
	\caption{The coefficient and exponent in the power law describing the distance from the exact value within which a percentage of the result lay with respect to number of iterations of the Hadamard test, for calculating the normalised trace of matrices of varying sizes. Each point is a different matrix that was tested, representing a range of dimensions. The log scale in the matrix dimensions reflects the fact that the size of the AJL matrices grows exponentially with increasing $n$. Error bars represent 95\% confidence intervals for the curve fitting.}
	\label{Fig:hadamard_convergence_trace}
\end{figure}

We also evaluated the rate of convergence of the Hadamard test when measuring the Markov trace by stochastic sampling of bases. For simplicity, we took a normalised regular trace instead, $\operatorname{\rm tr}(\Phi)/\operatorname{dim}(\Phi)$, which corresponds to a Markov trace where $\lambda_l=1$ for all $l$ and has a value of 1 when acting on the identity. In measuring the normalised trace we used a randomly selected basis vector for each iteration of the Hadamard test, then averaged the outputs as for the regular Hadamard test. We tested it on arbitrarily selected AJL matrices spanning a range of values for the parameters. For each, we measured the rate of convergence similarly to \Figref{Fig:hadamard_convergence_single} and performed a power law fit to it using MatLab's Curve Fitting Toolbox, recording the coefficient and the exponent against the dimension of the matrix measured for each percentile. A total of 42 matrices were measured in this manner. The results of these tests are in \Figref{Fig:hadamard_convergence_trace}.

As shown in \Figref{Fig:hadamard_convergence_trace}, the power law describing the rate of convergence does not vary perceptibly as the size of the matrix is increased from 2 dimensional to 70 dimensional, aside from random fluctuations which do not present any net increase or decrease. This means that, using a quantum computer to evaluate the Markov trace using the Hadamard test, the number of iterations required for a given accuracy is independent of the size of the matrix being measured.

For comparison, we also measured the convergence for measuring a single, randomly selected diagonal element from a randomly selected AJL matrix. We made 20 such measurements. The mean of the measured coefficients and powers was taken and recorded in Table~\ref{tab:hadamard_results}.

\begin{table}
\caption{Average rates of convergence for given percentiles in measuring a matrix via the Hadamard test. The number of iterations of the Hadamard test is denoted by $N$, and the expression gives the distance from the exact value within which the specified percentage of tests lie. The uncertainties indicate the standard deviation of the coefficient and exponent across the tests.}
\begin{tabular}{c c c}
	\hline
	\hline 
	Percentile & Single Element & Normalised Trace \\ 
	\hline 
	50\% & $0.94(\pm0.22)N^{-0.53\pm0.04}$ & $1.18(\pm0.13)N^{-0.51\pm0.02}$ \\ 
	 
	68\% & $1.29(\pm0.16)N^{-0.53\pm0.03}$ & $1.42(\pm0.07)N^{-0.50\pm0.01}$ \\ 
	
	95\% & $1.95(\pm0.23)N^{-0.50\pm0.02}$ & $2.23(\pm0.13)N^{-0.50\pm0.01}$ \\ 
	 
	99.7\% & $2.76(\pm0.50)N^{-0.49\pm0.04}$ & $3.11(\pm0.28)N^{-0.50\pm0.02}$ \\ 
	\hline
	\hline 
\end{tabular}
\label{tab:hadamard_results}
\end{table}

From our results in Table~\ref{tab:hadamard_results}, the Hadamard test converges at a rate proportional to $\frac{1}{\sqrt{N}}$. Higher percentiles have higher leading coefficients, as would be expected. From our results, the calculation of the trace by the Hadamard test incurs a small overhead compared to the calculation of a single element, with a slightly larger coefficient, but otherwise their rates of convergence are similar.

The expressions in Table~\ref{tab:hadamard_results}, in particular the coefficients, can be used to estimate confidence intervals for the output of the Hadamard test. For instance, when taking the normalised trace of a matrix using 10,000 iterations of the Hadamard test for each of the real and imaginary components, there is a 95\% probability that the output is within 0.022 of the exact value. To obtain an order of magnitude greater accuracy, it is necessary to perform 100 times more iterations of the Hadamard test.

\subsection{An Exact Algorithm}\label{sec:exact_ajl}

The AJL algorithm as described is completely general, allowing for the evaluation of the Jones polynomial at the $k$'th root of unity, for any integer $k\ge3$, using any quantum computing model. However, for implementation into a topological quantum computer, braids and weaves which approximate the gates required to perform the Hadamard test are necessary. The results of any such computation will be limited by how closely the braids approximate the target matrices.

However, the Jones polynomial has connections with the topological quantum field theories which underpin anyon models \cite{witten1989,Anyon_Computing_Freedman,Freedman_TQC_Jones_Polynomial,Jones_Polynomial_TQC}, so it might be reasonable to suspect that there should exist a more facile algorithm connecting anyons and the Jones polynomial.

For Fibonacci anyons, such a connection does make itself apparent for the $k=5$ case \cite{Shor2008a}. We have compared the Fibonacci braiding matrices, as derived in Section \ref{sec:fib_braid}, to the braiding matrices for the AJL algorithm, as derived in Section \ref{sec:algorithm_derive}. For $k=5$, we find that these two braiding matrices were equal, up to a sign and the chirality of the braid.

Specifically, consider $n=4$, with four strands, and four Fibonacci anyons. Let the AJL matrices be denoted as $b_i$ and the Fibonacci braiding matrices be denoted $\sigma_i$. We found that $b_i=-\sigma_i$ exactly. Considering the AJL matrices $\Theta_j(4,5)$ and the Fibonacci braiding matrices $\sigma_j$, we found that $\Theta_j(4,5) = -\sigma_j$. Similar relationships hold for smaller $n$, and we conjecture that they also hold for larger $n$.

This exact result leads itself to a special implementation of the AJL algorithm. Consider the case where we wish to investigate the plat closure of a braid with four strands, such as those presented in \Figref{Fig:knot_plat_braids}. We select $k=5$, such that $t=e^{2\pi i/5}$ and $d=2\cos(\pi/5)=\phi$, where $\phi=\frac{1+\sqrt{5}}{2}$ is the golden ratio. Create two pairs of Fibonacci anyons from the vacuum. We will label the initial state where each pair fuses to vacuum $\ket{0}$, and this is the first vector in our basis. Now perform the braid described by the braidword of the knot being investigated using the Fibonacci anyons, tracing that knot with the $2+1$ dimensional worldlines of the anyons. Then fuse each anyon pairwise.

The operation performed on the anyonic system by the braiding is exactly equal to the product of AJL matrices $\Phi(\rho(B))$ up to a sign. The probability that the $\ket{0}$ state is measured is
\begin{equation}
\text{Pr}(\ket{0}) = \abs{\bra{0}\Phi(\rho(B))\ket{0}}^2.
\label{eqn:exact_probability}
\end{equation}
This probability can be measured by repeated braiding and measurement. This is sufficient to find the magnitude of the bracket polynomial of the plat closure, by \eqref{eqn:plat2}. From \eqref{eqn:Jones2} we can state that the magnitude of the Jones polynomial at the 5'th root of unity is given by
\begin{equation}
\abs{V_{B^{\rm pl}}(e^{2\pi i/5})} = \phi^{\tfrac{n}{2}-1}\sqrt{\operatorname{Pr}(\ket{0})}.
\label{eqn:exact_jones}
\end{equation}

This simple algorithm, where the knot created by braiding Fibonacci anyons relates directly to the Jones polynomial of that knot, underscores the deep connections the Jones polynomial has with topological quantum field theory, and motivates the investigation of the Jones polynomial in this topological quantum computer.

This calculation can be taken further by considering another anyon model and finding how it relates to the AJL matrices. Consider the Majorana zero modes that may be used for realising the Ising anyon model \cite{Sarma2015a}, which has two non-trivial anyons conventionally denoted by $\sigma$ and $\psi$ and the vacuum denoted by ${\bf 0}$. These anyons are governed by the fusion rules,
\begin{equation}
\begin{aligned}
\sigma \otimes \sigma &= {\bf 0}\oplus  \psi \\
\psi \otimes \psi &= {\bf 0}\\
\sigma \otimes \psi &= \sigma\\
{\bf 0}\otimes \alpha &= \alpha, 
\end{aligned}
\label{eqn:ising_fusion}
\end{equation}
where $\alpha$ may be ${\bf 0}$, $\psi$ or $\sigma$ and the anyon $\sigma$ is not to be confused with the braid matrix $\sigma_{j,\text{Ising}}$.

With regards to the F and R moves, the only non-trivial arrangements have all the anyons starting as the non-Abelian anyon $\sigma$. In the basis where the first two $\sigma$ anyons fuse to the vacuum or $\psi$ respectively, the F and R matrices are \cite{Pachos_TQC_Book,Anyon_Computing_Nayak}
\begin{align}
F &= \frac{1}{\sqrt{2}} \begin{pmatrix}
1&1\\1&-1
\end{pmatrix}, \label{eqn:ising_fmatrix}\\
R &= e^{-\pi i/8} \begin{pmatrix}1&0\\0&i\end{pmatrix}.\label{eqn:ising_rmatrix}
\end{align}
This allows us to calculate the braid matrices for the Ising anyon model up to 3 strands, which are
\begin{align}
\sigma_{1,\text{Ising}} = R &= e^{-\pi i/8} \begin{pmatrix}1&0\\0&i\end{pmatrix},\label{eqn:ising_b1}\\
\sigma_{2,\text{Ising}} = FRF &= \frac{1}{\sqrt{2}} \begin{pmatrix}e^{\pi i/8}&e^{-3\pi i/8}\\e^{-3\pi i/8}&e^{\pi i/8}\end{pmatrix}.\label{eqn:ising_b2}
\end{align}

Calculating the AJL matrices for the $n=3$, $k=4$ case, similarly to the calculation in \eqref{eqn:AJL_braid_example}, yields
\begin{align}
\Theta_1(3,4) = e^{\pi i/8}\begin{pmatrix}1&0\\0&-i\end{pmatrix},\\
\Theta_2(3,4) = \frac{1}{\sqrt{2}}\begin{pmatrix}e^{-\pi i/8}&e^{3\pi i/8}\\e^{3\pi i/8}&e^{-\pi i/8}\end{pmatrix}.
\end{align}

Further calculations in the $n=4$ case shows that a similar relationship holds. Thus we may state that $\Theta_j(n,4)=\sigma_{j,\text{Ising}}^{-1}$ for $n\le4$, and we conjecture that it holds for higher $n$ as well. This means that braiding Ising anyons can be used for finding the magnitude of the Jones polynomial at the 4'th root of unity.

It seems probable that the result is generic such that the AJL matrices $\Theta_j(n,k)$ for each value of $k$ generate (the elementary braid matrices of) a certain anyon model associated with a column in Fig.~(\ref{Fig:ajl_matrices}), that could be used to find the exact value of the Jones polynomial at the specific roots of unity. Hence, we could generalise \eqref{eqn:exact_jones} to arbitrary anyon Model $k$ via
\begin{equation}
\abs{V_{B^{\rm pl}}(e^{2\pi i/k})} = d^{\tfrac{n}{2}-1}\sqrt{\operatorname{Pr}(\ket{0}_k)},
\label{eqn:exact_jones_general}
\end{equation}
where $d$ is given by \eqref{eqn:AJL_constants} and $\operatorname{Pr}(\ket{0}_k)$ is the probability of measuring the state where all anyon pairs fuse to vacuum in the anyon model corresponding to $k$ after performing the braid.

Of interest is the quantity $d=d_k=2\cos(\pi/k)$. For $k=4$, $d_4=d_\sigma=\sqrt{2}$, which is the quantum dimension of the $\sigma$ anyon in the Ising anyon model. For $k=5$, $d_5=\phi$, which is the quantum dimension of the $\tau$ anyon of the Fibonacci anyon model. The Models 6 and 7, see Fig.~(\ref{Fig:ajl_matrices}), have $d_6=\sqrt{3}$ and $d_7 \approx 1.8019$, respectively. Generically, Model $k$ anyons interpolate between Abelian anyons with $d_2=1$ and $\lim_{k\to\infty}d_k=2$.

We shall make the conjecture that this quantity $d_k$ is in general equal to the quantum dimension of a certain non-Abelian anyon of a particular anyon model whose braid matrices are straightforwardly linked to $\Theta_j(n,k)$. 
Since $d_7$ is not a square root of an integer, we anticipate the Model 7 to be capable of universal topological quantum computation, similarly to the (Fibonacci) Model 5 case, and in contrast to the (Ising) Model 4 case. A more thorough evaluation, beyond the scope of this work, will be needed to confirm whether or not this conjecture is true, but the strong connections between the Jones polynomial and the topological quantum field theory from which anyons arise are concretely apparent in these examples.

The existence of a similar algorithm for finding the Jones polynomial was implied by \cite{Shor2008a,Pachos_TQC_Book,Anyon_Computing_Freedman}, and it was mentioned briefly in \cite{Jones_Polynomial_TQC}. Here we have explicitly presented the algorithm in a direct and facile manner for Fibonacci and Ising anyons, shown its connection to the AJL algorithm, and (in Sec.~\ref{sec:results}) demonstrated its application to several simple knots for Fibonacci anyons.

As an algorithm for evaluating the Jones polynomial, it is fairly limited. It can only evaluate the Jones polynomial at the point $t=e^{2\pi i/5}$ for Fibonacci anyons, at $t=i$ for Ising anyons and generically at $t=e^{2\pi i/k}$ for Model $k$ anyons, and even then it only yields the magnitude of the Jones polynomial, because the phase is inaccessible to direct quantum measurement. Thus the general AJL algorithm, the Hadamard test and the compilation of braids approximating operations are still necessary to find values of the Jones polynomial outside these particular cases, and those methods are needed to use topological quantum computing for arbitrary quantum algorithms.

\section{Intermediate Summary}\label{sec:intermediate_summary}

Before detailing our numerical implementation of the topological quantum computer and using it to perform the AJL algorithm, we shall summarise the essential steps for using a topological quantum computer. Before any computation can begin, certain necessary preprocessing steps must be taken:

\begin{enumerate}
	\item Determine the braiding matrices for the chosen anyon model (Sec.~\ref{sec:fib_braid}) and define how many anyons are in a qubit (Sec.~\ref{sec:fib_compute}).
	\item Construct single qubit braids approximating the required operations to a desired accuracy, by brute force or another method (Sec.~\ref{sec:compile}). The minimum length of the braids scales as $\mathcal{O}(\log(1/\epsilon))$.
	\item Construct controlled operations by a technique such as the weaving method by  \cite{Bonesteel_Braid_Topologies}, using phase gates to correct for unintended phase differences (Sec.~\ref{sec:compile}).
\end{enumerate}

After this preprocessing is complete, a quantum algorithm can be performed. In particular, we are investigating the AJL algorithm for finding the Jones polynomial:

\begin{enumerate}
	\item Convert the knot or link $K$ under investigation into either the plat closure or trace closure of a braid $B$. The number of strands defines the parameter $n$.
	\item Classically compute the writhe $w(K)$ of the knot.
	\item Select an integer value for the parameter $k$, which defines the point $t=e^{2\pi i/k}$ where the Jones polynomial will be evaluated, as well as the related constant $d=2\cos(\pi/k)$.
	\item Find the AJL matrices for each elementary braid in $B$ for the given $n$ and $k$. Optionally decompose the AJL matrices into blocks.
	\item Compile controlled versions of these AJL matrices for the quantum computer. Also compile the Hadamard gate, $-\pi/2$ phase gate and the NOT gate.
	\item In the quantum computer, perform the Hadamard test (Sec.~\ref{sec:hadamard}), where the controlled operations corresponding to each elementary braid in $B$ are performed sequentially. Repeat $N$ times for the real component and $N$ times for the imaginary component, and take the average. The result will converge at a rate proportional to $1/\sqrt{N}$.
	\begin{enumerate}
		\item For the trace closure, for each iteration randomly select a basis for the target qubit(s) with weighting $\lambda_l$. The states can be initialised using NOT gates.
		\item For the plat closure, simply use the $\ket{0}$ state.
	\end{enumerate}
	\item Multiply the above result by $d^{n-1}$ for the trace closure, \eqref{eqn:AJL_kauffman_3}, or $d^{\tfrac{n}{2}-1}$ for the plat closure, \eqref{eqn:plat2}, obtaining the Kauffman bracket polynomial.
	\item Multiply the Kauffman bracket polynomial by $(-t^{-3/4})^{-w(K)}$ to obtain the Jones polynomial at the point $t$.
\end{enumerate}

The magnitude of the Jones polynomial at $t=e^{2\pi i/5}$ for the plat closure of a braid can also be found by performing the same braid with Fibonacci anyons, measuring the probability of obtaining $\ket{0}$, and using \eqref{eqn:exact_jones}.

\begin{figure}
\begin{algorithm}[H]
	\caption{Initialisation of the quantum computer to a register with $n$ qubits.}
	\label{alg:initialise}
\begin{algorithmic}[1]
	\State ${\tt qubits} \gets n$ 
	\State ${\tt state} \gets \operatorname{zeros}(2^n,1)$  \\
     \Comment{Column vector of zeros} 
	\State ${\tt state}[1] \gets 1$\\
	\Comment{First element is 1. Corresponds to $\ket{00\dots0}$.}
	\State ${\tt braidMatrix} \gets \mathbb{I}_{2^n}$
\end{algorithmic}
\end{algorithm}
\end{figure}

\section{Numerical Implementation}\label{sec:simulation}

To demonstrate the action of a Fibonacci anyon topological quantum computer, we have created a simple simulation in MatLab. This simulation, rather than considering the physical mechanisms involved with the anyons, deals purely with their exchange statistics.

In brief, the simulation creates a column vector representing the initialised qubit register. It then multiplies elementary braiding matrices together to form a matrix corresponding to the whole braiding operation. Then it multiplies that matrix onto the state vector to produce a final vector, which is used to construct a probability distribution from which one of the bases is chosen. This basis is returned as a bitstring, which corresponds to the state each qubit is measured to be in, which in turn corresponds to the fusion outcomes.

\subsection{Simulator Code}\label{sec:code}

Before performing any simulations, the simulator produces the elementary Fibonacci braiding matrices for one and two qubits, as described in Section \ref{sec:fib_braid}, using predetermined sequences of F and R moves and predetermined basis states. The elementary matrices and their inverses for a single qubit are recorded in $\tt elemOne$, and the elementary matrices and their inverses for two qubits are recorded in $\tt elemTwo$. These lists are indexed such that, if negative indices are wrapped around from the end of the list, then the negative indices correspond to the inverses of the corresponding positive indices (eg. $\sigma_2^{-1}$ is found at index $-2$).

The first step in quantum computation is to initialise a set of $n$ qubits, each in the $\ket{0}$ state. In the simulator, this is performed by generating a column vector. The braid operator is also initialised as the identity, because no braiding has taken place yet. The initialisation procedure is described in Algorithm \ref{alg:initialise}.

The next part of the quantum computer is performing braiding on single qubits and on pairs of qubits. First we will consider single qubit braids.

To perform a quantum computation, we have a list $\tt braidword$ of elementary braiding operations to perform, where each element of the list is an integer, positive or negative depending on the direction of the braid. We wish to perform this braid on the qubit in position ${\tt pos}$. The simulator's procedure for single qubit braids is described in Algorithm \ref{alg:1qubit}. Note that, as discussed in Section \ref{sec:fib_braid}, to apply the braiding operations in chronological order the matrices are right-multiplied onto each other.

\begin{figure}
\begin{algorithm}[H]
	\caption{Single qubit braid, performing $\tt braidword$ on qubit $\tt  pos$.}
	\label{alg:1qubit}
	\begin{algorithmic}[1]
		\State ${\tt matrix} \gets \mathbb{I}_2$
		\For {$i\gets1:\operatorname{length}({\tt braidword})$}
			\State ${\tt matrix} \gets {\tt matrix} * {\tt elemOne}[{\tt braidword}[i]]$
			\Comment{Find the matrix corresponding to the action of the specified braidword}
		\EndFor
		\State $ {\tt matrix} \gets \mathbb{I}_{2^{{\tt  pos-1}}} \otimes {\tt matrix} \otimes \mathbb{I}_{2^{{\tt qubits-pos-1}}}$
		\Comment{Apply tensor products to expand the operation to apply to the whole register.}
		\State ${\tt braidMatrix} \gets {\tt braidMatrix} * {\tt matrix}$
		\Comment{Append this braid to the overall braid.}
	\end{algorithmic}
\end{algorithm}
\end{figure}

The procedure for a two-qubit braid is slightly more complicated, due to the potential of leakage into non-computational basis states. We apply the braid specified by $\tt braidword$ to the adjacent qubits in positions $\tt pos$ and $\tt pos+1$. However, the elementary braiding matrices for the two qubit braid are 13 dimensional, while the computational subspace for two qubits is only 4 dimensional. It would be impractical for the simulator to track all the non-computational states, and would make it difficult to expand the simulator to arbitrary numbers of qubits. A deeper investigation into the effects of these non-computational states and leakage is beyond the scope of this work. For our simulation, we will apply the heuristic solution of discarding the non-computational states after performing the two qubit braid by truncating the matrix. Provided that leakage is small for the braids, then the effect of this truncation should also be small. The procedure for two qubit braids is described in Algorithm \ref{alg:2qubit}.

To determine the extent of the leakage error, we will use a variant of the operator norm. The regular operator norm describes the maximum amount by which a matrix can increase the norm of a vector, and is given by
\begin{equation}
\vert\vert A \vert\vert = \sqrt{\operatorname{maxEigenvalue}(A^{\dagger} A)}.
\label{eqn:operator_norm2}
\end{equation}

We instead want to find the smallest factor by which a matrix will change the norm of a vector. By repeating the derivation of the regular operator norm, it can readily be shown that this modified operator norm can be computed by
\begin{equation}
\vert\vert A \vert\vert_{\rm small} = \sqrt{\operatorname{minEigenvalue}(A^{\dagger} A)}.
\label{eqn:modified_operator_norm}
\end{equation}

Leakage error results in a transfer of probability density from the computational states into the non-computational states. Normally, unitary operators preserve normalisation, so $\abs{\abs{U}}$ and $\abs{\abs{U}}_{\rm small}$ both equal 1. However, by truncating the matrix, the matrix only approximates a unitary operator. By truncating the elements in non-computational states, this operator would cause the norm of some vectors to be reduced, thereby reducing the value of the modified operator norm \eqref{eqn:modified_operator_norm}. The probability of leakage into a non-computational state, which we use as the metric for the leakage error, is

\begin{equation}
{\rm leakage} = 1 - \abs{\abs{A}}_{\rm small}^2. \label{eqn:leakage}
\end{equation}

\begin{figure}
\begin{algorithm}[H]
	\caption{Two qubit braid, performing $\tt braidword$ on qubits $\rm pos$ and $\rm pos+1$}
	\label{alg:2qubit}
	\begin{algorithmic}[1]
		\State ${\tt matrix} \gets \mathbb{I}_{13}$
		\For {$i\gets1:\operatorname{length}({\tt braidword})$}
			\State ${\tt matrix} \gets {\tt matrix} * {\tt elemTwo}[{\tt braidword}[i]]$
			\Comment{Find the matrix corresponding to the action of the specified braidword}
		\EndFor
		\State ${\tt matrix} \gets {\tt matrix}[1:4,1:4]$
		\Comment{Truncate matrix to the first $4\times4$ elements}
		\State ${\tt leakage} \gets 1-\operatorname{minEigenvalue}({\tt matrix}^{\dagger}*{\tt matrix})$
		\Comment{Find the leakage error}
		\State \textbf{print} ${\tt leakage}$
		\State ${\tt matrix} \gets \mathbb{I}_{2^{{\rm pos-1}}} \otimes {\tt matrix} \otimes \mathbb{I}_{2^{{\rm qubits-pos-2}}}$
		\Comment{Apply tensor products to expand the operation to apply to the whole register.}
		\State ${\tt braidMatrix} \gets {\tt braidMatrix} * {\tt matrix}$
		\Comment{Append this braid to the overall braid.}
	\end{algorithmic}
\end{algorithm}
\end{figure}

In order to initialise the quantum computer into a particular initial state other than $\ket{00\dots0}$, as is usual for many algorithms, it is necessary to apply braids corresponding to NOT gates after the application of all the other braiding. This is because this initialisation operation needs to be left-multiplied onto the state vector, which means it must be right-multiplied onto the braiding matrix, which requires the `initialisation' to, counter-intuitively, be applied last.

Although not used in our quantum algorithm, the SWAP gate can be implemented in this simulator by right-multiplying the appropriate SWAP matrix to $\tt braidMatrix$, with the implicit understanding that this represents the exchange of whole four-anyon qubits.

Finally, once all the braids necessary for the quantum computation have been performed, it is necessary to measure the state. Physically, the anyons would be fused sequentially and the fusion outcomes would be recorded. This sequence of fusion outcomes corresponds to the states of the qubits, which may either be $\ket{0}$ or $\ket{1}$, unless a non-computational state is measured. This measured state can be directly represented as a bitstring, corresponding to the state of each qubit. The simulator performs this measurement as in Algorithm \ref{alg:measure}.

Of note is the possible return value of `error'. In performing two qubit operations, leakage can cause the normalisation of the state vector to become less than unity. The `error' return value represents measuring a non-computational state.

\begin{figure}
\begin{algorithm}[H]
	\caption{Measurement of the quantum computer}
	\label{alg:measure}
	\begin{algorithmic}[1]
		\State ${\tt state} \gets {\tt braidMatrix} * {\tt state}$
		\\ \Comment{Apply the braiding operation}
		\For {$i\gets1:\operatorname{length}({\tt state})$}
			\State ${\tt probability}[i] \gets \abs{{\tt state}[i]}^2 + \operatorname{sum}({\tt probability}[1:i-1])$
		\\ 	\Comment{Build probability distribution as a cumulative sum of the magnitude squared of the state vector components}
		\EndFor
		\State ${\tt r} \gets \operatorname{random}(1)$
		\\ \Comment{Use a random number between 0 and 1 to search the probability distribution}
		\State ${\tt bits} \gets \text{`error'}$
		 \Comment{Default case is return an error. Will happen if ${\tt r}>\abs{{\tt state}}^2$.}
		\For {$i\gets1:\operatorname{length}({\tt probability} )}$
			\If {${\tt r} \le {\tt probability}[i]$}
				\State ${\tt bits} \gets \operatorname{Decimal2Binary}(i-1)$
			\\	\Comment {We have randomly picked this basis state}
				\State \textbf{break}
			\EndIf
		\EndFor
		\State \Return ${\tt bits}$
	\end{algorithmic}
\end{algorithm}
\end{figure}

After a measurement has been performed, the qubits have all been fused together. The quantum computer must be initialised again, as per Algorithm \ref{alg:initialise}, in order to continue computation.

This simulator allows for the simulation of a Fibonacci anyon topological quantum computer down to the detail of which elementary braids are performed. It can be adapted to any anyon model simply by changing the sets of elementary braiding matrices to those which correspond to that anyon model. A classical front-end is used to decide which braids to perform and to process the output of the quantum computer.

For the sake of simplicity, this simulator lacks a few of the more technically challenging aspects of such a quantum computer. It does not account for physical errors that might occur in a system of physical anyons. It also handles non-computational states in a heuristic manner, which does not capture any interactions between non-computational states and computational states.

Furthermore, this is a relatively naive simulation of a quantum computer, where we simply multiply together matrices. While more efficient and subtle methods of simulating quantum computers exist, this method was chosen for its simplicity and the transparency of its implementation, which are both important for this pedagogical demonstration.

Notwithstanding these simplifications, this simulator is capable of simulating universal quantum computation by performing braiding of anyons.

\subsection{Simulation of AJL Algorithm}\label{sec:results}

\subsubsection{General Procedure}

To use the quantum computer to perform the AJL algorithm, braids and weaves approximating the requisite gates had to be compiled. Weaves for the AJL matrices between $k=4$ and $k=13$ were found as in Section \ref{sec:braid_convergence}. Weaves up to 15 elementary weaving operations in length were searched. For $\Theta_1(2,k)$ and the first block of $\Theta_2(3,k)$, the approximations had errors between 0.0011 and 0.0157, with the exception of $\Theta_1(2,10)$, which had an exact solution up to an overall phase.

These weaves were constructed into controlled gates by the method in Section \ref{sec:two_qubit_braids}. For the phase correction for these gates, weaves of length up to 15 elementary weaving operations were searched to obtain a phase gate corresponding to each AJL weave.

To measure the second, scalar block in the $n=3$ AJL matrices, phase gates with phases corresponding to that element were created by searching weaves up to 15 elementary weaving operations long.

These AJL weaves, due to the equivalence between certain AJL matrices, are sufficient to perform the AJL algorithm on the trace closure of any braid with 2 or 3 strands and on the plat closure of any braid with 4 strands, for $k$ between 4 and 13.

\begin{figure}
	\includegraphics[width=\linewidth]{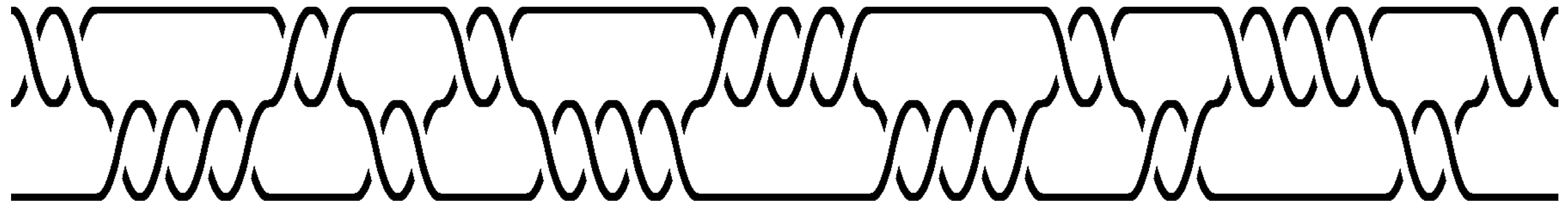}
	\caption{A weave approximating the $-\pi/2$ phase gate, $\left(\begin{smallmatrix}1 & 0 \\ 0 & -i\end{smallmatrix}\right)$, with an error of 0.0045. Time points to the right in this diagram. The braid consists of 36 elementary braiding operations and has the braidword $\sigma_{2}^{2} \sigma_{1}^{-4} \sigma_{2}^{-2} \sigma_{1}^{2} \sigma_{2}^{2} \sigma_{1}^{4} \sigma_{2}^{-4} \sigma_{1}^{-4} \sigma_{2}^{2} \sigma_{1}^{-2} \sigma_{2}^{4} \sigma_{1}^{2} \sigma_{2}^{2}$.}
	\label{Fig:phase_weave}
\end{figure}

Necessary for the Hadamard test was the Hadamard gate and the $-\pi/2$ phase gate (which is the inverse of the $\pi/2$ phase gate). These gates were found by searching weaves up to 18 elementary weaving operations in length. The braid for the Hadamard gate, in \Figref{Fig:hadamard_weave}, had an error of 0.003, and the braid for the phase gate, in \Figref{Fig:phase_weave}, had an error of 0.0045.

Also used is the weave for the NOT gate, for initialisation of states, and the injection weave, for the construction of controlled gates. These weaves are in \Figref{Fig:NOT_weave} and \Figref{Fig:inject_weave} respectively.

The following procedure was used to perform the AJL algorithm in our simulation of the quantum computer. The knot under investigation was specified with a braidword, the number of strands $n$, and whether the knot was the trace or plat closure of the braid. The parameter $k$ was also specified.

Given this $n$ and $k$, we retrieved the relevant AJL weaves, and the corresponding phase corrections. We concatenated the AJL weaves into a single longer weave in the order specified by the knot's braidword, then used the method in Sec.~\ref{sec:two_qubit_braids} to convert that into a controlled operation, with the first qubit being the control qubit. The corresponding phase corrections were then applied to the control qubit. This concatenation was done, rather than applying individual controlled operations for each braid in the knot, because adjacent injection weaves would cancel each other out, and thus would be redundant. The phase gates on the first qubit commute with the controlled operations, so they were applied afterwards.

\begin{figure*}
	\includegraphics[width=\linewidth]{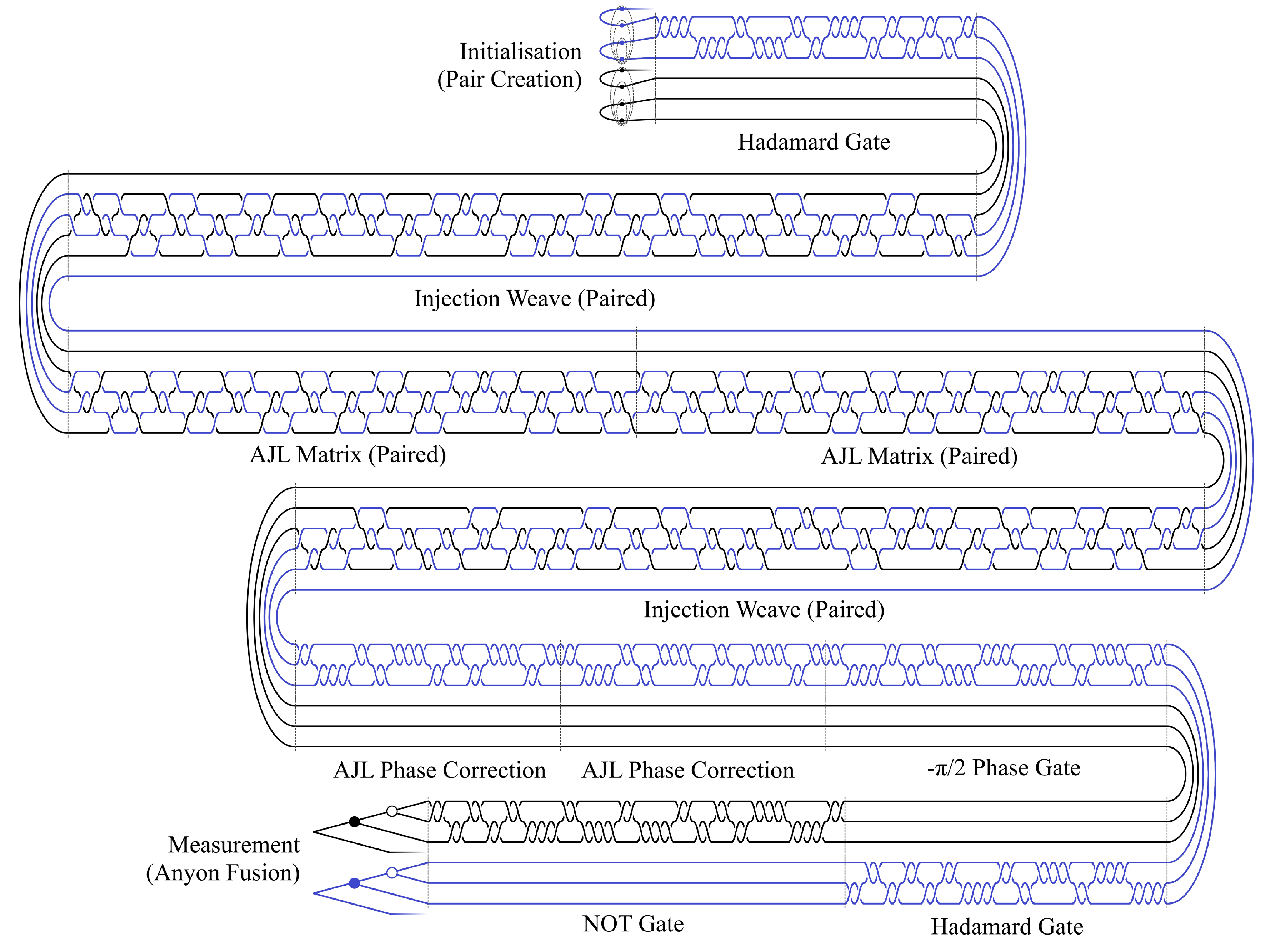}
	\caption{A braid typical of a computation in the AJL algorithm. In this particular braid, the imaginary component of the second diagonal element in the AJL matrix product corresponding to the trace closure of the positive Hopf link with $k=6$ is being measured. It has a leakage error of $3.6\times10^{-6}$. The first qubit is coloured blue. The second qubit is coloured black. The segments marked as `paired' indicate that the weave has been modified such that two anyons are being moved instead of a single anyon. The state of the anyons at the hollow circles in the measurement step determine the state of the qubit. The target operation for the AJL Matrix is $\Theta_1(2,6)^{-1}=\left(\begin{smallmatrix}0.7071-0.7071i&0\\0&0.2588+0.9629i		\end{smallmatrix}\right)$. The weave presented performs the operation $e^{4.1364i}\left(\begin{smallmatrix}0.7026-0.7115i&-0.0091+0.0022i\\0.0017-0.0092i&0.2527+0.9675i\end{smallmatrix}\right)$. To correct for the phase factor, the AJL Phase Correction weave (approximately) implements a ${-4.1364}$ phase gate.}
	\label{Fig:big_weave}
\end{figure*}

This larger weave was used as the controlled gate within the Hadamard test, with the whole controlled gate performed as a single two-qubit weave. The rest of the Hadamard test used the weaves corresponding to the Hadamard gate and $-\pi/2$ phase gate. When measuring the Markov trace, the state of the second qubit was determined randomly based on a distribution weighted by $\lambda_l$ (\eqref{eqn:lambda}). The qubit could be initialised to the $\ket{1}$ state by applying the weave approximating the NOT gate. A representative braid corresponding to a computation is given in \Figref{Fig:big_weave}.

After all weaving had been performed, the state was measured. If the first bit of the returned bitstring was 0, then the Hadamard test returned 1. If the first bit was 1, then the Hadamard test returned $-1$. This was performed for a specified number of iterations for each of the real and imaginary outputs, and the mean of the outputs for each component was taken. This number was the approximation of the Markov trace for the trace closure or the first matrix element for the plat closure.

The writhe of the knot was calculated classically, and then the appropriate factors were multiplied to the result of the Hadamard test to give an approximation to the Jones polynomial at the point $t=e^{2\pi i/k}$.

The Hadamard test is stochastic in nature. The results in Table~\ref{tab:hadamard_results} were used to estimate the confidence intervals for a given output of the Hadamard test. For the confidence interval for that point in the Jones polynomial, the figure for the Hadamard test was multiplied by $d^{n-1}$ for the trace closure or $d^{\tfrac{n}{2}-1}$ for the plat closure. In our results, we reported the 95\% confidence interval for each data point in the Jones polynomial as error bars.

Because our quantum computer is a classical simulation, we were able to access information that would not be measurable in a real quantum computer. As a measure of comparison to the stochastic results, we directly read the components of the state vector and the probability of measuring each qubit in a given state, bypassing the random nature of Algorithm \ref{alg:measure}. This was used to precisely determine the expectation value of the Hadamard test for a given set of weaves in the quantum computer. This, in turn, gave the output of the stochastic measurements in the limit of an infinite number of iterations. This quantity is not affected by the stochastic nature of the regular measurements, but it is affected by how closely the given weaves approximate the intended operations. As such, this metric, although inaccessible to a real quantum computer, provides a measure of the quality of the weaves, and indicates where the measurement will converge to.

Because the knots under investigation were relatively simple, we verified the output of the quantum AJL algorithm against the known analytical solution for the Jones polynomial.

For the braids we have generated, the difference between the exact value of the Jones polynomial and the result given by braiding in the limit of infinite iterations was typically of the order of 0.01. As such, for our measurements in the AJL algorithm, we used 10,000 iterations of the Hadamard test, which gave a precision of the same order of magnitude. More iterations would be superfluous for our weaves, for it would give greater precision than accuracy.

The time complexity of the algorithm as performed by this simulated quantum computer was quantified by counting the number of elementary braiding operations performed. If it is assumed, in a physical implementation, that each elementary braid takes some fixed amount of time and that the computer spends most of its time braiding, then the number of elementary braiding operations taken to run an algorithm is directly proportional to the time a physical quantum computer would spend running the algorithm.

\begin{figure}
	\includegraphics[width=0.8\linewidth]{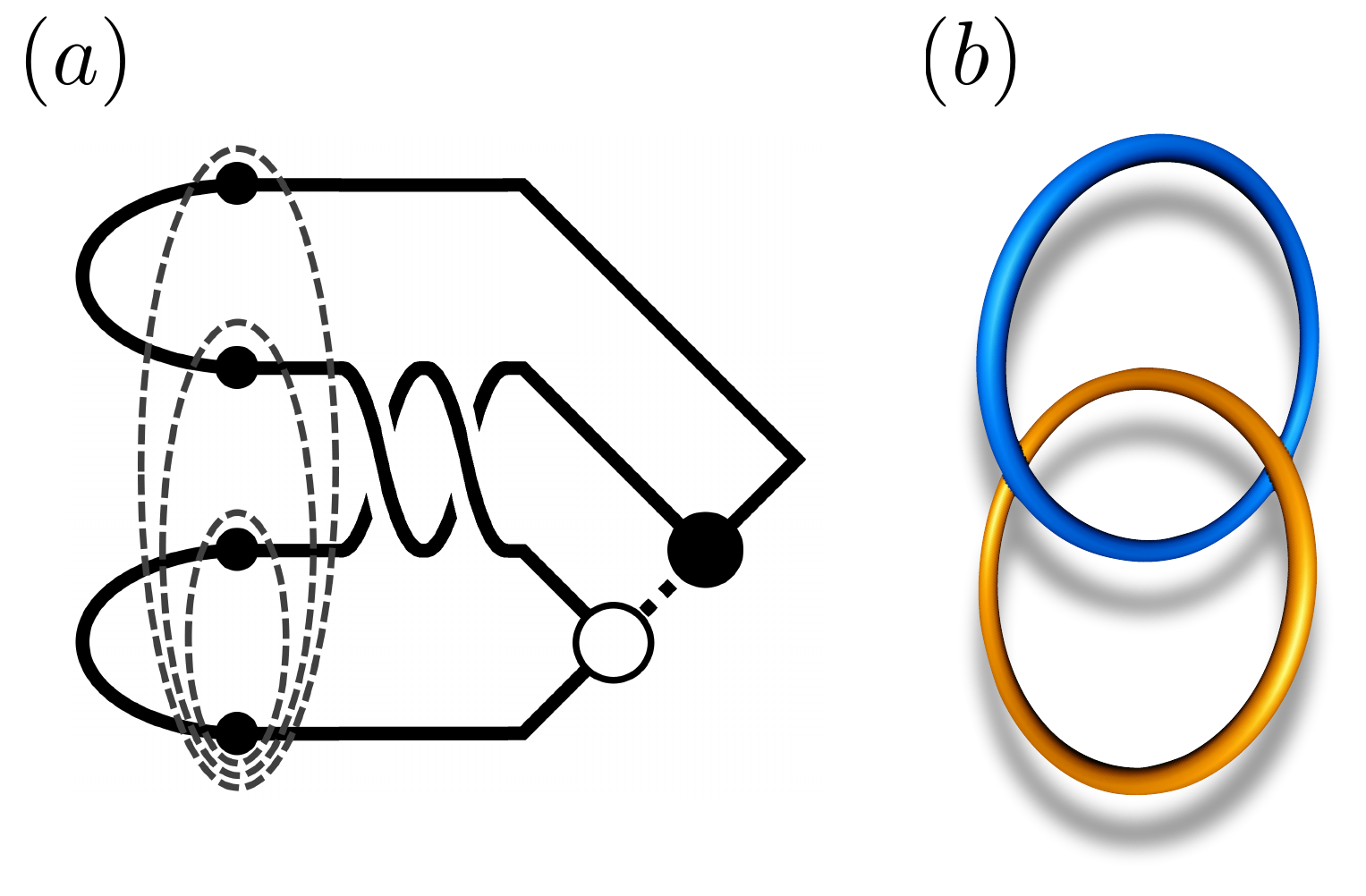}
	\caption{The braid (a) used to obtain the magnitude of the Jones polynomial for the Hopf link at $k=5$ using Fibonacci anyons. The result of the anyon fusion marked by the open circle is measured to determine whether the $\ket{0}$ state was measured. Note that this braid (the worldline knot of anyons) is topologically equivalent to the Hopf link (physical knot) shown in (b).}
	\label{Fig:exact_braid_example}
\end{figure}

For the special case of evaluating the magnitude of the Jones polynomial when $k=5$, a simpler procedure was used. For all the knots tested, they can be expressed as the plat closure of a braid with four strands, and since a single qubit contains four anyons the quantum computer was initialised to have a single qubit. A single qubit braid exactly matching the braidword corresponding to the physical knot was applied. Measurement was performed, and the process was repeated for a desired number of iterations. The ratio of the number of times $\ket{0}$ was measured to the number of iterations was taken to be the measured value of $\operatorname{Pr}(\ket{0})$, which was then applied to \eqref{eqn:exact_jones}. A representative braid for this computation is in \Figref{Fig:exact_braid_example}.

Because this was a Bernoulli experiment, where the number of $\ket{0}$ outcomes were simply counted, the confidence interval for the measured $\operatorname{Pr}(\ket{0})$ was approximated using a method for determining binomial confidence intervals, namely the Wilson score interval \cite{Binomial_Intervals}, which can be calculated as follows. We empirically measure a probability of $p$ over $N$ iterations. We want to find the $(1-\alpha)\times100$\% confidence interval (so 95\% would have $\alpha=0.05$). The corresponding quartile is
\begin{equation}
z = \operatorname{probit}(1-\alpha/2) = \sqrt{2} \operatorname{erf}^{-1}(1-\alpha), \label{eqn:quartile}
\end{equation}
where $\operatorname{erf}^{-1}$ is the inverse error function. Define the relocated center estimate
\begin{equation}
p' = \left(p+\frac{z^2}{2N}\right)/\left(1+\frac{z^2}{N}\right),
\end{equation}
and the corrected standard deviation
\begin{equation}
s' = \left(\sqrt{\frac{p(1-p)}{N}+\frac{z^2}{4N^2}}\right)/\left(1+\frac{z^2}{N}\right).
\end{equation}
The lower and upper bounds of the Wilson score interval are
\begin{align}
w^{-} = p' - z s', \\
w^{+} = p' + z s'.
\end{align}
The upper and lower bounds for the 95\% confidence intervals were calculated for $\operatorname{Pr}(\ket{0})$. Then those values were applied in \eqref{eqn:exact_jones} and reported alongside the measured magnitude of the Jones polynomial in brackets as the 95\% confidence interval for the magnitude of the Jones polynomial. We performed 1,000,000 iterations per knot.

As for the AJL algorithm, we compared this stochastic result to the exact value obtained by a classical evaluation of the AJL algorithm, and the value of the quantum computer in the limit of infinite iterations determined by directly reading the state. These two comparison figures were equal to each other, to within machine precision, as expected, so only one such number was reported for each knot.

\subsubsection{Positive Hopf Link}

The positive Hopf link is a Hopf link oriented such that it has positive writhe. Its Jones polynomial is $-t^{5/2}-t^{1/2}$, \eqref{eqn:hopf_jones}, and its writhe is $+2$.

\begin{figure}
	\includegraphics[width=\linewidth]{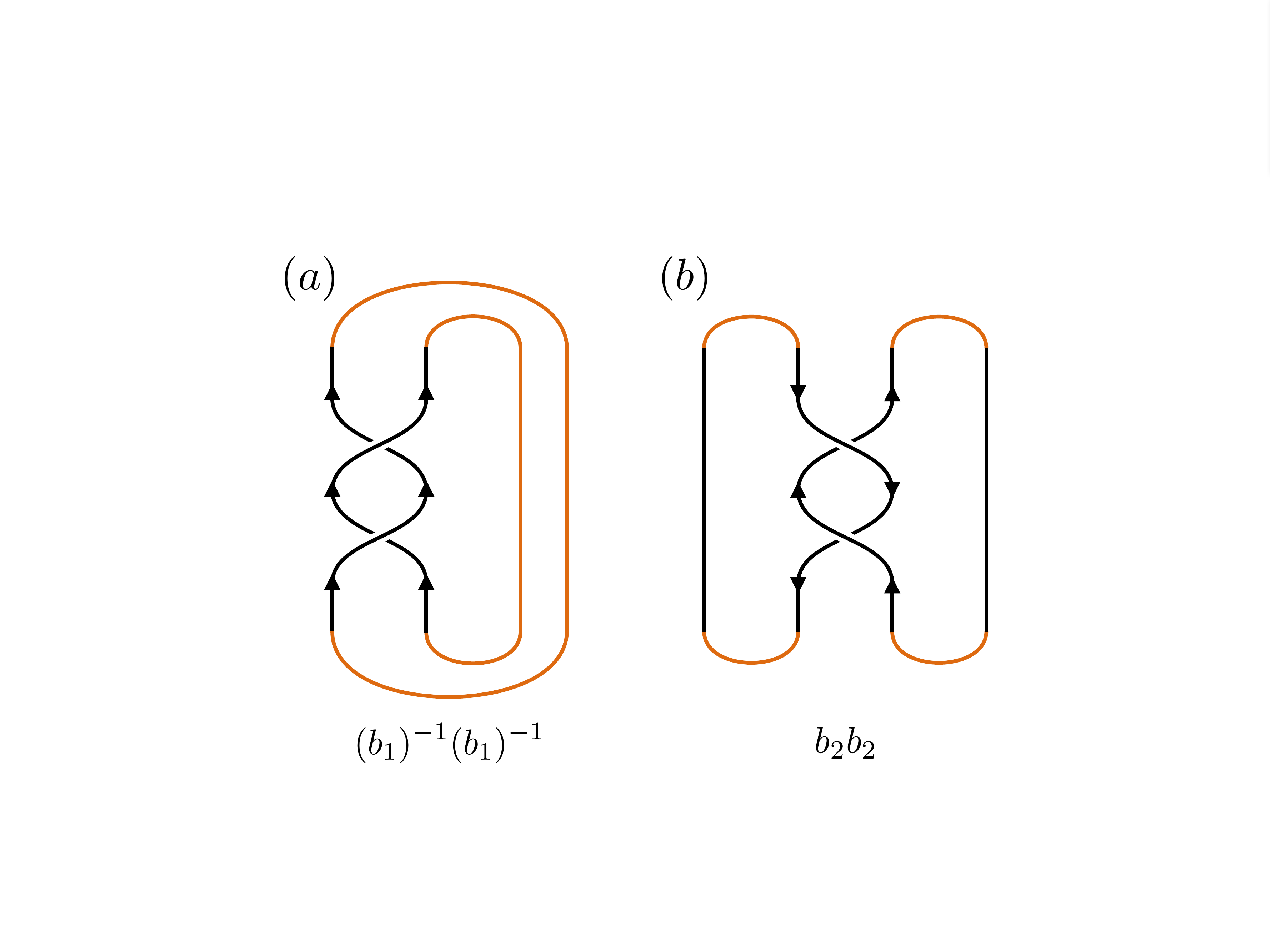}
	\caption{Oriented braid closures of the positive Hopf link, with the (a) trace and (b) plat closures.}
	\label{Fig:positive_hopf_braid_oriented}
\end{figure}

The trace and plat closures of braids corresponding to the postive Hopf link are shown in \Figref{Fig:positive_hopf_braid_oriented}. The trace closure has $n=2$ with braidword $b_1^{-1}b_1^{-1}$. The plat closure has $n=4$ with braidword $b_2 b_2$.

Because this Jones polynomial has square roots, care must be taken as to which square root is used. We investigated the points $t=e^{2\pi i/k}$. For the AJL algorithm, we have defined $t = A^{-4}$, where $A=ie^{-\pi i/2 k}$. This means that the square root of $t$ is $t^{1/2} = A^{-2} = -e^{\pi i/k}$, which is not the principal square root $e^{\pi i/k}$ but instead its negative.

Because software such as MatLab assumes the principal square root when a square root is taken, the exact solution to the Jones polynomial is plotted as $t^{5/2}+t^{1/2}$ here.

The outputs of our quantum computer simulations for the AJL algorithm for the positive Hopf link are shown in \Figref{Fig:jones_polynomials_hopf}(a),(b). Evaluating the trace closure took 93,437,024 elementary braiding operations and the plat closure took 88,240,000 elementary braiding operations.

The output of our quantum computer for determining the magnitude of the Jones polynomial at $k=5$ was 0.621 (0.617,0.626), compared to the exact value of 0.618. This took 2,000,000 elementary braiding operations.

\subsubsection{Negative Hopf Link}

The negative Hopf link is a Hopf link oriented such that it has negative writhe. Its Jones polynomial is $-t^{-5/2}-t^{-1/2}$, \eqref{eqn:hopf_jones_negative}, and its writhe is $-2$.

\begin{figure}
	\includegraphics[width=\linewidth]{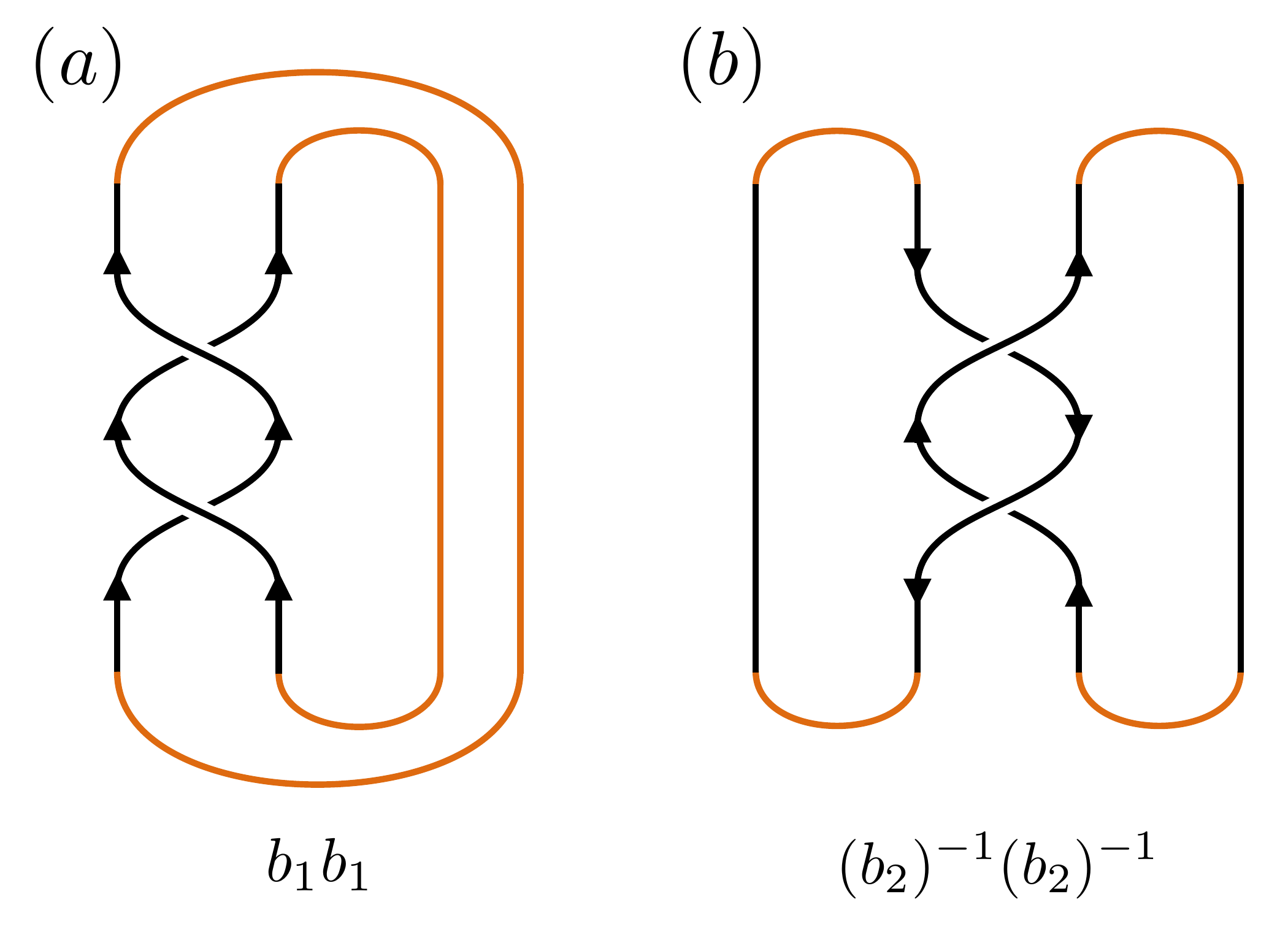}
	\caption{Oriented braid closures of the negative Hopf link, with the (a) trace and (b) plat closures.}
	\label{Fig:negative_hopf_braid_oriented}
\end{figure}

The trace and plat closures of braids corresponding to the negative Hopf link are shown in \Figref{Fig:negative_hopf_braid_oriented}. The trace closure has $n=2$ with braidword $b_1 b_1$. The plat closure has $n=4$ with braidword $b_2^{-1} b_2^{-1}$.

For reasons discussed for the positive Hopf link, because of the square roots in the Jones polynomial the plotted exact solution is actually $t^{-5/2}+t^{-1/2}$.

The outputs of our quantum computer simulations for the AJL algorithm for the negative Hopf link are shown in \Figref{Fig:jones_polynomials_hopf}(c),(d). Evaluating the trace closure took 93,446,000 elementary braiding operations and the plat closure took 88,240,000 elementary braiding operations.

The output of our quantum computer for determining the magnitude of the Jones polynomial at $k=5$ was 0.619 (0.614,0.624), compared to the exact value of 0.618. This took 2,000,000 elementary braiding operations.

\subsubsection{Left Trefoil}

The Jones polynomial of the left trefoil knot is $-t^{-4}+t^{-3}+t^{-1}$, \eqref{eqn:Ltrefoil_jones}, and its writhe is $-3$.

\begin{figure}
	\includegraphics[width=\linewidth]{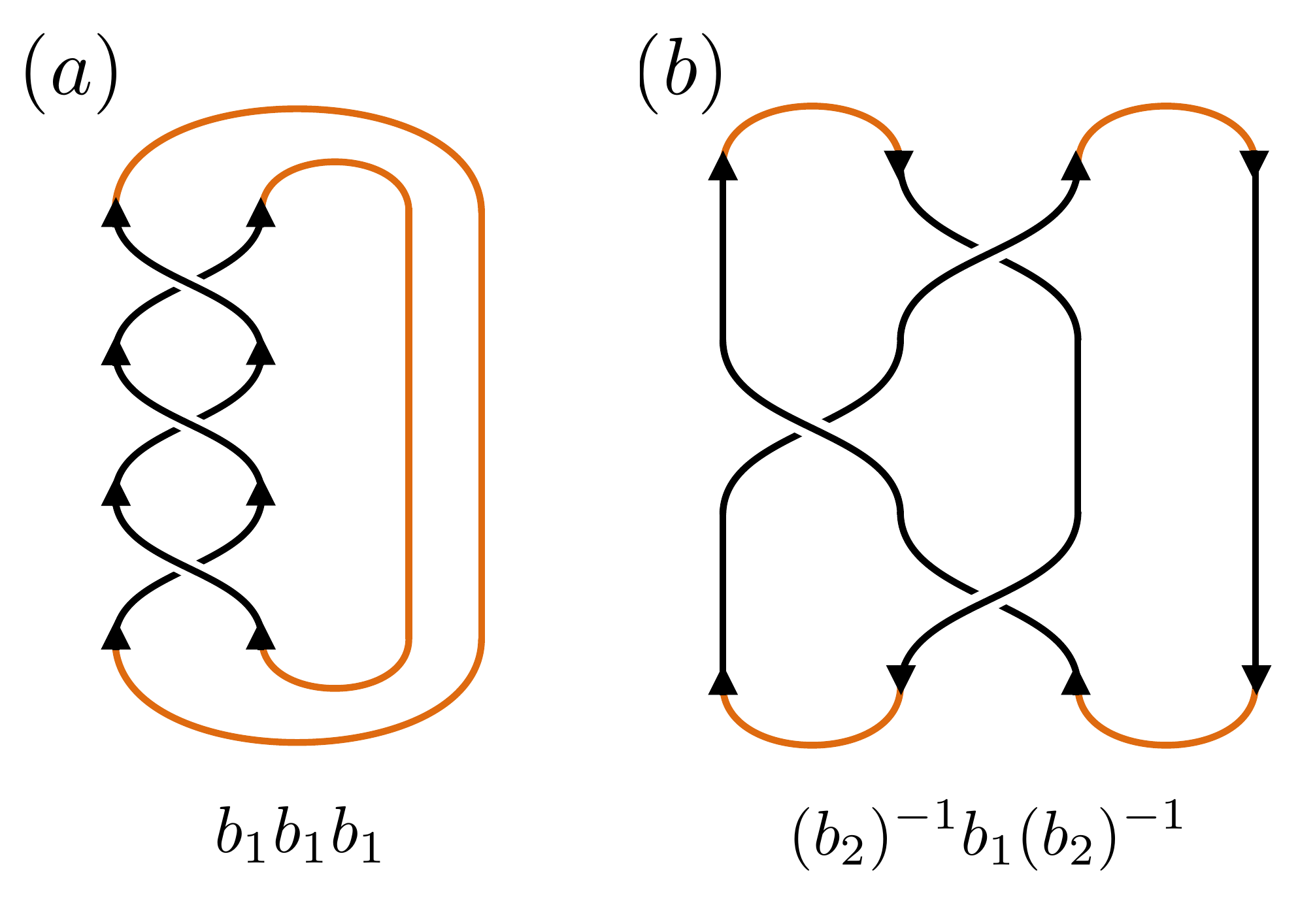}
	\caption{Oriented braid closures of the left trefoil, with the (a) trace and (b) plat closures.}
	\label{Fig:left_trefoil_braid_oriented}
\end{figure}

The trace and plat closures of braids corresponding to the left trefoil are shown in \Figref{Fig:left_trefoil_braid_oriented}. The trace closure has $n=2$ with braidword $b_1 b_1 b_1$. The plat closure has $n=4$ with braidword $b_2^{-1} b_1 b_2^{-1}$.

The outputs of our quantum computer simulations for the AJL algorithm for the left trefoil are shown in \Figref{Fig:jones_polynomials_trefeight}(a),(b). Evaluating the trace closure took 109,359,840 elementary braiding operations and the plat closure took 104,160,000 elementary braiding operations.

The output of our quantum computer for determining the magnitude of the Jones polynomial at $k=5$ was 1.543 (1.541,1.544), compared to the exact value of 1.543. This took 3,000,000 elementary braiding operations.

\subsubsection{Right Trefoil}

The Jones polynomial of the right trefoil knot is $-t^{4}+t^{3}+t$, \eqref{eqn:Rtrefoil_jones}, and its writhe is $+3$.

\begin{figure}
	\includegraphics[width=\linewidth]{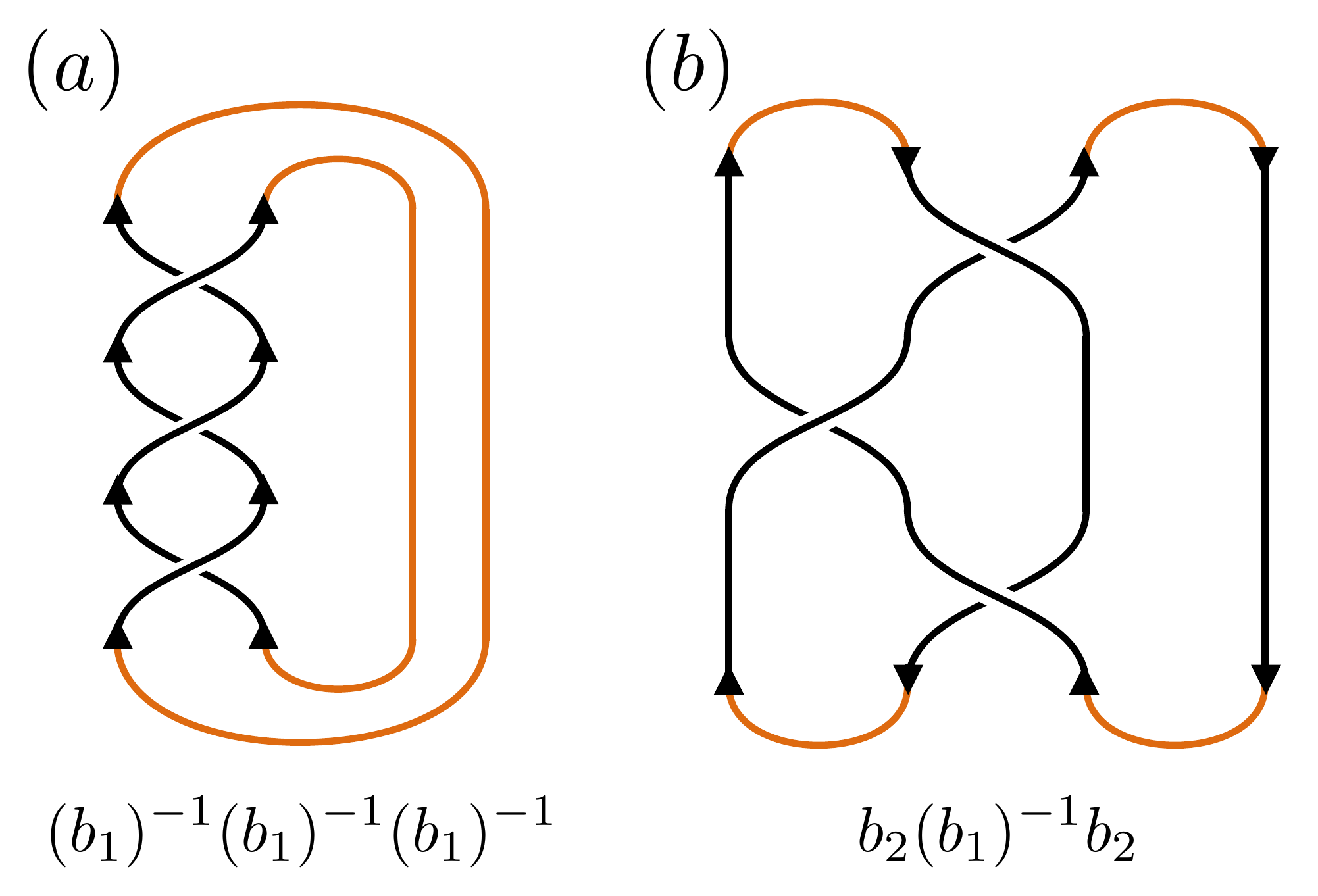}
	\caption{Oriented braid closures of the right trefoil, with the (a) trace and (b) plat closures.}
	\label{Fig:right_trefoil_braid_oriented}
\end{figure}

The trace and plat closures of braids corresponding to the right trefoil are shown in \Figref{Fig:right_trefoil_braid_oriented}. The trace closure has $n=2$ with braidword $b_1^{-1} b_1^{-1} b_1^{-1}$. The plat closure has $n=4$ with braidword $b_2 b_1^{-1} b_2$.

The outputs of our quantum computer simulations for the AJL algorithm for the right trefoil are shown in \Figref{Fig:jones_polynomials_trefeight}(c),(d). Evaluating the trace closure took 109,350,688 elementary braiding operations and the plat closure took 104,160,000 elementary braiding operations.

The output of our quantum computer for determining the magnitude of the Jones polynomial at $k=5$ was 1.543 (1.541,1.544), compared to the exact value of 1.543. This took 3,000,000 elementary braiding operations.

\subsubsection{Figure-Eight Knot}

The Jones polynomial of the figure-eight knot is $t^2 - t + 1 - t^{-1} + t^{-2}$, \eqref{eqn:eight_jones}, and its writhe is $0$.

\begin{figure}
	\includegraphics[width=\linewidth]{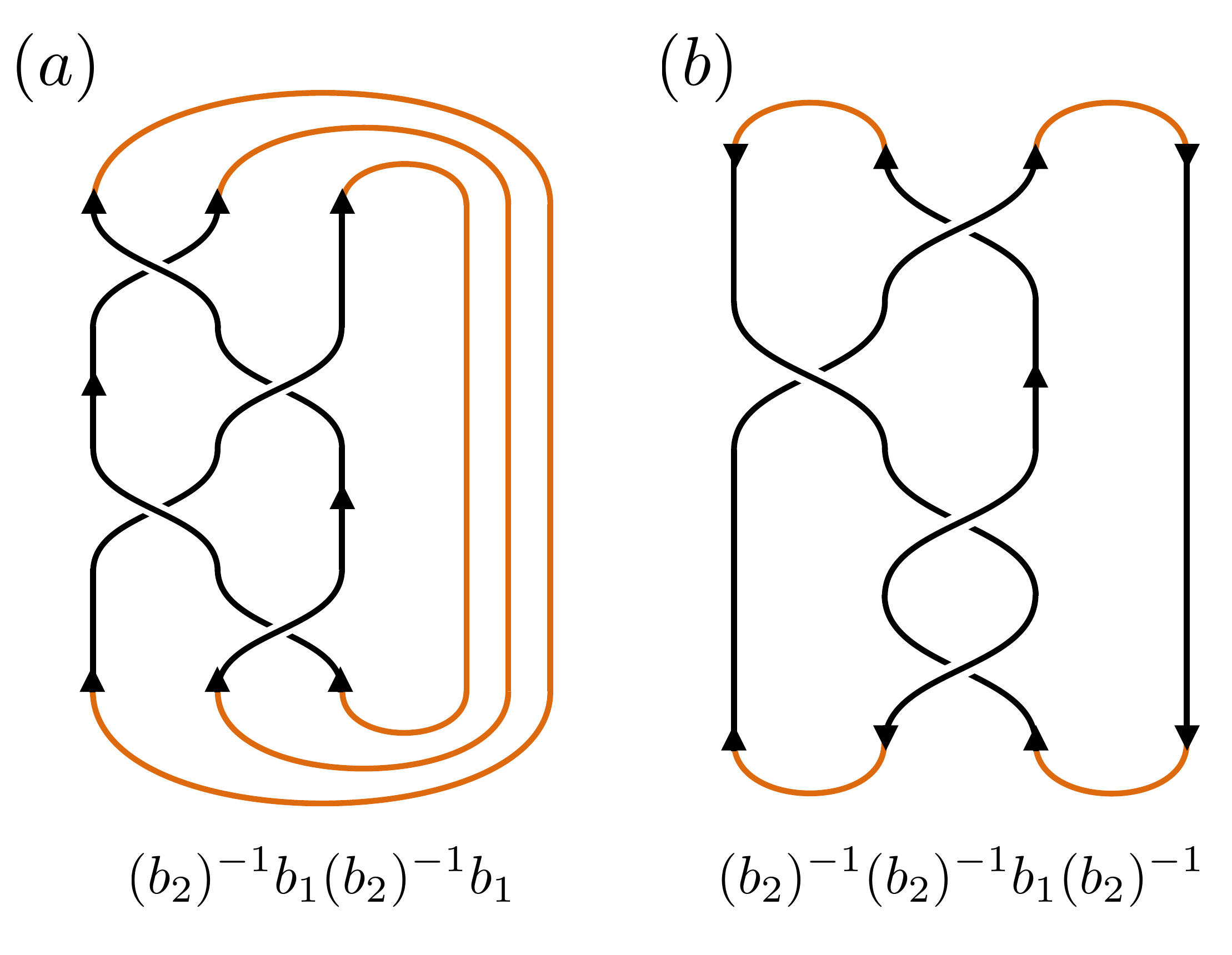}
	\caption{Oriented braid closures of the figure-eight knot, with the (a) trace and (b) plat closures.}
	\label{Fig:Figure_eight_braid_oriented}
\end{figure}

The trace and plat closures of braids corresponding to the figure-eight knot are shown in \Figref{Fig:Figure_eight_braid_oriented}. The trace closure has $n=3$ with braidword $b_2^{-1} b_1 b_2^{-1} b_1$. The plat closure has $n=4$ with braidword $b_2^{-1} b_2^{-1} b_1 b_2^{-1}$.

For the $n=2$ cases, it was possible to express the AJL matrices as a single $2\times2$ matrix. This is not possible for $n=3$ when $k\ge5$. The AJL matrices for $n=3$ are decomposed into a $2\times2$ matrix, which is evaluated normally, and a scalar component corresponding to the third element, which is evaluated by the variant of the Hadamard test for scalars, as in \Figref{Fig:hadamard_scalar}, where the controlled operation is replaced by a single qubit phase gate.

For evaluating the Markov trace, which test is used is dependent on which path is randomly selected for each iteration. Some iterations run the regular $2\times2$ matrix Hadamard test, while the other iterations run the scalar Hadamard test.

\begin{figure*}
	\includegraphics[width=.86\linewidth]{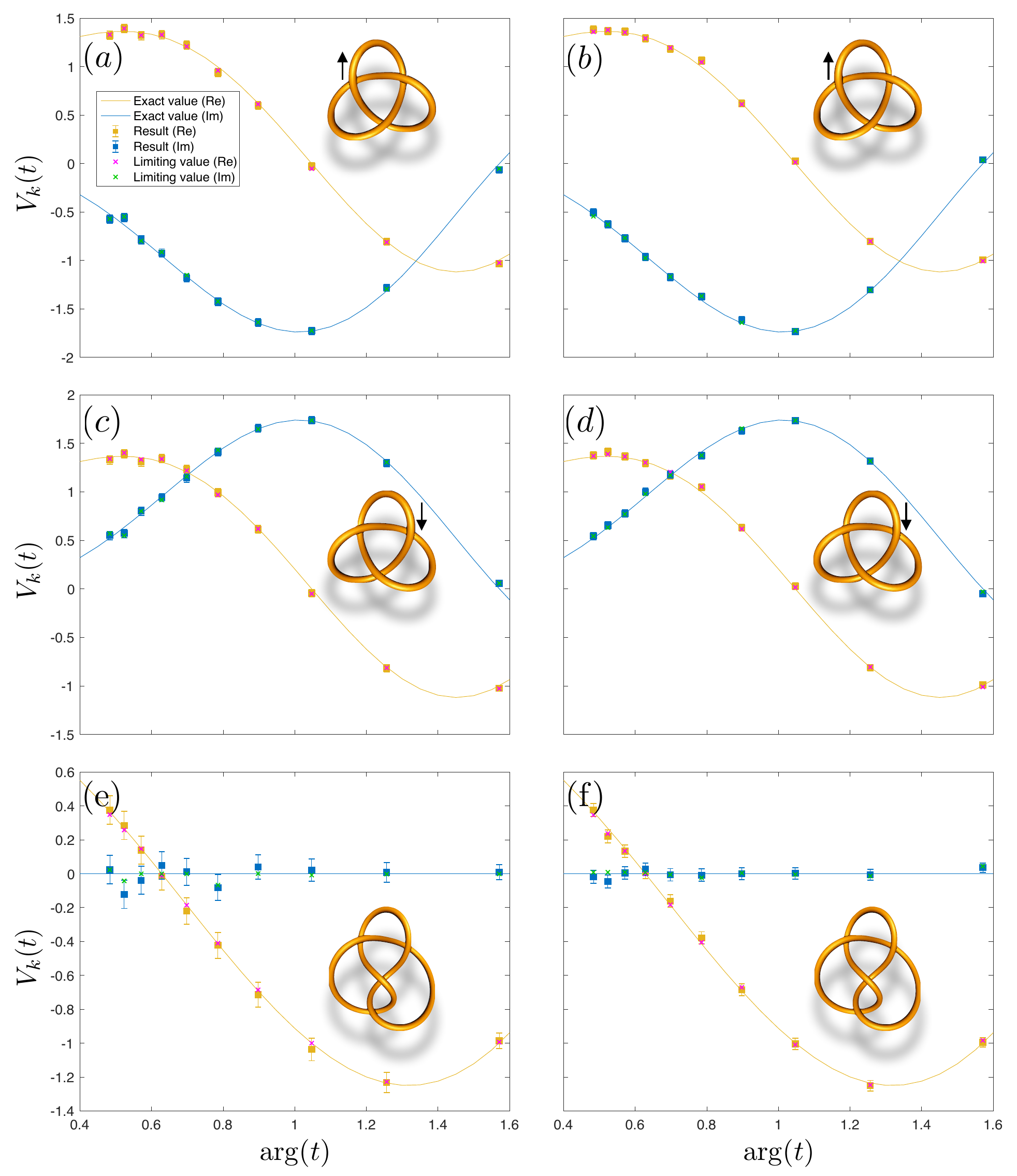}
	\caption{Results from the determination of the Jones polynomial of the left (a),(b) and right (c),(d) trefoils and the figure-eight knot (e),(f), evaluated with both the trace (a),(c),(e) and plat (b),(d),(f) closures of a braid. The horizontal axis shows the complex phase of the point $t$ at which the Jones polynomial is evaluated. Square markers with error bars are the results obtained stochastically from the Hadamard test and the crosses mark the limiting value for $N\to \infty$. Real and imaginary components are measured separately.}
	\label{Fig:jones_polynomials_trefeight}
\end{figure*}

The outputs of our quantum computer simulations for the AJL algorithm for the figure-eight knot are shown in \Figref{Fig:jones_polynomials_trefeight}(e),(f). Evaluating the trace closure took 85,217,804 elementary braiding operations and the plat closure took 120,480,000 elementary braiding operations.

The output of our quantum computer for determining the magnitude of the Jones polynomial at $k=5$ was 1.235 (1.232,1.238), compared to the exact value of 1.236. This took 4,000,000 elementary braiding operations.

\begin{figure*}
	\includegraphics[width=0.86\linewidth]{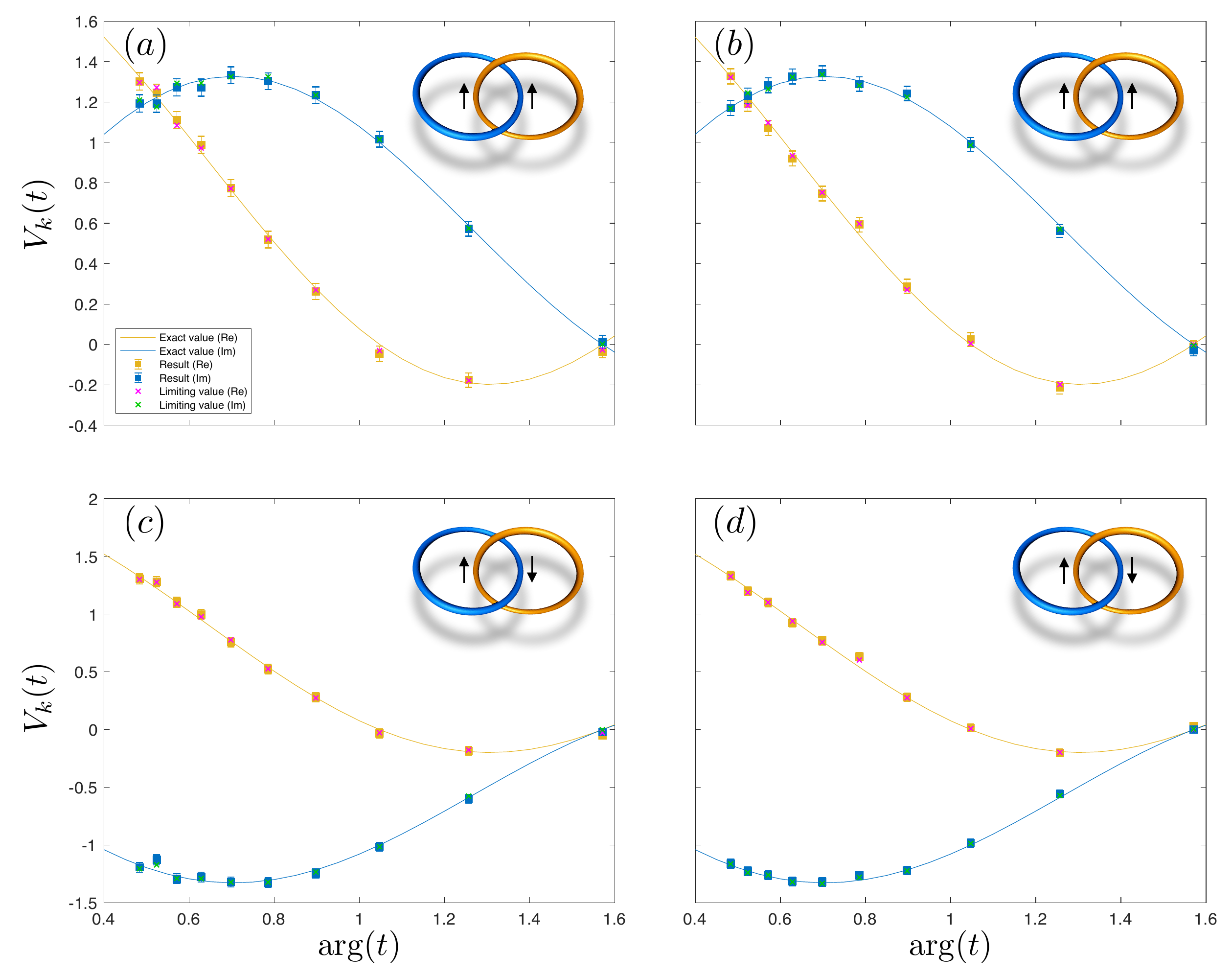}
	\caption{Results from the determination of the Jones polynomial of the positive (a),(b) and the negative (c),(d) Hopf links, evaluated with both the trace (a), (c) and plat (b),(d) closures of a braid. The horizontal axis shows the complex phase of the point $t$ at which the Jones polynomial is evaluated. Square markers with error bars are the results obtained stochastically from the Hadamard test and the crosses mark the limiting value for $N\to \infty$. Real and imaginary components are measured separately.}
	\label{Fig:jones_polynomials_hopf}
\end{figure*}

\subsection{Discussion}

Convergence of the AJL algorithm to the Jones polynomial is determined by multiple factors. The precision is determined primarily by the number of iterations of the Hadamard test, with more iterations and repeated measurements resulting in a convergence at a rate proportional to $1/\sqrt{N}$.

The precision is also influenced by the size of the knot. The output of the quantum computer is scaled by $d^{n-1}$ or $d^{\tfrac{n}{2}-1}$ for trace or plat closures respectively, where $2>d>1$ for $k>3$. The breadth of the confidence interval, or the uncertainty produced by the stochastic nature of the measurement, is also scaled by that factor, meaning knots with more strands and thus higher $n$ would have larger uncertainties in the measurement of the Jones polynomial, all else being equal. This is why the error bars for the trace closure of the figure-eight knot which has $n=3$, shown in \Figref{Fig:jones_polynomials_trefeight}(e), are approximately 0.2 wide while other knots with $n=2$ have error bars under 0.1 in width.

The accuracy of the results of the AJL algorithm, or any algorithm, performed with a topological quantum computer is dependent on the accuracy of the braids used to approximate the various unitary operators. Even with an infinite number of iterations of the AJL algorithm, where the uncertainty due to stochastic variations has been reduced to zero, the results will only be as accurate as the braids used to approximate the unitary operators. To achieve more accurate results, it is necessary to compile more accurate braids for all the operators used in the quantum algorithm. These braids have a length proportional to $\log(1/\epsilon)$, where $\epsilon$ is the error of the approximation.

Suppose you wish to use the quantum AJL algorithm to determine the Jones polynomial for a knot at a point to within $\epsilon$ of the exact value. The length of the braids scales as $\log(1/\epsilon)$, while the number of iterations for the Hadamard test scales as $1/\epsilon^2$. As such, the number of elementary operations needed to perform the AJL algorithm, and thus the time complexity, will scale as $\mathcal{O}((1/\epsilon)^2\log(1/\epsilon))$.

The logarithmic factor is intrinsic to the topological quantum computer and independent of the algorithm used. The quadratic factor is specific to the AJL algorithm, and reflects the discrete stochastic nature of measurement in the quantum computer. This quadratic factor would be expected to appear in other quantum algorithms in which the measured quantity is the probability amplitudes of the states, because the standard deviation of the binomial distribution scales as $\sqrt{N}$ so the standard deviation of a ratio derived from discrete trials would scale as $1/\sqrt{N}$. However, for other quantum algorithms where the measured quantity is a particular state which is observed with high probability, as for Shor's factorisation algorithm \cite{Shor}, then a smaller (possibly constant) factor would replace the quadratic factor.

Consider the plat closure of the positive Hopf link, as in \Figref{Fig:jones_polynomials_hopf}(b), where the points are within approximately 0.05 of the exact values. It took 88,240,000 elementary braiding operations (with 200,000 measurements and 800,000 anyon pairs created) to achieve this accuracy. Suppose the experiment is repeated with a target accuracy of $10^{-3}$ for each point. This would require approximately 500,000,000 measurements, 2,000,000,000 anyon pairs created and 509,000,000,000 elementary braiding operations. This is comparable to the amount of braiding necessary to use Fibonacci anyons to evaluate Shor's factorisation algorithm for a 128-bit number, and would likely take several hours if implemented using electrons in the quantum Hall effect \cite{Shor_TQC_Resources}.

To achieve a precision of around $10^{-16}$, which is comparable to a classical computer working with 64 bit floating point numbers, would require approximately $5\times10^{34}$ measurements and $3\times10^{38}$ elementary braiding operations, which would be unrealistic in any practical system. This quadratic scaling of the quantum AJL algorithm means that, while still efficient in the technical sense, it is impractical if very high precisions are desired.

The number of elementary braiding operations needed to compute the AJL algorithm increases linearly with the length of the braidword, and thus the number of crossings, representing the knot under investigation, since each term in the braidword adds an extra operator to the computation, which adds a constant number of elementary braiding operations. This linear complexity is a significant improvement over the exponential complexity of the more direct algorithm presented in Section \ref{sec:invariants}, albeit at the cost of only being able to evaluate the Jones polynomial at a discrete number of points.

An exception to this trend is observable in the data for the Jones polynomial calculated from the trace closure. The Hopf link, with a braidword length of 2, took 93 million elementary braiding operations to evaluate. The trefoil, with a braidword length of 3, took 109 million elementary braiding operations to evaluate. The figure-eight knot, with a braidword length of 4, took only 85 million elementary braiding operations to evaluate. The reason for this anomalously short evaluation time was because approximately one third of the iterations for the figure-eight knot computed the much shorter scalar variant of the Hadamard test, which involves no controlled operations.

For the plat closure, the Hadamard test was performed only on $2\times2$ matrices for the different knots, so a direct comparison is possible. The Hopf link took 88 million elementary braiding operations, the trefoil took 104 million, and the figure-eight knot took 120 million. Each extra braidword element added 16 million elementary braiding operations; a constant rate of change consistent with the expected linear growth.

However, this linear complexity applies to the AJL algorithm performed classically as well. Each added element in the braidword multiplies an extra matrix, and for a fixed matrix dimension matrix multiplication is a constant time operation. Simply looking at the complexity of the AJL algorithm with respect to the number of crossings may give the impression that the AJL algorithm is efficiently solvable classically.

This is only true for fixed $n$. With a small number of strands, as used in this paper, the matrices are small in size, so their multiplication is relatively efficient, so for bounded $n$ (for example, knots representable as braids with four strands or less) the time complexity of the classically performed AJL algorithm is indeed linear. This smallness is what made classical simulation in this paper feasible.

However, for arbitrary $n$, the dimension of the matrices involved grows exponentially with $n$. An arbitrary knot or link may require an arbitrarily large number of strands in its braid. In particular, if a link contains $m$ loops, its braid representation requires a minimum of $m$ strands for a trace closure or $2m$ strands for a plat closure. This exponential growth of matrix dimension makes classical computation of the matrix products inefficient.

Quantum computation allows for these matrix products to be efficiently computed as the composition of quantum gates, applied one after the other. The gates will need to span more qubits to compensate for the higher dimensions of the matrices, but since the dimension of the matrices represented by the quantum gates also grows exponentially with the number of qubits encompassed, the number of qubits needed grows only linearly with $n$. The complexity of these gates increases only as a polynomial function of $n$ \cite{Aharonov_Jones_Algorithm1}, so the time complexity of the quantum algorithm will grow only as a polynomial function of $n$, not exponentially, which allows the quantum AJL algorithm to be implemented efficiently, in the technical sense.

Although the AJL algorithm provides the Jones polynomial at only a discrete set of points, it may in principle be possible to use it to determine the full Jones polynomial using curve fitting. It is possible to determine the upper and lower bounds on the powers in the Jones polynomial of a knot without computing the Jones polynomial itself \cite{Jones_polynomial_breadth}, from which a model polynomial for curve fitting can be constructed. The additional constraints that the Jones polynomial must have integer coefficients and is equal to exactly 1 at $e^{2\pi i/3}$ would also assist in curve fitting. The disadvantage of this approach is that for any large knots the Jones polynomial would be a very high order polynomial, which could make curve fitting impractical.

Even without the whole polynomial, useful information can still be extracted from the points provided by the AJL algorithm. It can be used to distinguish knots, for if two knots have different values of the Jones polynomial at any point, then they must have different Jones polynomials, and thus be inequivalent knots. It can be used to determine whether a knot is chiral or achiral, because the Jones polynomial of achiral knots such as the figure-eight knot at the roots of unity will have no imaginary component. And if only the value of the Jones polynomial at a single point is needed, as in topological quantum field theories, then the AJL algorithm is sufficient.

The magnitude of the Jones polynomial at $t=e^{2\pi i/5}$, as evaluated with Fibonacci anyons without approximations, contains less information than the points obtained from the AJL algorithm, but is much faster to compute. Because no braids approximating quantum gates are used, since all braiding performs the desired operation exactly, the time complexity of the algorithm with respect to the desired precision scales only as $\mathcal{O}((1/\epsilon)^2)$, and each iteration requires only exactly as many elementary braiding operations as there are in the knot's braidword. This allows for higher precision in the same amount of time compared to the quantum AJL algorithm. Although there is less information provided by this single point, it can still be used to help distinguish inequivalent knots.

For indicating the error bars for the outputs of the AJL algorithm, we used the empirically obtained average data from Table~\ref{tab:hadamard_results}. While this is adequate for demonstrating approximate average behaviour of the percentiles, it does not capture the variability between trials in \Figref{Fig:hadamard_convergence_trace}. A more accurate, albeit more complicated, method of estimating the uncertainty in the outputs of the AJL algorithm would be to analytically derive them using binomial confidence intervals such as the Wilson score interval. This would produce a separate asymmetric confidence interval for the real and imaginary outputs of the Hadamard test. However, determining the behaviour of this elliptical confidence region under rotation in the complex plane, as from multiplication by $(-t^{-3/4})^{-w(K)}$, would have added an unnecessary layer of complexity to this pedagogical demonstration. The use of the empirically determined average behaviour was adequate for our purposes.

\section{Conclusions}\label{sec:conclusion}

Topological quantum computation is a rich field of study due to its potential fault tolerance imparted by the topological nature of the underlying systems. Fibonacci anyons are a model of particular interest due to their capacity to perform universal quantum computation by braiding alone. Here we have discussed the operations braiding performs, how a quantum computer can be defined for Fibonacci anyons, and how to construct arbitrary unitary operations from the braiding of Fibonacci anyons. We have also provided an explanation of the AJL algorithm for determining the Jones polynomial, and how it may be applied as a quantum algorithm.

By performing simulations in MatLab, we have explicitly demonstrated the braiding operations required to perform quantum computation with Fibonacci anyons. We have shown how topological quantum computers can be used to compute the Jones polynomial at roots of unity and discussed how to generalise these principles to generic quantum algorithms. 

We have also demonstrated a connection between Fibonacci anyons and the Jones polynomial at $t=e^{2\pi i/5}$, and similarly for Ising anyons and the Jones polynomial at $t=i$, and have conjectured that other anyon models with similar relations exist. Since such specific problems can be solved exactly for many knots and links using only four non-Abelian anyons corresponding to a single topological qubit, they serve as ideal proof of concept experiments of topological quantum computers. Specifically, using Fibonacci anyons the magnitude $|V_{\rm Hopf}(e^{i 2\pi /5})|=(1+\sqrt{5})\sqrt{\operatorname{Pr}(\ket{0}_5)}/2=(\sqrt{5}-1)/2$ of the Jones polynomial of the Hopf link at the fifth root of unity can be obtained using only two elementary physical braiding operations and thereafter measuring the annihilation probability of the anyons upon fusing them, see Figs~\ref{Fig:exact_braid_example} and \ref{Fig:fib_braid}. Similarly for Ising anyons measuring the magnitude $|V_{\rm Hopf}(e^{i2\pi /4})|=\sqrt{2}\sqrt{\operatorname{Pr}(\ket{0}_4)}=0$ corresponds to the non-trivial outcome that after only two elementary braids that return the anyons to their original positions there is a vanishing probability for the anyons to fuse back to vacuum, where as if no braiding is done this probability is one. Extending these results to arbitrary knots or links is straightforward. Similar relations are anticipated to hold for a broad variety of anyon models. 

With this practical introduction into the usage of topological quantum computers, we are looking forward to further work performed in this field. An explicit quantum circuit for performing the AJL algorithm for knots with arbitrary numbers of strands could be designed. The use of anyon braiding to compute the magnitude of the Jones polynomial at specific roots of unity could be explored further. Other quantum algorithms, such as Shor's algorithm, could be performed using topological quantum computation. Braiding statistics for anyon models besides the Fibonacci and Ising anyon models could be calculated to allow topological quantum computation with such models as well. Quantum algorithms could be performed in simulators capturing the full physics of topological quantum computers, and integrated with quantum error correcting codes. And physical experiments will continue to search for non-Abelian anyons suitable for topological quantum computation, so that once found, computation can be performed with such anyons and the results compared with theoretical models.

\begin{acknowledgements}
We acknowledge support from the Australian Research Council (ARC) via Discovery Project No. DP170104180.
\end{acknowledgements}


%

\end{document}